\documentclass[a4paper,fleqn]{cas-sc}
\usepackage[numbers]{natbib}
\usepackage{bigints}
\usepackage{mathtools} 
\DeclarePairedDelimiter\norm\lVert\rVert
\def\tsc#1{\csdef{#1}{\textsc{\lowercase{#1}}\xspace}}
\tsc{WGM}
\tsc{QE}
\tsc{EP}
\tsc{PMS}
\tsc{BEC}
\tsc{DE}

\begin{document}
\let\WriteBookmarks\relax
\def\floatpagepagefraction{1}
\def\textpagefraction{.001}
\shorttitle{The synchronized dynamics of time-varying networks}
\shortauthors{Dibakar Ghosh et~al.}

\title [mode = title]{The synchronized dynamics of time-varying networks}                      

\author[1]{Dibakar Ghosh }
\cormark[1]
\address[1]{Physics and Applied Mathematics Unit, Indian Statistical Institute, 203 B. T. Road, Kolkata-700108, India}
\cormark[1]
\fnmark[1]

\author[2,3]{Mattia Frasca }
\address[2]{Dipartimento di Ingegneria Elettrica Elettronica e Informatica, University of Catania, Catania, Italy}
\address[3]{Istituto di Analisi dei Sistemi ed Informatica “A. Ruberti”, Consiglio Nazionale delle Ricerche (IASI-CNR), 00185 Roma, Italy}
\cormark[1]

\author[4,5]{Alessandro Rizzo }
\address[4]{Department of Electronics and Telecommunications, Politecnico di Torino, Turin, Italy}
\address[5]{Office of Innovation, New York University Tandon School of Engineering, Brooklyn, NY, 11201, USA}
\cormark[1]

\author[6]{Soumen Majhi }
\address[6]{Department of Mathematics, Bar-Ilan University, Ramat-Gan 5290002, Israel}

\author[1]{Sarbendu Rakshit }

\author[7,8]{Karin Alfaro-Bittner }
\address[7]{Departamento de F\'isica, Universidad T\'ecnica Federico Santa Mar\'ia, Av. Espa\~na 1680, Casilla 110V, Valpara\'iso, Chile}
\address[8]{Universidad Rey Juan Carlos, Calle Tulip\'an s/n, 28933 M\'ostoles, Madrid, Spain}
\fnmark[2]

\author[8,9,10]{Stefano Boccaletti }
\address[9]{Moscow Institute of Physics and Technology, 9 Institutskiy per., Dolgoprudny, 141701 Moscow, Russia}
\address[10]{CNR - Institute of Complex Systems, Via Madonna del Piano 10, I-50019 Sesto Fiorentino, Italy}

\cortext[cor1]{These Authors equally contributed to the Manuscript}

\fntext[fn1]{dibakar@isical.ac.in}
\fntext[fn2]{Corresponding Author: k.alfaro.bittner@gmail.com }
\begin{abstract}
Over the past two decades, complex network theory provided the ideal framework for investigating the intimate
relationships between the topological properties characterizing the wiring of connections among a system's unitary components
and its emergent synchronized functioning.
An increased number of setups from the real world found therefore a representation in term of graphs, while more and more sophisticated methods
were developed with the aim of furnishing a realistic description of the connectivity patterns under study.
In particular, a significant number of systems in physics, biology and social science features 
a time-varying nature of the interactions among their units.
We here give a comprehensive review of the major results obtained by contemporary studies on the emergence
of synchronization in time-varying networks. In particular, two paradigmatic frameworks will be described in details. 
The first encompasses those systems where the 
time dependence of the nodes' connections is due to adaptation, external forces, or any other process affecting each of the
links of the network. The second framework, instead, corresponds to the case in which the structural evolution of the graph is  due
to the movement of the nodes, or agents, in physical spaces and to the fact that interactions may be ruled by space-dependent laws in a way that connections are
continuously switched on and off in the course of the time.
Finally, our report ends with a short discussion on promising directions and open problems for future studies.
\end{abstract}


\maketitle
\tableofcontents
\section{Introduction}

\subsection{Time-varying networks and synchronization}

Complex networks are nothing but collections of nodes (or vertices) connected by links (or edges) that form a connectivity wiring featuring specific and rich topological properties. Such mathematical objects provide actually representations for modeling many real-world, distributed, systems \cite{albert2002statistical,newman2003structure,boccaletti2006complex}. At the same time, they also yield a comprehensive framework to investigate the rise of collective behaviors emerging from the interaction of a large number of dynamical units, in systems whose functioning occurs at microscopic scales (like, for instance, metabolic and genetic networks), or in systems working at mesoscales (as the human brain), or even in systems, such as human societies or infrastructure networks, which organize at global scales.

One of the most prominent features of real-world networks is that the interactions among the components are not fixed in time, but they have an explicit temporal nature. They may be adaptive (and, therefore, their strength changes in time), or they may even be suppressed in some moments and activated in others.
Collective dynamics emerging in time-varying systems, such as consensus~\cite{olfati2007consensus,consensusBauman}, disease spreading~\cite{kohar2013emergence}, process of chemotaxis~\cite{tanaka2007general}, and many others, are found, on the other hand, in various areas and sectors, including functional brain networks~\cite{valencia2008dynamic}, power transmission systems~\cite{sachtjen2000disturbances}, person-to-person communication~\cite{onnela2007structure}, wireless sensor networks~\cite{sivrikaya2004time}, as well as many biological networks like metabolic, protein-protein interaction networks, and gene-regulatory systems~\cite{przytycka2010toward,lebre2010statistical,kukkillaya2007inferring}.

\par Properly modeling processes such as mutation in biological systems~\cite{han2004evidence}, synaptic plasticity in neuronal networks~\cite{destexhe2004plasticity}, or adaptation in social or financial market dynamics~\cite{anderson2018economy} would then require accounting for time-varying networks whose evolution may take place over characteristic time scales that are even commensurate with those of the node dynamics. In neuronal systems, existing neuronal interactions may not be active for all time, and new links may form over time. This makes the framework of temporal networks most suitable for modeling neuronal communication, as the passage of chemical molecules and electrical signals between neurons can be mimicked by a temporary edge between them that is switched on when these flows are active~\cite{holme2012temporal}. There have been few studies incorporating the time-varying character of connections in this context~\cite{valencia2008dynamic,fallani2008persistent,bassett2011dynamic}. For instance, Ref.~\cite{fallani2008persistent} studied persistent patterns of interconnection in time-varying cortical networks in humans during a simple motor act extracted from a set of high-resolution electroencephalography (EEG); Ref.~\cite{valencia2008dynamic} analyzed the dynamical evolution of functional brain networks in time-frequency space; and  Ref.~\cite{bassett2011dynamic} identified significant modular structure in human brain function during learning over a range of temporal scales.
So, a shift from a static to a dynamic neuronal interaction scenario is essential for further understanding neuronal communication.

\par On the other hand, complex networks are the prominent candidates to describe the occurrence of synchronization, in many areas of science~\cite{albert2002statistical,boccaletti2006complex}. Such a collective state was first observed by Huygens in weakly coupled clock pendula~\cite{hugenii1986parisiis}, and later described in a variety of systems, ranging from fireflies in the forest~\cite{buck1976synchronous}, animal gaits~\cite{collins1993coupled}, descriptions of the heart~\cite{torre1976theory,guevara1988phase}, improved understanding of brain seizures~\cite{netoff2002decreased}, nonlinear optics~\cite{vanwiggeren1998communication,roy1994experimental}, and meteorology~\cite{duane1999co}. In particular, it has been shown that even chaotic oscillators can synchronize under suitable coupling functions and/or network architectures~\cite{pecora1990synchronization,boccaletti2002synchronization}. Recent investigations have sought to characterize how oscillatory elements coupled according to a large scale network architecture are impacted by the choice of the interaction topology and the corresponding spectral properties of the network~\cite{lee2005synchronization,barahona2002synchronization,hong2002synchronization,jost2001spectral,jalan2003self}. A recent review on various aspects of synchronization in coupled systems and networks is available in Ref.~\cite{boccalettibook}.

\par Synchronization of populations of dynamical units has attracted researchers from diverse fields such as physics, mathematics, biology, ecology and engineering~\cite{pikovsky2003synchronization,winfree1967biological,osipov2007synchronization,arenas2008synchronization,nishikawa2010network}. Synchronization processes indeed are at the basis for the emergence of coherent global behaviors in both normal and abnormal brain functions~\cite{varela2001brainweb}, and play a crucial role in determining the food web dynamics in ecological systems~\cite{berlow1999strong}. So far, synchronized behaviors~\cite{boccaletti2002synchronization} have been mostly studied in the limit of static networks, e.g., networks whose wiring of connections is fixed with the emphasis focusing on how the complexity in the overall topology influences the propensity of the coupled units to synchronize~\cite{lago2000fast,nishikawa2003heterogeneity,pecora1998master,barahona2002synchronization}. In particular, it has been established that proper weighting procedures in static complex networks are able to greatly enhance the appearance of synchronized behavior~\cite{motter2005enhancing,hwang2005synchronization,chavez2005synchronization}. However, lately there have been efforts to incorporate a time-varying nature of the interactions leading to evolving networks. In one way such time variations represent the evolution of interactions over time. In another way they can be helpful in representing discontinuities in interactions, where the nodes interact only for limited time. Such time-varying interactions are commonly found in social networks, communication, biological systems, spread of epidemics, computer networks, world wide web, engineering systems, etc., and have been shown to result in significantly different emergent phenomena~\cite{valencia2008dynamic,fallani2008persistent,belykh2004blinking,amritkar2006synchronized,stilwell2006sufficient,tang2010small,gautreau2009microdynamics,stehle2010dynamical,mondal2008rapidly,masuda2013temporal,choudhary2014taming,ExperimentBhowmick}.

\par In various works, the case of time-varying networks has been taken into account~\cite{belykh2004blinking,stilwell2006sufficient,skufca2003communication,porfiri2006random,boccaletti2006synchronization,sorrentino2008adaptive}, among which most of the researches are prone to the fast switching case, i.e., the time scale of the variation in networks is much shorter than that of the oscillator dynamics. Systems under different time scales of network variation may exhibit very different synchronous behaviors, where the role of time scales for network synchronization could be of crucial importance. However, the case of a network evolution which takes place over characteristic time scales that commensurate with those of the node dynamics characterizes many situations, such as synaptic plasticity in neuronal networks~\cite{destexhe2004plasticity}, social or financial market adaptation dynamics~\cite{anderson2018economy,arthur2018economy,blume2005economy}, or mutation processes in biological systems~\cite{han2004evidence}. In these situations, the time scale competition between local dynamics and network evolution becomes of utmost importance and, for this reason, is thus focused in this review as well.

\par  A temporal progression of links is an inherent feature also of several natural and artificial networks~\cite{pastor2007evolution}, and a static approximation to such systems is valid only when the changes in links occur over extremely long time scales. For instance, Ref.~\cite{kataoka2003dynamical} describes the case in which a so-called function dynamics gives rise to networks that evolve according to a dynamical system. Major advances have been made in the analysis of such time-varying networks. In social interaction networks~\cite{wasserman1994social}, the social relationship or communication between pairs of individuals changes continuously, and so links are continuously created or terminated or changed over time. There are applications where the coupling strengths and even the network topology can evolve in time. A large volume of literature has focused on temporal networks whose connectivity and coupling strengths vary over time~\cite{zanette2004dynamical,stojanovski1997sporadic,ito2001spontaneous}. In this context, recent researches have also focused on the emergence of synchronization in a time-varying complex network~\cite{mondal2008rapidly,belykh2004connection,kohar2014synchronization,lu2005time,frasca2008synchronization,majhi2017synchronization,levis2017synchronization}.
\par Many earlier approaches have studied the stability of the synchronized state in time-invariant networks by linearizing the dynamical equations. The master stability function approach~\cite{pecora1998master} relates the spectral properties of the graph Laplacian of the network to synchrony of supported oscillators. It is shown that the spectrum of the graph Laplacian can be used to assess stability of the controlled system. This technique has been used in the study of synchronization stability on arbitrary network architectures~\cite{barahona2002synchronization}. Considering certain time-varying coupled network architectures, Stilwell et al. specifically built a novel concept of fast switching stability criterion~\cite{stilwell2006sufficient}. They adopted a mathematical machinery from the field of switched systems which was not typically used in the synchronization community, and extracted stability criteria. Such approaches have enabled the analysis of stability of large class of synchronized oscillators. These local stability results can only be valid for small perturbations, and here small could actually be infinitesimal in some cases. It has been shown that if the connections change quite rapidly, then the network can be essentially modeled as the aggregate of the interactions over time~\cite{belykh2004blinking,so2008synchronization}. However, if the Laplacian matrices at different times do not commute, the spread of transverse Lyapunov exponents decreases and for coupled R\"{o}ssler oscillators the stability range of the time-varying underlying network is larger than that of for the time-average underlying network~\cite{amritkar2006synchronized}. Long lasting interactions slowed down diffusion in such networks and the slow eigenmodes of the effective Laplacian matrix were shown to be affected more as compared to fast eigenmodes~\cite{masuda2013temporal}.

\subsection{Organization of the report}
In order to describe all the above mentioned cases (and many more for which the emergence of an organized dynamics is essentially ruled by time-dependent interactions among the system's units), we have decided to organize this report as follows. 

In Chapter~\ref{sec:secII} we give a short account of the fundamental definitions and mathematical concepts that will be then used along the entire paper. Far from offering an exhaustive discussion on the formal and mathematical frame for temporal networks, the section intends to cover some basic aspects and key notions of graph theory which needed proper extensions for being applicable to temporal networks, and to introduce the formalism that will be then adopted along all other sections. 

In Chapter~\ref{sec:monolayer}  we focus on the case in which synchronization emerges among {\it static} network nodes. This latter means that either the nodes are not embedded in a physical space or, if they are, their coordinates remain fixed in time. In any case, the nodes represent \emph{dynamical systems}, and the attribute static should be referred to the way in which they influence the mechanisms of link evolution. In this context, the wiring of connections may experience temporal changes of any type, as a response to specific mechanisms or processes (such as plasticity and adaptation,
or external modulations and control, or forcing of any kind), which are taking place in parallel with the node intrinsic dynamics. We will describe both the case of monolayer networks and that of multilayer and hypernetwork structures.

In Chapter~\ref{sec:secV}, we deal with the complimentary case of synchronization in systems of mobile agents.  A typical example is a temporal proximity graph describing a population of agents,
each one of them being equipped with a communication system with limited range.
During their motion, agents then communicate only if they are located at a physical distance shorter than the so-called sensing radius, so that the network of interactions is determined also by the characteristics of the agent motion.

Finally, in Chapter~\ref{sec:secVII}, we give our short conclusive viewpoint on the subject, delineate open problems,
and offer hints and ideas for perspective future studies in the field. 


\section{Graph theoretical preliminaries}
\label{sec:secII}
In this Chapter, we give some fundamental definitions and concepts that are essential for our entire review. They cover some basic
aspects of graph theory and some key notions on temporal networks.

Preliminarily, let us notice that the term `network' is often used to refer to the physical system, whereas the term `graph' to refer to the mathematical representation of a network. In the following we will not consider this distinction and use the two terms equivalently.

\subsection{Networks and time-dependent networks}

A network $\mathcal{G}$ is mathematically represented by a pair of two sets, $\mathcal{V}$ as the set of vertices/nodes, and $\mathcal{E}$ as the set of edges/links among nodes. In short, $\mathcal{G}(\mathcal{V},\mathcal{E})$. The set of edges is formed by ordered pairs of nodes, such that $(v_i,v_j)\in \mathcal{E}$ indicates the existence of a link from node $v_i \in \mathcal{V}$ (or, shortly, $i$) to node $v_j\in \mathcal{V}$ (or, shortly, $j$).

The cardinality of the set $\mathcal{V}$ is usually indicated with $N$ and represents the number of nodes in the network. Instead, the cardinality of $\mathcal{E}$ is usually indicated as $L$ and represents the number of links in the structure.

The network is said to be \emph{undirected} if $(v_i,v_j)\in \mathcal{E}$ implies that also $(v_j,v_i)\in \mathcal{E}$. Otherwise, the network is \emph{directed}.
 
The network is said \emph{simple}, if there are no multiple edges connecting the same pair of nodes and there are no self-loops, i.e., links starting and ending in the same node. Unless otherwise specified, in the following we will always refer to simple networks.

Networks may also be weighted and, in this case, they are represented by a triplet of sets, i.e., $\mathcal{G}(\mathcal{V},\mathcal{E},\mathcal{W})$, where $\mathcal{W}$ is a set of weights, one associated to each link of the network.

According to the notation introduced, the description of an unweighted network $\mathcal{G}(\mathcal{V},\mathcal{E})$ is tantamount to specifying the two sets $\mathcal{V}$ and $\mathcal{E}$, which can be equivalently done by providing a list of nodes and a list of edges.

An alternative representation makes use of the \emph{adjacency matrix} $\mathcal{A}$, that is, an $N \times N$ matrix where the generic element $\mathcal{A}_{ij}$ is given by: $\mathcal{A}_{ij}=1$ if $(v_i,v_j)\in \mathcal{E}$, and $\mathcal{A}_{ij}=0$ otherwise. Since we are considering simple graphs, then $\mathcal{A}_{ii}=0$, $\forall i=1, 2,\ldots,N$. From the adjacency matrix, it is often convenient to define the Laplacian matrix $\mathcal{L}$ with elements given by: $\mathcal{L}_{ij}=-\mathcal{A}_{ij}$ for $i \neq j$, and $\mathcal{L}_{ii}=\sum\limits_{j=1}^N\mathcal{A}_{ij}$.

As the focus of this review is on networks that evolve in time, we have to preliminarily notice that these networks have been referred to with different names such as temporal networks, time-dependent networks, time-varying networks, and so on. These networks are structures where the set of nodes, or the set of edges, or both, depend on time, i.e., $\mathcal{V}=\mathcal{V}(t)$, $\mathcal{E}=\mathcal{E}(t)$ or both $\mathcal{V}=\mathcal{V}(t)$ and $\mathcal{E}=\mathcal{E}(t)$. Here $t \in [0,t_{max}]$, where $t_{max}$ is the lifetime or observation period. Without lack of generality, we will restrict the attention to the case where only the set of edges depend on time (note, in fact, that the case of a time-dependent set of nodes can be re-conducted to a static scenario by properly defining an augmented set of nodes, including all the nodes existing at same time in the observation period of the network).

Let us, then, consider a time-dependent network $\mathcal{G}(\mathcal{V},\mathcal{E}(t))$ and discuss how one can represent it. Although there are many ways to represent a time-dependent network, two widely adopted descriptions are the \emph{event-based} and the \emph{snapshot} representation.

\begin{figure}
\centering
\includegraphics[scale=0.3]{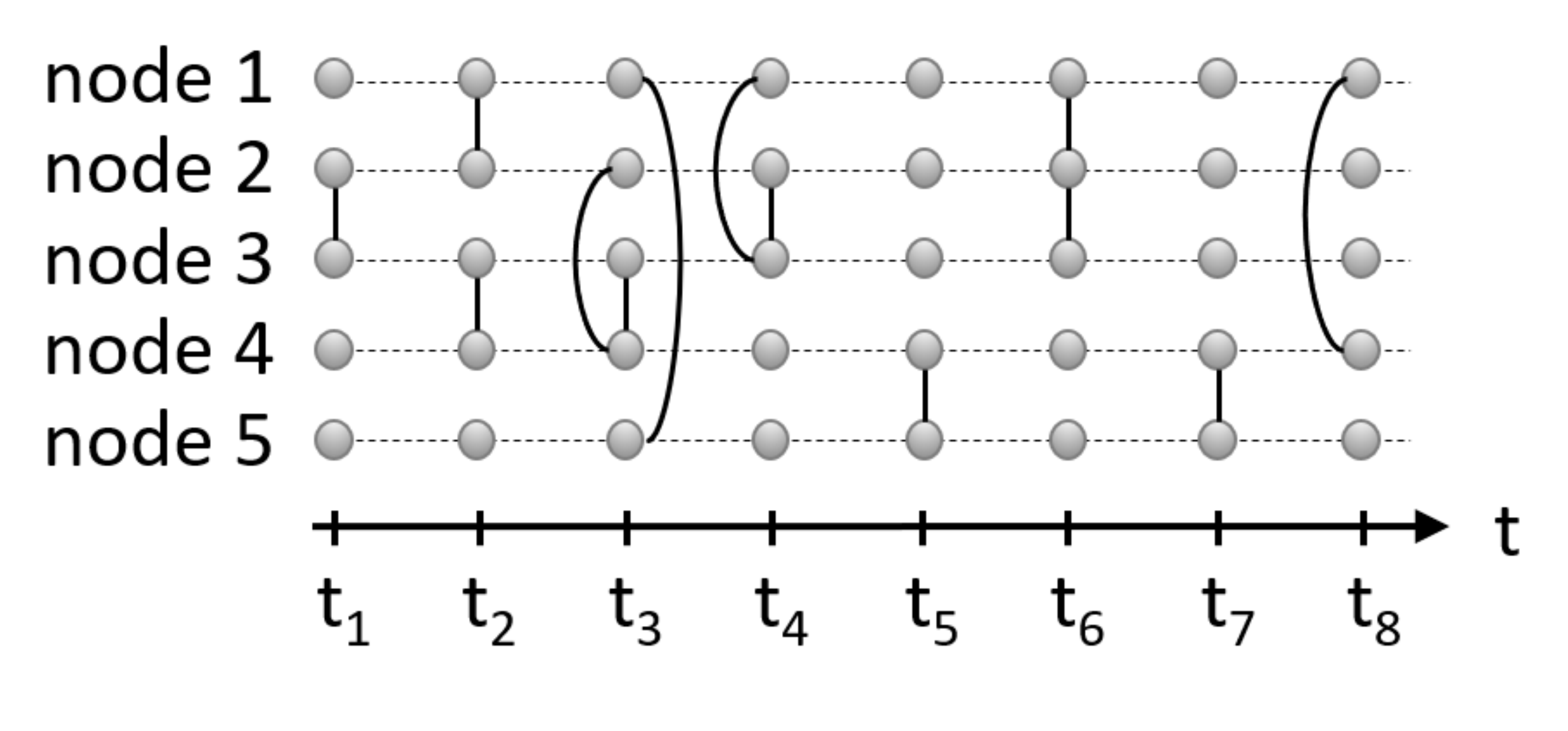}
\caption{\label{fig:secIImodelA} A temporal network corresponding to the following event-based representation: $\{(2,3,t_1)$, $(1,2,t_2)$, $(3,4,t_2)$, $(1,5,t_3)$, $(2,4,t_3)$, $(3,4,t_3)$, $(1,3,t_4)$, $(2,3,t_4)$, $(4,5,t_5)$, $(1,2,t_6)$, $(2,3,t_6)$, $(4,5,t_7)$, $(1,4,t_8)\}$.}
\end{figure}

In the event-based representation, the temporal network is described by giving the time-ordered list of events. Let us index the events with $h=1, 2,\ldots,n_{ev}$, where $n_{ev}$ is the number of events in the observation window $[0, t_{max}]$. Each event consists of an ordered triplet $(v_i^{(h)},v_j^{(h)},t_h)$, where $v_i^{(h)}$ is the starting node and $v_j^{(h)}$ the ending node of a link $(v_i^{(h)},v_j^{(h)})$ that is created at time $t_h$ and ceases to exist at time $t_{h+1}$. The temporal network is fully characterized when the list of all events is provided, i.e., $\{(v_i^{(h)},v_j^{(h)},t_h); h=1, 2,\ldots,n_{ev})\}$. The event-based representation can be viewed as an extension, to the time-varying case, of the representation of a classical network in terms of a list of edges. Here, being the links dependent on time, also the time at which the link occurs has to be specified. An example of a temporal network described by an event-based representation is shown in Fig.~\ref{fig:secIImodelA}.

\begin{figure}
\centering
\includegraphics[scale=0.3]{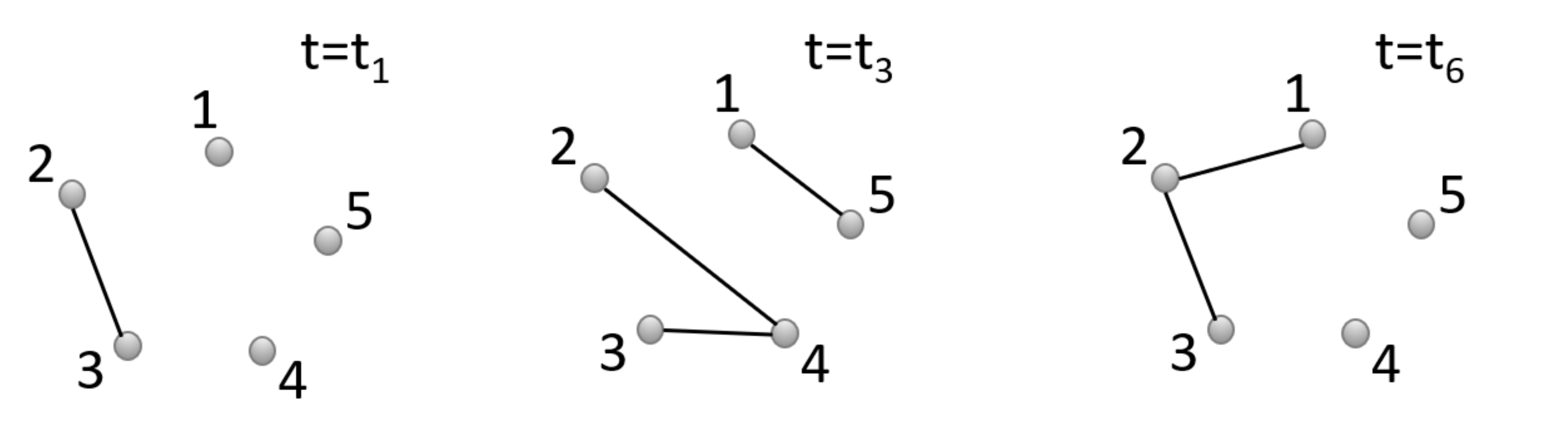}
\caption{\label{fig:secIImodelB} Three snapshots of the temporal network in Fig.~\ref{fig:secIImodelA}.}
\end{figure}

\sloppy
In the snapshot representation, the temporal network is described as a discrete-time sequence of networks: $\mathcal{G}=\{\mathcal{G}(t_1),\mathcal{G}(t_2),\ldots,\mathcal{G}(t_{max})\}$. This corresponds to give a sequence of adjacency matrices $\mathcal{A}=\mathcal{A}(t)=\{\mathcal{A}(t_1),\mathcal{A}(t_2),\ldots,\mathcal{A}(t_{max})\}$, where the generic element $\mathcal{A}_{ij}(t)$ of this time-dependent adjacency matrix is equal to one, when nodes $i$ and $j$ are connected by a link at time $t$. An example of the snapshot representation is shown in Fig.~\ref{fig:secIImodelB}. Equivalently, a set of Laplacian matrices $\mathcal{L}(t)=\{\mathcal{L}(t_1), \mathcal{L}(t_2),\ldots,\mathcal{L}(t_{max})\} $ can be used to characterize the evolution of connectivity over time.

\begin{figure}
\centering
\includegraphics[scale=0.4]{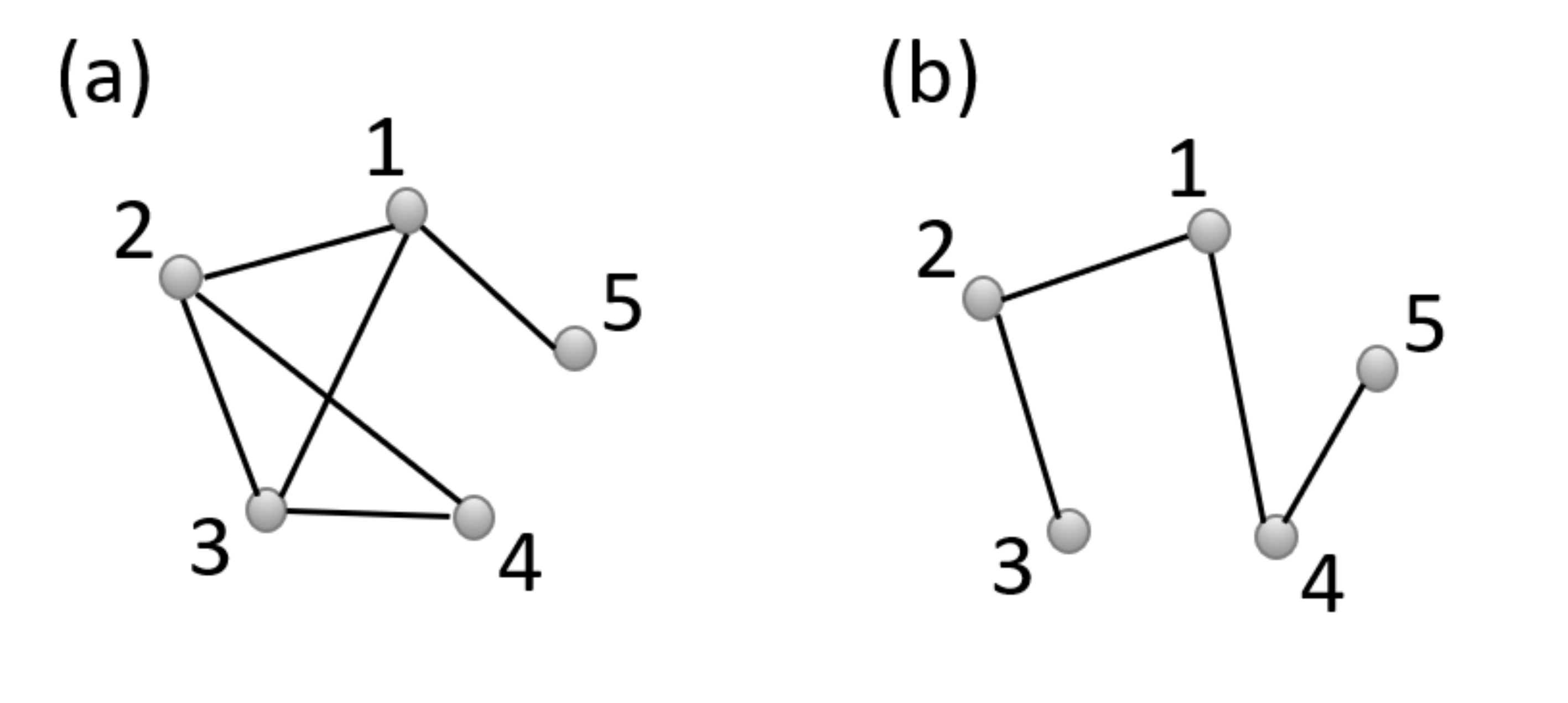}
\caption{\label{fig:secIImodelC} Coarse-grained description of the temporal network in Fig.~\ref{fig:secIImodelA} with $T_w=4$: (a) $t_1\leq t \leq t_4$; (a) $t_5\leq t \leq t_8$.}
\end{figure}

If a temporal network is described with an event-based representation and the time is discretized, then an equivalent snapshot representation can be given. Notice, however, that the snapshot representation is sometimes constructed in such a way that it provides a \emph{coarse-grained} description of the event-based representation. This is the case when a time window of length $T_w$ is defined and $\mathcal{A}_{ij}(t_h)$ is set equal to one if one or more links from $i$ to $j$ occurred at any time $t \in [t_h, t_h+T_w)$ (alternatively, if a weighted network representation is adopted, $\mathcal{A}_{ij}(t_h)$ can be set equal to the number of events linking nodes $i$ and $j$ in the time window $[t_h, t_h+T_w)$). An example of a coarse-grained description of the temporal network of Fig.~\ref{fig:secIImodelA} with $T_w=4$ is shown in Fig.~\ref{fig:secIImodelC}.

As the focus of this review is on collective dynamical behavior, in the following we will mostly resort to use the snapshot representation. In fact, this seems a more convenient representation when coupled dynamical units are dealt with, as the coupling terms are often written with reference to the adjacency matrix (or analogously to the Laplacian matrix). In the next section, we will show some examples of how the network, and so the adjacency matrix, may depend on time. We will see that different types of processes may regulate this time-dependence, thus generating diverse models of temporal networks that are particularly useful in the study of collective dynamical behaviors emerging from time-varying interaction mechanisms.

\subsection{Models of time-dependent networks}

In the previous section, we have seen that in the snapshot representation a temporal network may be described in terms of a time-varying adjacency matrix $\mathcal{A}=\mathcal{A}(t)$. However, links may depend on time in different ways. Although they can be a direct function of time, it is often the case that they are function of another process that in turn explicitly depends on time. Notice that different models of temporal networks are obtained as soon as the time-dependence of $\mathcal{A}$ is better characterized. With a slight abuse of notation, in such cases we will indicate $\mathcal{A}=\mathcal{A}(\mathbf{\sigma}(t))$ where $\mathbf{\sigma}(t)$ is a generic function of time used to model how $\mathcal{A}$ evolves in time. To make a few examples, in the case of blinking networks $\mathcal{A}=\mathcal{A}(\mathbf{s}(t))$, where $\mathbf{s}(t)$ is a sequence of binary values generated by a stochastic process; for activity driven networks, instead, $\mathcal{A}=\mathcal{A}(\mathbf{a}(t))$, where $\mathbf{a}(t)$ is the stack vector of the activities at the nodes; in temporal proximity graphs and in metapopulation models $\mathcal{A}=\mathcal{A}(\mathbf{y}(t))$, where $\mathbf{y}$ is the stack vector of the agent positions in a continuous space in the case of temporal proximity graphs or discrete in the case of metapopulation models, where the agents are located in the nodes of a backbone network modeling interconnections among subpopulations. Another possibility is represented by adaptive networks, where the evolution of links depends on the node states. In this case, the coefficients of $\mathcal{A}$ are dynamical variables themselves, whose specific rule of evolution contributes to determine the behavior of the entire system. All these models of temporal networks will be briefly described in the following.

\subsubsection{Blinking networks}

We begin with a general model of temporal networks where the presence of a link between two nodes of the network at each time $t$ is the result of a stochastic process. In this model, known as \emph{blinking network}, at each time $t$ each pair of nodes has a given probability to be connected (in general, different from pair to pair), and, as time evolves, links will be randomly activated and de-activated.

The model is formally described as follows~\cite{hasler2013dynamics,hasler2013dynamics2}. Let $\mathbf{s}(t):[0,\infty)\in \{0,1\}^L$ be a piecewise constant function that takes the constant binary vector value $s^h=(s_1^h,\ldots,s_L^h)$ for $t \in [t_{h-1},t_h)$. The sequence of binary vectors $s^h$ with $h=1,2,\ldots$ represents the \emph{switching sequence}, determining at each time $t_h$ which links exist in the network, or equivalently, which links are \emph{switched on}. In particular, when $s_i^h=1$ then link $i \in \mathcal{E}(t_h)$, whereas when $s_i^h=0$ then
link $i \notin \mathcal{E}(t_h)$. The switching sequences are considered instances of a stochastic process $S^h$, $h=1,2,\ldots$, where the random vectors $S^h$ are independent and identically distributed, with $p_s$ being the probability that $S^h$ assumes the value $s \in \{0,1 \}^L$. According to this model, hence, the temporal network has a number $L$ of links which are independently switched on and off. Consequently, the adjacency matrix can be expressed as function of the switching sequence, i.e., $\mathcal{A}=\mathcal{A}(\mathbf{s}(t))$. Synchronization in blinking networks has been studied with particular attention to the behavior that is obtained when the time scale of the stochastic process is faster than that of the dynamics of the units, a scenario which clarifies the origin of the name `blinking networks'~\cite{hasler2013dynamics,hasler2013dynamics2}.

This model includes several specific cases which are particularly important in the study of synchronization. The first case we discuss is the \emph{on-off coupling}. In this case, the temporal network has $L$ links which are simultaneously turned on or off. In the blinking network model this corresponds to have only two binary vector values with non-zero probability, i.e., $s_{on}=[1,...,1]^T$ with probability $p_{on}$, and $s_{off}=[0,...,0]^T$ with probability $p_{off}=1-p_{on}$. In this specific case, the adjacency matrix can be rewritten as $\mathcal{A}(\mathbf{s}(t))=\varepsilon(t)\mathcal{A}_{b}$, where $\mathcal{A}_{b}$ represents the backbone structure whose $L$ links are simultaneously turned on or off, and $\varepsilon(t)=\{0,1\}$ with $\varepsilon(t)=0$ with probability $p_{on}$ and $\varepsilon(t)=1$ with probability $p_{off}$.

The second interesting case is a temporal network where at each time $t$ links are generated according to the Erd\"os-R\'enyi model for (static) random networks. Each snapshot of the temporal network, therefore, represents a structure that can be modeled as an Erd\"os-R\'enyi network. This model corresponds to a blinking network where $L=N(N-1)/2$ (all possible pairs are considered) and the probability that a component of the vector $S^h$ is equal to one, indicated as $\Pi(S_i^h=1)$, is independent from $i$ and $h$, i.e., $\Pi(S_i^h=1)=p$, where $p$ is the wiring probability of the Erd\"os-R\'enyi model.

Finally, we notice that the blinking model incorporates also temporal networks where there exists a fixed backbone that does not change in time, whereas the other links depend on time. In this case, $\Pi(S_i^h=1)=1$, $\forall i \in \mathcal{E}_b$, where $\mathcal{E}_b \subset \mathcal{E}$ is the set of the edges of the backbone structure.

\subsubsection{Activity driven networks}
\label{sec:adnmodel}
Activity driven networks (ADNs) have been introduced in 2012 by Perra et al.~\cite{perra2012activity} to model the concurrent evolution, at comparable time scales, of link formation and node dynamics. This regime is typical of many real-world phenomena. For example, in the contemporary, hyper-connected world, humans can travel around the world at the same speed at which epidemics incubate and spread, favoring the inception of pandemics~\cite{y2018charting}.

Activity driven models constitute a parsimonious alternative to connectivity-driven models, where interactions are based on spatial proximity \cite{porfiri2006random,frasca2006dynamical,frasca2008synchronization,frasca2012spatial} and a motion and interaction model should be coupled to the node dynamics model. In ADNs, in fact,  the temporal interaction pattern of each node is dictated by a single parameter, called \em activity potential \em (sometimes shortened in just \em activity\em), which quantifies the attitude of the node to generate connections over time. More precisely, the activity potential of a node is defined as the ratio between the number of interactions made by the node and the total number of interactions occurring in the network during a given time interval. In its original incarnations, activity potentials of nodes are constant and are obtained as independent and identically distributed realizations of stochastic variables. The study of several temporal networks representative of socio-technical systems of different nature have led to conjecture that nodes potentials  are often distributed as power laws~\cite{perra2012activity}.

Considering a network of $N$ nodes labeled with $i=(1, 2,\ldots, n)$, where each node has constant activity potential $a_i$, the link formation dynamics of an ADN over a discrete time interval $ [t, t+\Delta t]$ is exemplified as follows:
\begin{enumerate}
    \item the ADN is initialized to be fully disconnected;
    \item each node $i$ becomes \em active \em with probability $a_i \Delta t$. An active node forms $m$ undirected links (with $m$ constant integer) with nodes drawn at random from a uniform distribution;
    \item
    all the links are removed, discrete time is updated, and the process is resumed from the first step.
\end{enumerate}

While the instantaneous instances of an ADN consists of mostly disconnected networks, the union of all such instances has a degree distribution that asymptotically scales like the distribution of the activity potentials of nodes, for large network sizes and times~\cite{perra2012activity,perra2014PRL}.

\begin{figure}[t]
\begin{center}
\includegraphics[scale=0.4]{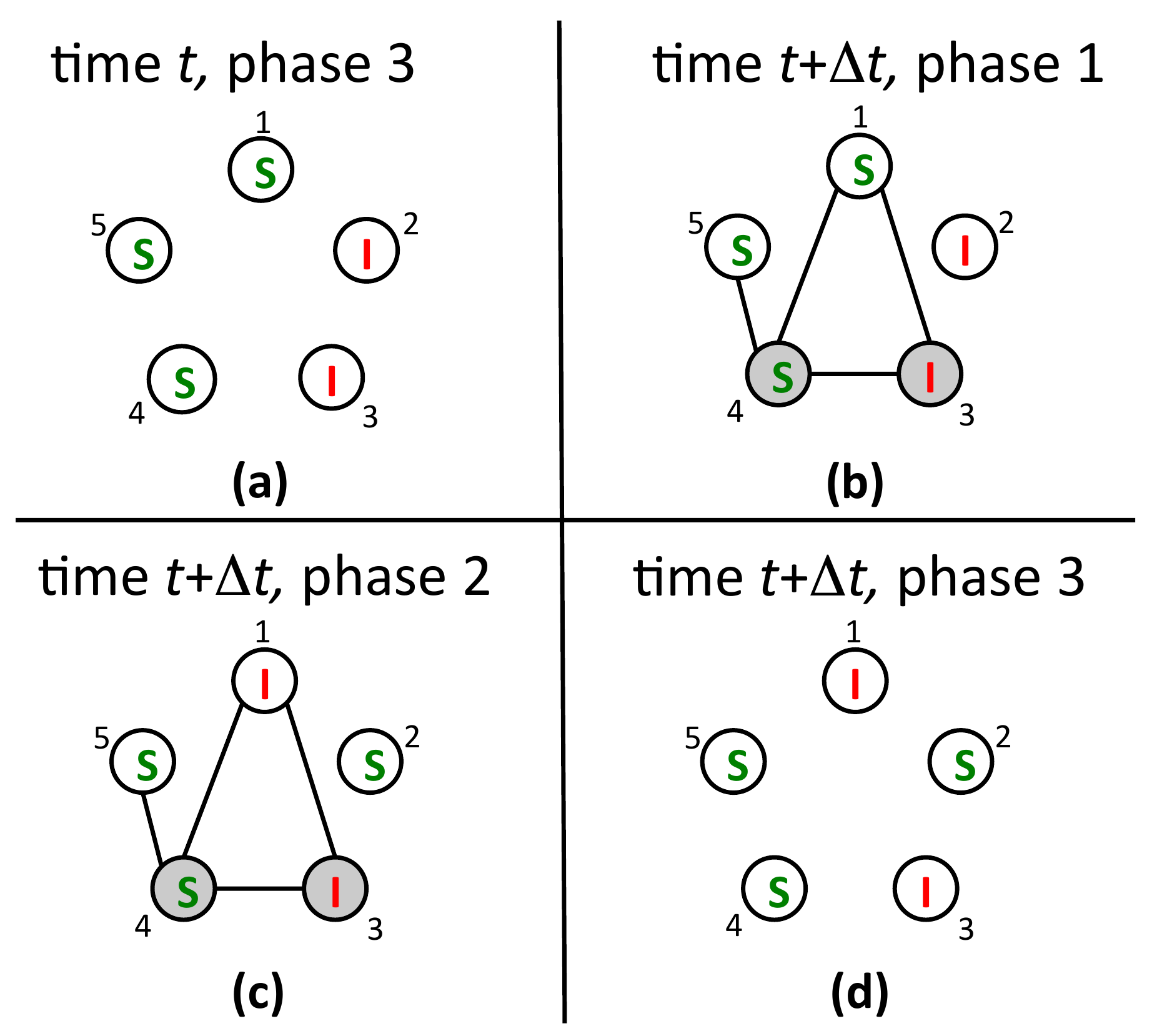}
   \end{center}
    \caption{An SIS epidemic model evolving on an ADN with $N=5$ nodes and $m=2$ links per active node. Nodes' health states are encircled, and active nodes are shaded. (a) At the last phase of time $t$, the ADN is disconnected and nodes 2 and 3 are infected. Between $t$ and $t+\Delta t$: (b) nodes 3 and 4 become active and contact nodes 4 and 1, and 5 and 1, respectively; (c) the epidemic process evolves, so that node 3 infects node 1, nodes 2, 4, and 5 remain in the susceptible state, and node 2 recovers; and (d) time $\Delta t$ has elapsed and all the network edges are removed before a new time increment is initiated. \\{\it{Source:}} Reprinted figure from Ref.~\cite{rizzo2016network} \textcopyright~2016, with permission from Elsevier.}
   \label{fig:ADN-model}
\end{figure}

Besides the link formation process, a dynamical system located at each node can coevolve at a comparable time scale, either according to a continuous-time formalism, or to a discrete-time one. Such a dynamics can be dictated, moreover, by either a deterministic or a stochastic process. While the dynamical systems located on nodes can independently evolve in time, the temporary formation of a link provides to connected process the opportunity to communicate, for instance exchanging information, or through diffusion mechanisms.  Hence, such a communication may occur between steps 2 and 3 of the link formation process described above. To highlight such co-evolution mechanism without introducing a cumbersome mathematical formalism, we illustrate the evolution of an ADN with $N=5$ nodes where a Susceptible-Infected-Susceptible epidemic process (thus, evolving in discrete time through a stochastic process) coevolves with the network formation. In this example, illustrated in Fig.~\ref{fig:ADN-model}, nodes can transit from a susceptible state to an infected one with a certain probability only upon the formation of a contact with an infected node, whereas an infected node can reverse its state to susceptible autonomously, with a certain probability, without the occurrence of a contact.

The ADN formalism has been extended along different directions in the last decade. Behavioral traits in node dynamics, depending on global observables of the process unfolding upon the network have been considered to account for the effects of individual behavior on epidemic spreading~\cite{rizzo2014effect}. This model has then been used to tackle real epidemiological models, such as the Ebola Virus Disease in West Africa~\cite{rizzo2016network}, or the COVID-19 diffusion in Italy~\cite{parino2021modelling} and in the U.S.~\cite{behring2021adherence}. Further studies concentrated on a continuous-time, discrete-distribution approach to deal with the possibility of analytical treatment and avoid the confounds related with the choice of the sampling time~\cite{zino2016continuous,zino2017analytical} and the introduction of memory effects toward the study of self-excitement dynamics~\cite{zino2018modeling,zino2020analysis}. Further theoretical studies deal with the analysis of consensus~\cite{mistry2015committed,hasanyan2020leader,ogura2018distributed,zino2019consensus,mistry2015committed}, collective motion \cite{hasanyan2020analysis}, diffusion of innovation~\cite{rizzo2016innovation}, voter models~\cite{moinet2018generalized}, and synchronization of chaotic dynamics~\cite{buscarino2018synchronization}.

\subsubsection{Temporal proximity graphs}
\label{SecIISec:temporalproxgraph}
Another interesting class of synthetic temporal network models derives from the generalization to the time-varying case of spatial graphs. Spatial/geometric graphs are characterized by nodes located in a space equipped with a metric. An example is the random geometric graph that is obtained by considering nodes distributed uniformly in a random way in a two-dimensional Euclidean space and connecting two nodes if their relative distance is smaller than a given threshold, usually defined as the \emph{interaction/neighborhood radius}. Once nodes are allowed to move according to a motion law, then, the resulting graph is a time-varying one.

This approach is clearly general and can start from other types of spatial graphs, so that it yields a class of synthetic temporal networks, rather than a single model. Each member of this class of models is fully specified once the metric used in the space, the rule to set the links among the nodes, and the motion law are given. In this context, nodes/vertices of the network are often referred to as \emph{agents} to represent their capability to move in the space. Typical applications of these temporal networks arise in the context of transportation and mobility systems, mobile phone networks, multi-agent robotics, and epidemic modeling.

To stem our discussion to a specific example, which has been proved to be particularly effective in the study of synchronization (as we will see in Sec.~\ref{sec:chaoticoscimovingagents}), we now describe, in some more detail, temporal proximity graphs.

\begin{figure}
	\centering
	\includegraphics[scale=0.3]{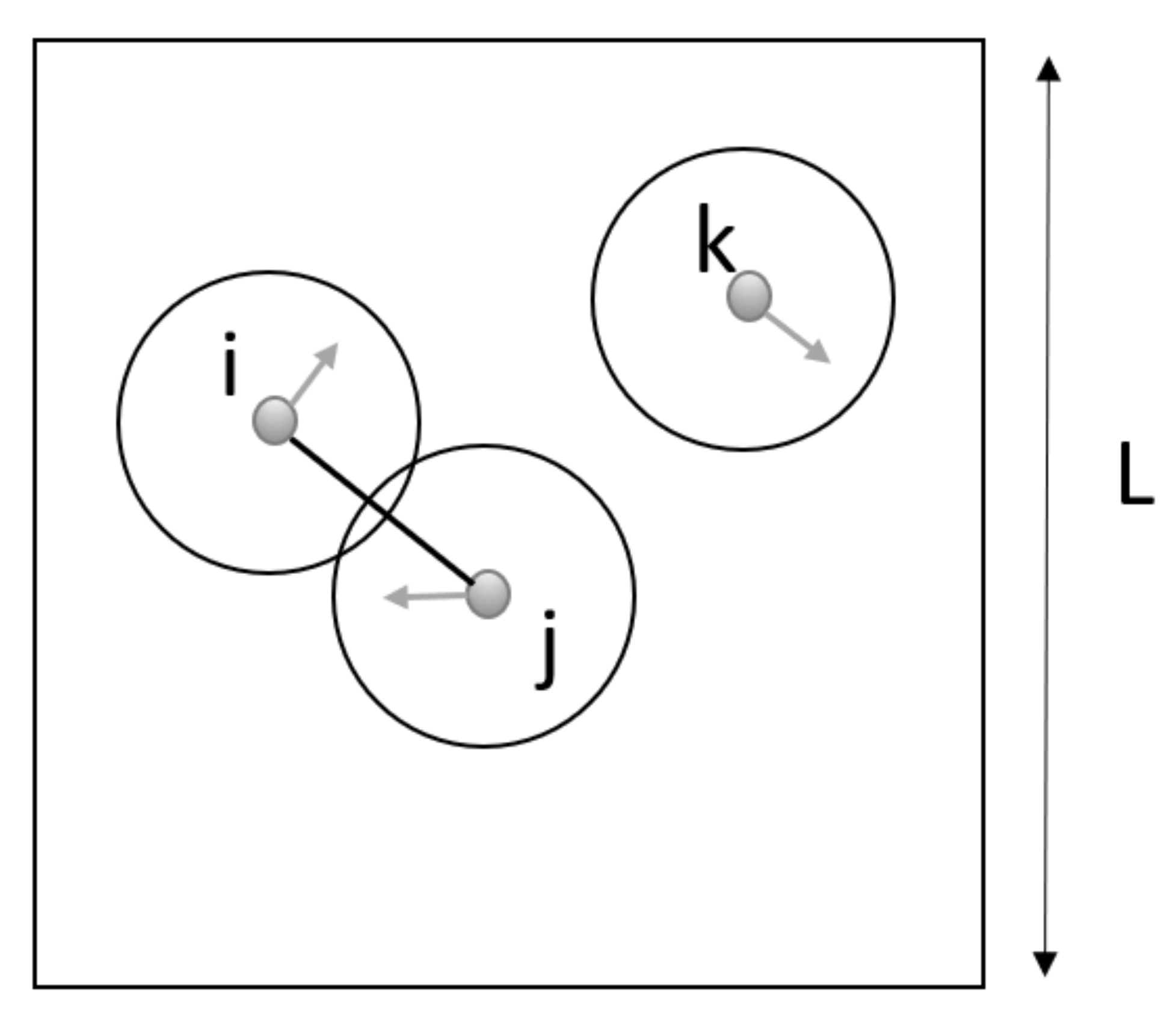}
	\caption{\label{fig:secIImetric} An example of a snapshot of a temporal proximity graph with $N=3$ agents moving in a square plane of size $L$. The black circles denote the sensing/interaction areas of the agents.
	}
\end{figure}

Let us consider $N$ agents located in a space, without lack of generality assumed to be a two-dimensional $L\times L$ square with periodic boundary conditions. We indicate the position of agent $i$ at time $t$ as $\mathbf{y}_i(t)=[y_{i,1}(t),y_{i,2}(t)]^T$ and consider two agents to be connected at time $t$ if their distance is less than the interaction radius $R$ (Fig.~\ref{fig:secIImetric}). At each time $t$, hence, the generic $ij$ element of the adjacency matrix $\mathcal{A}_{ij}(t)$ encoding the network connectivity is defined by:

\begin{equation}
\mathcal{A}_{ij}(t)=1 \Leftrightarrow \| \mathbf{y}_i(t) - \mathbf{y}_j(t) \| \leq R.
\end{equation}

Typically, the Euclidean norm is used, such that, in the two-dimensional case under analysis, we have:

\begin{equation}
\mathcal{A}_{ij}(t)=1 \Leftrightarrow \sqrt{\left (y_{i,1}(t)-y_{j,1}(t) \right )^2+\left ( y_{i,2}(t)-y_{j,2}(t) \right )^2 } \leq R.
\end{equation}

This model accounts for agents equipped with limited sensing/communication capabilities, as typically occurs in multi-agent systems~\cite{mesbahi2010graph}. One can think to agents as disks of radius $R$ that communicate, and hence \emph{interact}, with each other, only if they overlap at some time. For this reason, the model is also known as (temporal) $R-$disk proximity graph.

To fully characterize the model, the motion law needs to be also specified. The selection of the motion law is strictly related to the application considered. For instance, if the multi-agent system needs to be coordinated into a formation, then a specific control law to rule agent motion is required. Here, we consider a generic setup where agents move independently of the other units of the system as random walkers that eventually perform long distance jumps.

In more detail, one defines a jump probability $p_j \in [0,1]$, and considers the following rules for updating the agent positions when they perform a jump or when they do not. Let us start with the second case. In this case, the $i$-th agent moves with velocity $\mathbf{v}_i(t)$, having constant modulus $v$ and variable heading $\theta_i$, such that $\mathbf{v}_i(t)=ve^{j\theta_i(t)}$. The heading of agent $i$ is updated randomly at discrete time steps $t_k$, with $t_k-t_{k-1}=\tau_M$, such that one has

\begin{equation}
\left. \begin{array}{l} \mathbf{y}_i(t_k+\tau_M)=\mathbf{y}_i(t_k)+\tau_M\mathbf{v}_i(t_k),
\\
\theta_i(t_k)=\eta_i(t_k),
\end{array}
\right. \label{eq:randomwalkers}
\end{equation}

\noindent where $\eta_i(t_k)$ is an independent random variable chosen at each time $t_k$ with uniform probability in the interval $[-\pi,\pi]$. As periodic boundary conditions are assumed, the agent positions are considered modulus $L$.

In addition, the model includes the possibility that agents perform long-distance jumps with probability $p_j$. When such an event occurs, then the position of the agent $i$ performing this jump is updated as follows:

\begin{equation}
\left. \begin{array}{l} \mathbf{y}_i(t_k+\tau_M)=\xi_i(t_k),
\end{array}
\right. \label{eq:randomjumpers}
\end{equation}

\noindent where $\xi_i(t_k)$ is a vector of two independent random variable chosen at each time $t_k$ with uniform probability in the interval $[0,L]$. In summary, each agent with probability $1-p_j$ moves as a random walker performing a step of length $v\tau_M$ in an arbitrary direction, and with probability $p_j$ it jumps in an arbitrary position of the plane, thus performing a step of random length.

The jumping probability $p_j$ represents a control parameter for the system that tunes the type of motion and, consequently, the properties of the temporal network. If $p_j$ is set to one, then each snapshot of the temporal network exactly corresponds to an instance of the random geometric graph in the given plane. On the contrary, for $p_j \neq 1$, the model exhibits correlations among agent positions at successive time steps which, in turn, generate correlations among the links of temporal network snapshots.

The effect of $p_j$ on the structure of the temporal network can be unveiled by using different ways to extend classical measures for static networks to time-varying scenarios, as considered in Refs.~\cite{tang2010small} and~\cite{buscarino2008disease}. Quite interestingly, these different approaches point towards the same result, a small-world behavior emerging as a function of the parameter $p_j$. More in general, the problem of defining the proper measures to capture the topological characteristics of a temporal network is far from trivial and often open to many different solutions. This aspect, although very important, goes beyond the purpose of this report and we refer the reader to the book~\cite{masuda2020guide} for a detailed discussion on the topic.

\begin{figure}
\centering
\includegraphics[scale=0.4]{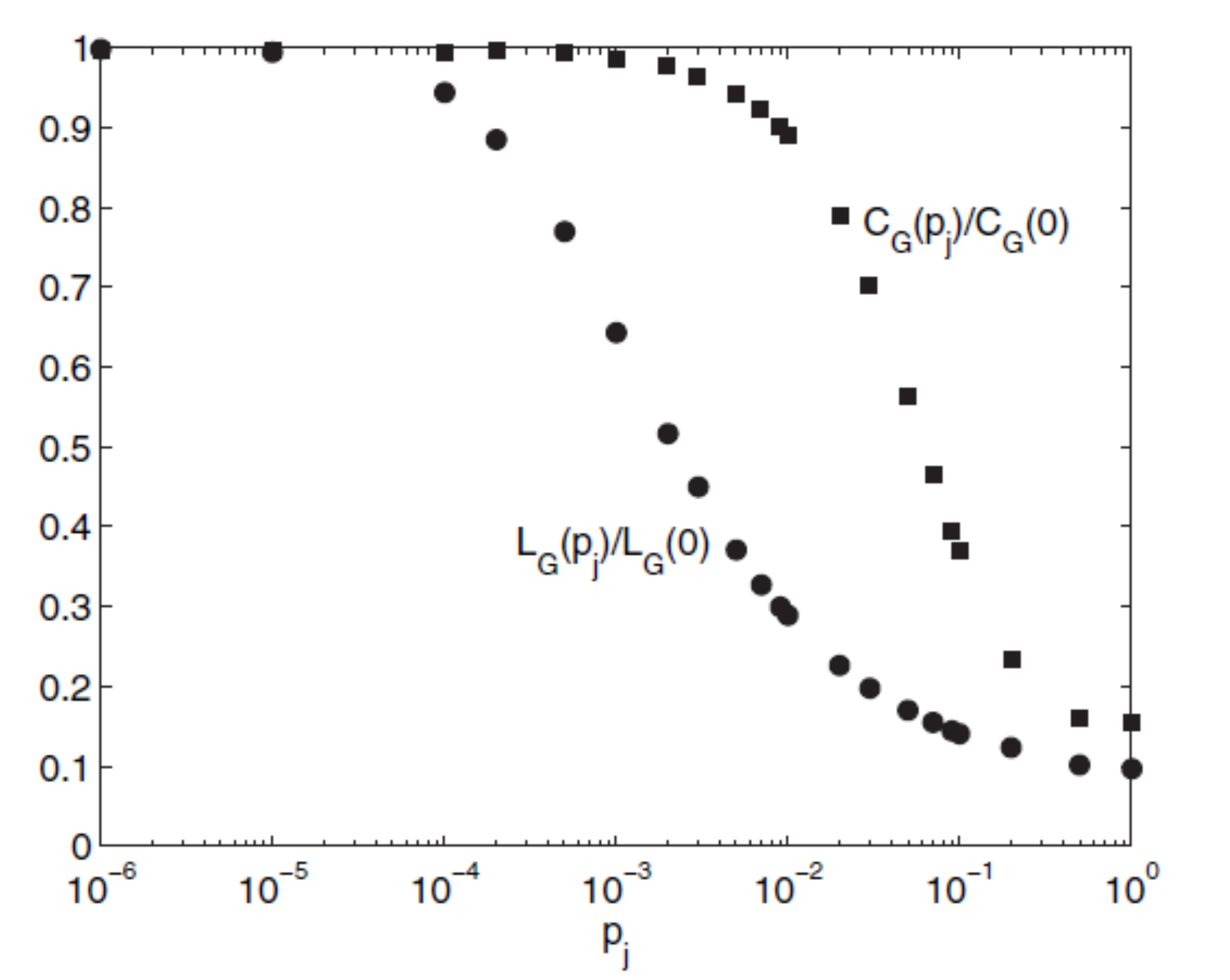}
\caption{\label{fig:secIIsmallworldTV} The small-world property in temporal proximity graphs may be tuned by the jump probability $p_j$, a characterizing parameter of the agent motion law. Results are obtained for a system with $N=1000$ agents, moving with speed modulus $v=0.1$, and distributed in a planar space with agent density $\rho=\frac{N}{L^2}=1$.
\\{\it Source:} Reprinted figure with permission from Ref.~\cite{buscarino2008disease}.
}
\end{figure}

To hallmark the small-world effect in the temporal proximity graph, following the approach presented in Ref.~\cite{buscarino2008disease}, let us define a new time-varying adjacency matrix, indicated as $\mathcal{A}_\tau(t)$, averaging the properties of the snapshot in a moving time window of length $\tau$. The generic $ij$ element of this matrix is defined as follows:
$\mathcal{A}_\tau(t)=1$ if $\mathcal{A}_\tau=1$, at least for one $t'$ with $t'=t,t-1,\ldots,t-\tau+1$, otherwise $\mathcal{A}_\tau(t)=0$. Then, the time-average value of the characteristic path length, $L_G$, and the clustering coefficient, $C_G$, as a function of $p_j$ are calculated for $\mathcal{A}_\tau(t)$. These parameters decrease with increasing $p_j$ (Fig.~\ref{fig:secIIsmallworldTV}), but quite interestingly there is an interval of values of $p_j$ where the clustering coefficient is still large and the characteristic path length is already small, indicating the presence of a small-world effect. The same conclusion is obtained by inspecting other network measures as done in Ref.~\cite{tang2010small}, where the Authors consider the average topological overlap of the neighbor set of a node between two successive graphs in the sequence, which provides a measure of the local connectivity of the nodes, and the characteristic
temporal path length, which, on the contrary, provides an indication of the average distance between two nodes. These two parameters again decrease with $p_j$ with a region clearly indicating the presence of a small-world property in the temporal network.

\subsubsection{Temporal spatial graphs with nearest neighbors interactions}

Another particularly relevant example of temporal spatial graphs is obtained when a topological rather than metric criterion is used to set the connections among agents at each time $t$. In this scenario, a fixed number $M$ of agents is defined and, at each time $t$, each agent links with exactly other $M$ agents, selecting, in particular, the $M$ agents at the closest distance from its actual position, that are sometimes named the \emph{topological
neighbors}.

\begin{figure}
\centering
\includegraphics[scale=0.3]{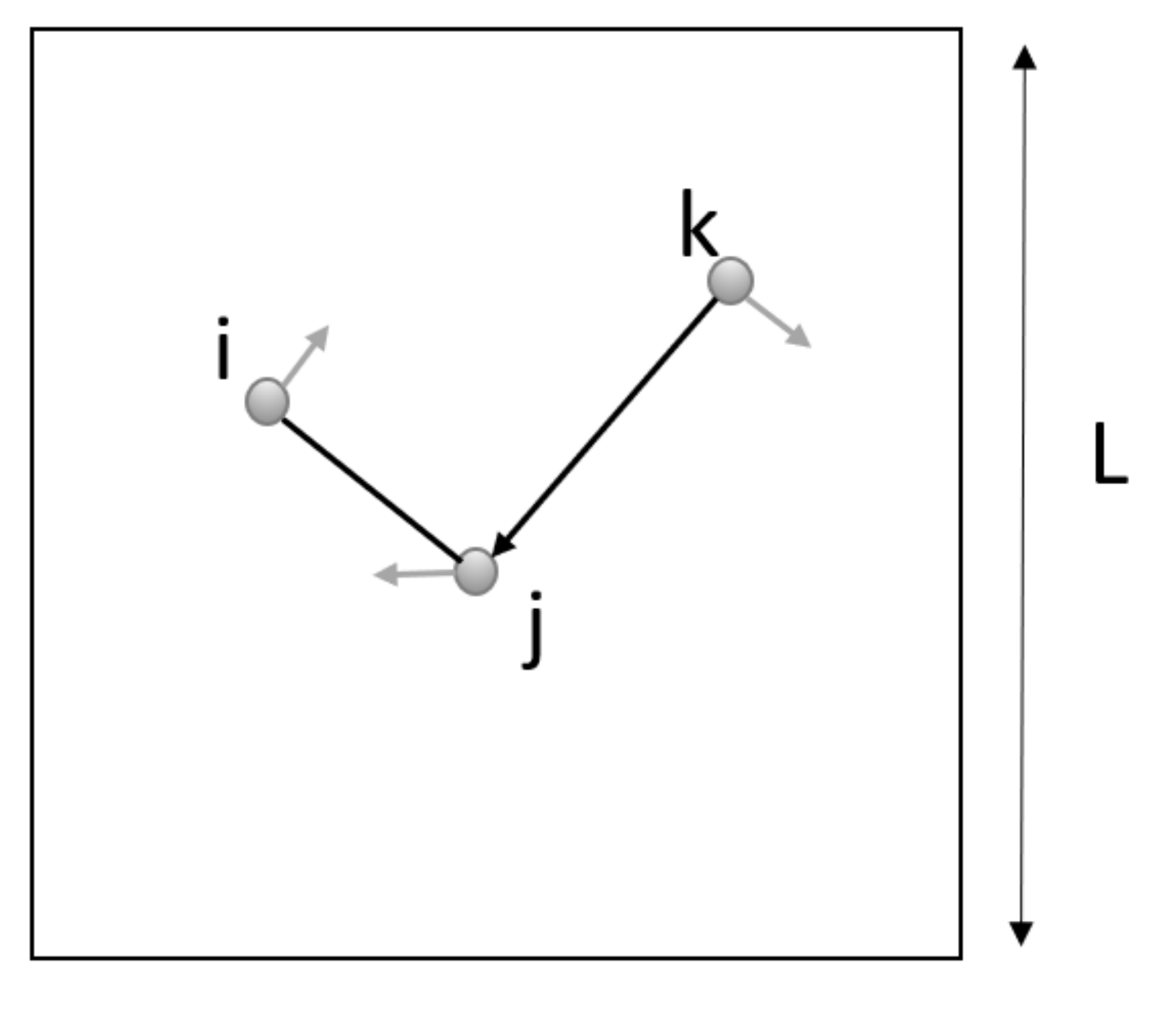}
\caption{\label{fig:secIItopological} An example of a snapshot of a temporal spatial graph with $N=3$ agents where connections are set according to a topological criterion.
}
\end{figure}

This rule produces adjacency matrices $\mathcal{A}(t)$ that, in the general case, are not symmetric, whereas those in the temporal proximity graphs discussed in the previous section are symmetric. An example with $M=1$ is illustrated in Fig.~\ref{fig:secIItopological}, which shows that agent $k$ is the nearest neighbor of agent $j$, but not vice-versa. In fact, the nearest neighbor of agent $j$ is agent $i$, such that $i$ and $j$ are connected by a bidirectional link, whereas $j$ and $k$ by a directed one. With the topological interaction rule, each agent always interacts with a fixed number of other units, regardless
of their geometric distances. In multi-agent systems, this has the benefit of generating snapshots that are always connected, but, in general, requires a more powerful communication system to reach units at an arbitrary distance.

Quite interestingly, in biological systems, both examples of use of the metric and the topological interaction rules are found. Models of animal flocking, which often have been the source of inspiration for engineers to design control protocols for multi-agent systems~\cite{olfati2006flocking,zhu2012flocking,wang2016synchronization}, have, in fact, shown that the metric interaction scheme is likely to be adopted in collective motions by groups with an high density such as in locust swarming~\cite{buhl2006disorder} and fish schooling~\cite{makris2009critical}, whereas the topological interaction scheme when the flock-mates can be perceived even at a large distance such as in bird flocks~\cite{ballerini2008interaction}.

\subsubsection{Face-to-face networks}

\begin{figure}
\centerline{\includegraphics[scale=0.4]{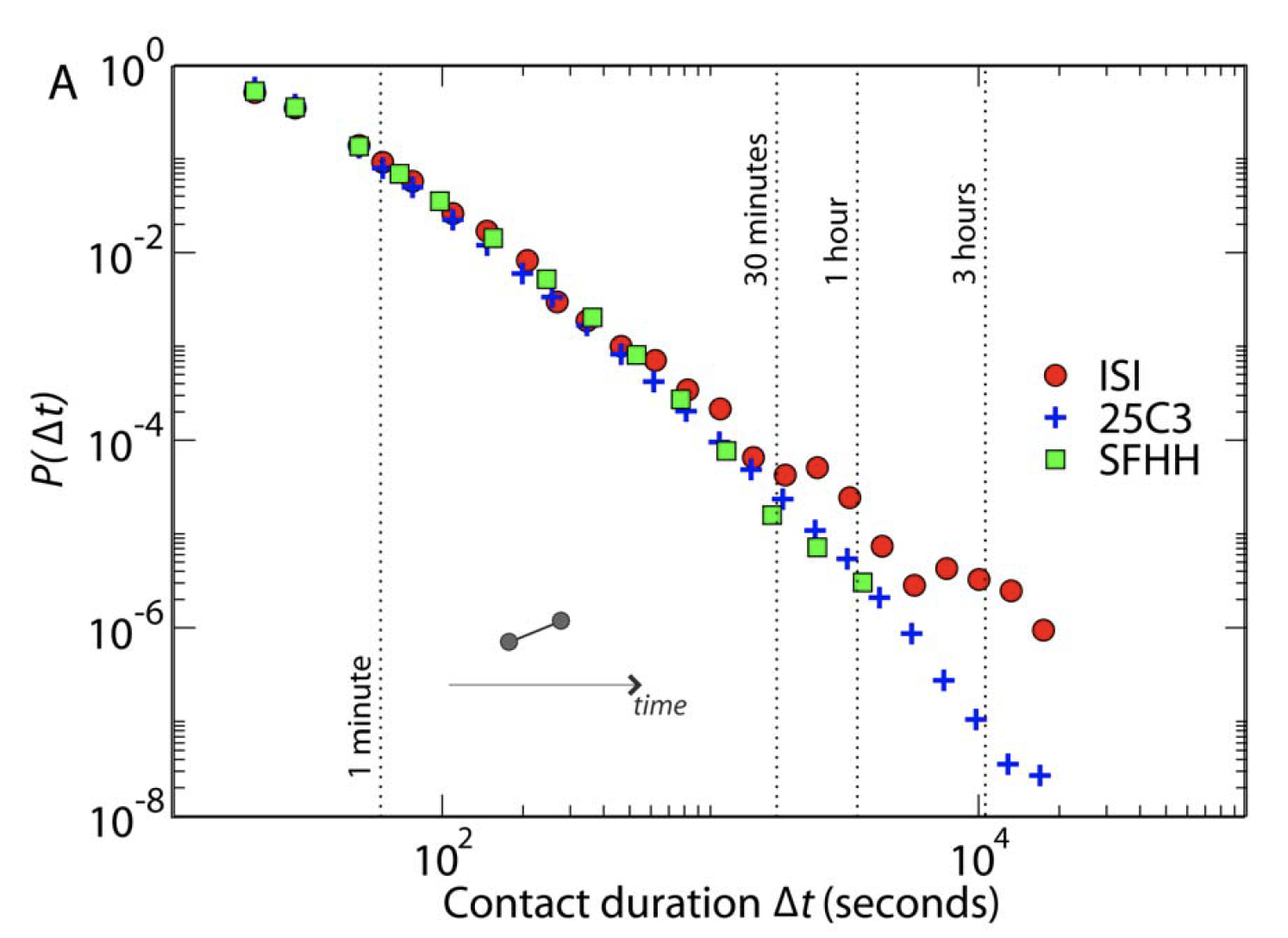}}
\caption{Probability distribution of duration of contacts between any two given persons in three different deployments of the SocioPatterns experiment.
\\ {\it Source}: Reprinted figure with permission from Ref.~\cite{cattuto2010dynamics}.
}\label{fig:face2face}
\end{figure}

Social networks constitute a salient example of time-varying networks, due to their continuous evolution, often concurrently with dynamical processes occurring on their nodes and exchanging information during interactions~\cite{wasserman1994social}. Notably, these systems are characterized by intermittent and rewiring links, multiple characteristic time scales, burstiness, formation of communities, and other complex behaviors, which questions the adequacy of traditional analyses relying on the strong hypothesis of Poisson distributed processes. Due to advances in technologies, several data collection strategies have been put forward to characterize these systems, making available diverse data sets on urban and long-range mobility cell phone calls, online interactions, and human proximity~\cite{zhao2011social}. In particular, the latter has been pursued since 2008 by the SocioPatterns collaboration~\cite{SocioPatterns} through the realization of low-cost  sensors able to record mutual proximity of their wearers by exchanging low-power radio packets. Sensors have been  distributed to attendee at gatherings such as schools, museums, or conferences, revealing common statistical properties and the coexistence of heterogeneous time scales, ranging from $20$ seconds to several hours, which entails bursty patterns of interaction. Moreover  the presence of super-connectors is observed, a concept equivalent to \em hubs \em in static networks~\cite{cattuto2010dynamics}. Figure~\ref{fig:face2face} illustrates the probability distribution of the duration of contacts between any two given persons in three different deployments of the SocioPatterns experiment, denoting the lack of a characteristic time scale and a striking similarity among the experiments.

Different models have been proposed to reproduce salient characteristics of face-to-face interactions. An agent-based model is proposed in Refs.~\cite{stehle2010dynamical,zhao2011social}. The model is constructed upon a population with a fixed number of individuals confined in a bounded two-dimensional space, where well-mixing conditions can be assumed. At any time instant, agents can be either isolated or belonging to a group; thus, the contact network is formed by disconnected cliques of different size. The contact networks evolve by letting each agent decide whether to join a group, if they are isolated, or leaving the group to which they belong. Probabilities of switching state (from isolated to grouped, and vice versa) obey to a memory effect, whereby they depend on the state of the agent and on the time an agent spent in its current state. A reinforcement mechanism is put forward such that agents that interact for long times are less likely to leave their group and, conversely, agents that are isolated for long times are less likely to join a group. This concept is similar to that of preferential attachment in complex networks~\cite{barabasi1999emergence}, and may have connections with Hebbian-like mechanisms at the underlying cognitive level. The proposed model is amenable to some analytical treatment and is able to qualitatively reproduce empirical results derived from experimental data~\cite{cattuto2010dynamics,scherrer2008description}.

\begin{figure}[t]
	\centerline{\includegraphics[scale=0.43]{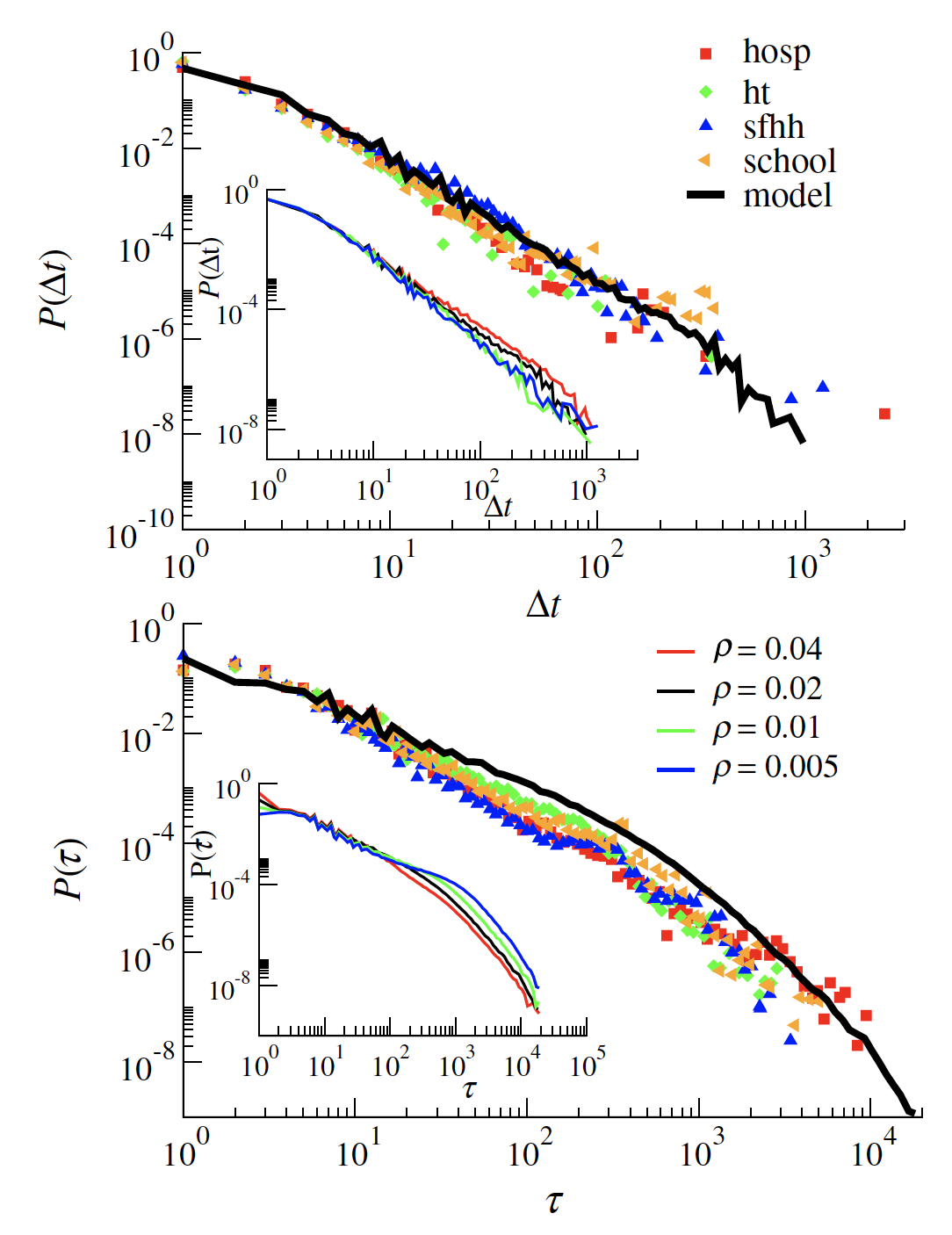}}
	\caption{Distribution of the contact duration,
		$P(\Delta t)$, (top) and distribution of the time interval between consecutive
		contacts, $P(\tau)$, (bottom) for various data sets and for the
		model in Ref.~\cite{starnini2013modeling}. Insets: Same distributions for the attractiveness
		model with different density. Symbols refer to empirical
		data; lines to results of the model, for different densities $\rho$.
		\\ {\it Source}: Reprinted figure with permission from Ref.~\cite{starnini2013modeling} \textcopyright ~2013 by the American Physical Society.
	}\label{fig:face2faceStarnini}
\end{figure}

A further salient agent-based model of face-to-face interactions has been proposed in Ref.~\cite{starnini2013modeling}. Therein, agents perform a biased random-walk and interact according to spatial proximity. Agents are supposed to have a heterogeneous level of \em attractiveness, \em which biases the random-walk of agents toward the most attractive ones. This is a typical phenomenon occurring in social, economic, and natural communities, where some individuals are able to attract most of the attention of the entire community. Differently from Refs.~\cite{stehle2010dynamical,zhao2011social}, agents can enter or exit an \em active \em state, where they are enabled to move and make connections. The resulting model is Markovian, and even in its simplicity, is able to capture many features of empirical and experimental data. Figure~\ref{fig:face2faceStarnini} illustrates the distribution of the contact duration (top) and the distribution of the time interval between consecutive contacts (bottom) for various datasets~\cite{SocioPatterns} and the proposed model, and for different population densities, $\rho$, denoting a great  agreement. Further investigations are carried out on the correlation between the number of different contacts and the temporal duration of such contacts, yielding to the observation of a super-linear ``hub-like'' behavior, whereby nodes with high degrees tend to spend more time in interactions with others than individuals with a lower number of connections, a phenomenon empirically observed in Ref.~\cite{cattuto2010dynamics}. The model is also able to reproduce an empirical phenomenon observed in human mobility, whereby the tendency of agents to interact with new agents decreases in time. This phenomenon manifests itself through a sub-linear increase of the number of different contacts of single individuals~\cite{song2010modelling}.

\subsubsection{Metapopulation models}

Metapopulation models consider an ensemble of individuals that are distributed in local populations (modeling, for instance, neighborhoods, cities, urban areas, or ecological habitats), and may migrate from one population to another. This system can be, hence, described by a network where the nodes are the local populations, within which all the agents/individuals interact each other, and the edges represent the migration routes that the individuals may follow. The system may equivalently described as a larger network where nodes now represent the individuals and links account for the interactions within the local populations. As the composition of these local populations change in time because of migration, then the network is effectively time-varying.

Metapopulation models are widely used in mathematical epidemiology~\cite{pastor2015epidemic,ball2015seven}, where they allow to capture the different spatial and temporal scales of epidemic spreading in a system of interconnected populations, in ecology~\cite{hanski1998metapopulation,hanski2004ecology}, where they describe interactions through migration of individuals among local habitats, in game theory~\cite{huia2007spatial,nagatani2018metapopulation}, and in synchronization dynamics~\cite{gomez2013motion}. From a theoretical perspective, they are also important to model
non-Poissonian distributions of inter-event times~\cite{fonseca2021}.

\subsubsection{Adaptive networks}
\label{subsec2:adaptativenetworks}
Adaptive networks are a class of temporal networks where structure and dynamical states coevolve~\cite{gross2009adaptive,sayama2013modeling}. In this case, the adjacency matrix depends on the state itself in a way that can be static, such that the generic element, for instance, can be written as function of the state of the nodes $i$ and $j$, i.e., $\mathcal{A}_{ij}(t)=\mathcal{A}_{ij}(\mathbf{x}_i(t),\mathbf{x}_j(t))$, or dynamic, such that the update law for the coefficients of $\mathcal{A}(t)$ need to be specified. This latter case can be formally expressed by indicating the weights as $\mathcal{A}_{ij}(t)=w_{ij}(t)$ and providing the equations for their dynamics that, considering, for instance, again the scenario where they depend on the states at the nodes $i$ and $j$, read as $\dot{w}_{ij}=f(\mathbf{x}_i(t),\mathbf{x}_j(t))$.

The idea of adaptive networks was pioneered by the concept of dynamic graphs introduced by Siljak~\cite{vsiljak2008dynamic} and characterized by weights that vary in time and obey to a differential equation. Dynamic graphs do no include the possibility for networks to grow, that is instead incorporated in the Holland's concept of complex adaptive systems~\cite{holland1992adaptation}. Recently, the adaptation and growth mechanisms of networks have been framed into the more general formalism of \emph{evolving dynamical networks} that also incorporates the evolution of the component dynamics~\cite{gorochowski2012evolving,delellis2010synchronization}.

Examples of adaptive networks are commonly found in nature. Flocks of birds or schools of fishes possess the ability of reshaping the structure of interactions by forming or suppressing interconnections among the individuals and adjusting their strength~\cite{delellis2010synchronization}. Further examples of real-world systems that can be modeled as adaptive networks include social systems, neural networks and other biological networks~\cite{sayama2013modeling}.

Adaptive networks are also of utmost importance in control engineering where they represent a way to embed a control law able to re-adapt the weight of a link or modulate the structure of interactions in order to achieve a desired state or collective behavior for the system~\cite{delellis2010synchronization}. For instance, one can consider a network starting from a configuration that does not synchronize and use adaptation mechanisms for the links that evolve the system towards a structure supporting synchronization. We will discuss this example and several others in Sec.~\ref{sec:monolayer}.

\subsection{Hypernetworks and multilayer networks}
\label{sec:secIIC}
\par
\begin{figure}
\centerline{\includegraphics[scale=0.180]{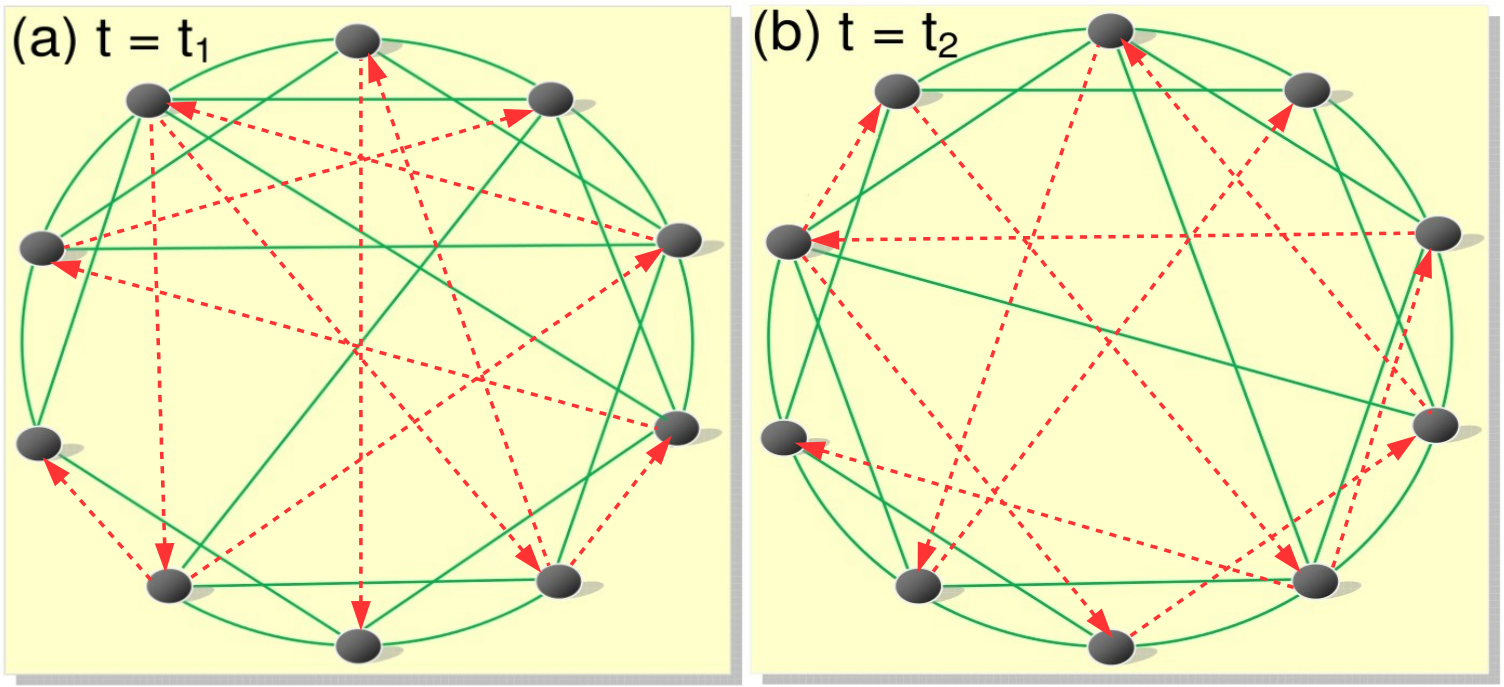}}
\caption{Schematic representation of time-varying connections at two different time instants: (a) $t_1$ and (b) $t_2$. The red dashed and green solid lines denote the different types of interactions forming the hypernetwork. \\ {\it Source}: Reprinted with permission from Ref.~\cite{rakshit2018emergence}.
}\label{fig_02}
\end{figure}

 Consider a family of networks $\mathscr{G}^{[\alpha]}=\big(V,E^{[\alpha]}\big)$, $\alpha=1,2,\dots,M$, where $V$ is a fixed set of nodes for each $\alpha$, and $E^{[\alpha]}\subseteq V\times V$ is a non-empty set of edges. If $E=\big\{ E^{[\alpha]}: \alpha=1,2,\dots,M \big\}$ is a family of edges, which represents various interaction types, then a  hypernetwork is a pair $\mathscr{H}=(V,E)$. Here each $E^{[\alpha]}$ corresponds to a different mode of interaction. We call each of these as a {\it tier}. Here, $M$ is the total number of tiers in the hypernetwork. For a time-varying hypernetwork, each tier $E^{[\alpha]}(t)$ of the network is a function of time.
The hypernetwork $\mathscr{H}(t)$ is said to be jointly connected, if the union of its frozen-time projected networks $\Big(\bigcup_{\beta=1}^L V_\beta~,~\bigcup\limits_t\bigcup_{\alpha=1}^M E^{[\alpha]}(t)\Big)$ constitutes a connected graph.
In a frozen-time, any tier $\big(V,E^{[\alpha]}\big)$ may have one or more disconnected components, but the frozen-time projected network should be connected. A schematic diagram illustrating time-varying interactions in a hypernetwork of $10$ nodes is shown in Fig.~\ref{fig_02}.

\par A multilayer network is a pair $\mathscr{M}=(\mathscr{G},\mathscr{C})$, where $\mathscr{G}=\big\{ \mathscr{G}_\beta=(V_\beta,E_\beta) : \beta\in\{1,2,\dots,L\} \big\}$ is a family of graphs each representing a layer and $\mathscr{C}=\big\{ E_{\beta_1\beta_2}\subseteq V_{\beta_1}\times V_{\beta_2} : \beta_1,\beta_2\in\{1,2,\dots,L\}, \beta_1\ne\beta_2 \big\}$ is the set of interconnections between the nodes of non-identical layers $\mathscr{G}_{\beta_1}$ and $\mathscr{G}_{\beta_2}$. The elements of $E_\beta$ are called intra-layer connections, while the elements of $\mathscr{C}$ are called crossed layers, where each element of $E_{\beta_1\beta_2}$ stands for an inter-layer connection. If a multilayer network is formed by the same number of vertices in each layer, and each node is only connected to its counterpart node in the rest of the layers, then it is known as multiplex network. Therefore, for multiplex networks, $|V_1|=|V_2|=\dots=|V_L|=N$ and $E_{\beta_1\beta_2}=\big\{ (v_i^{[\beta_1]},v_i^{[\beta_2]}): i=1,2,\dots,N \mbox{~and~}v_i^{[\beta_1]}\in V_{\beta_1},v_i^{[\beta_2]}\in V_{\beta_2} \big\}$, where $|V|$ denotes the cardinality of the vertex set $V$.

\par A multilayer hypernetwork is an ordered pair $\mathscr{M}_\mathscr{H}=(\mathscr{G},\mathscr{C})$, where $\mathscr{G}=\big\{ \mathscr{G}_\beta=(V_\beta,E_\beta):\beta\in\{1,2,\dots,L\} \big\}$ is a family of graphs, each representing a layer, in which $E_\beta=\big\{ E_\beta^{[\alpha]}:\alpha\in\{1,2,\dots,M\} \big\}$ is a family of hyperlinks for each tier $\alpha$. $\mathscr{C}=\big\{ E_{\beta_1\beta_2}\subseteq V_{\beta_1}\times V_{\beta_2}:\beta_1,\beta_2\in\{1,2,\dots,L\},~\beta_1\ne\beta_2 \big\}$ is the set of inter-layer connections between the nodes of non-identical layers $\mathscr{G}_{\beta_1}$ and $\mathscr{G}_{\beta_2}$. A time-varying hypernetwork is obtained when each tier $E_\beta^{[\alpha]}(t)$ is a function of time, but the inter-layer connections $\mathscr{C}$ are time-invariant.
The network $\mathscr{M}_\mathscr{H}(t)$ is said to be jointly connected if the union of its frozen-time projected network $\Big(\bigcup_{\beta=1}^L V_\beta~,~\bigcup\limits_t\bigcup_{\alpha=1}^M\bigcup_{\beta=1}^L E_\beta^{[\alpha]}(t)\bigcup\mathscr{C}\Big)$ constitutes a connected graph.
At any frozen-time, any tier $\big(V,E^{[\alpha]}\big)$ may have one or more disconnected components, but the necessary condition for achieving complete synchronization inside each layer is that the frozen-time projected network be jointly connected.

\begin{figure}[t]
\centerline{\includegraphics[scale=0.30]{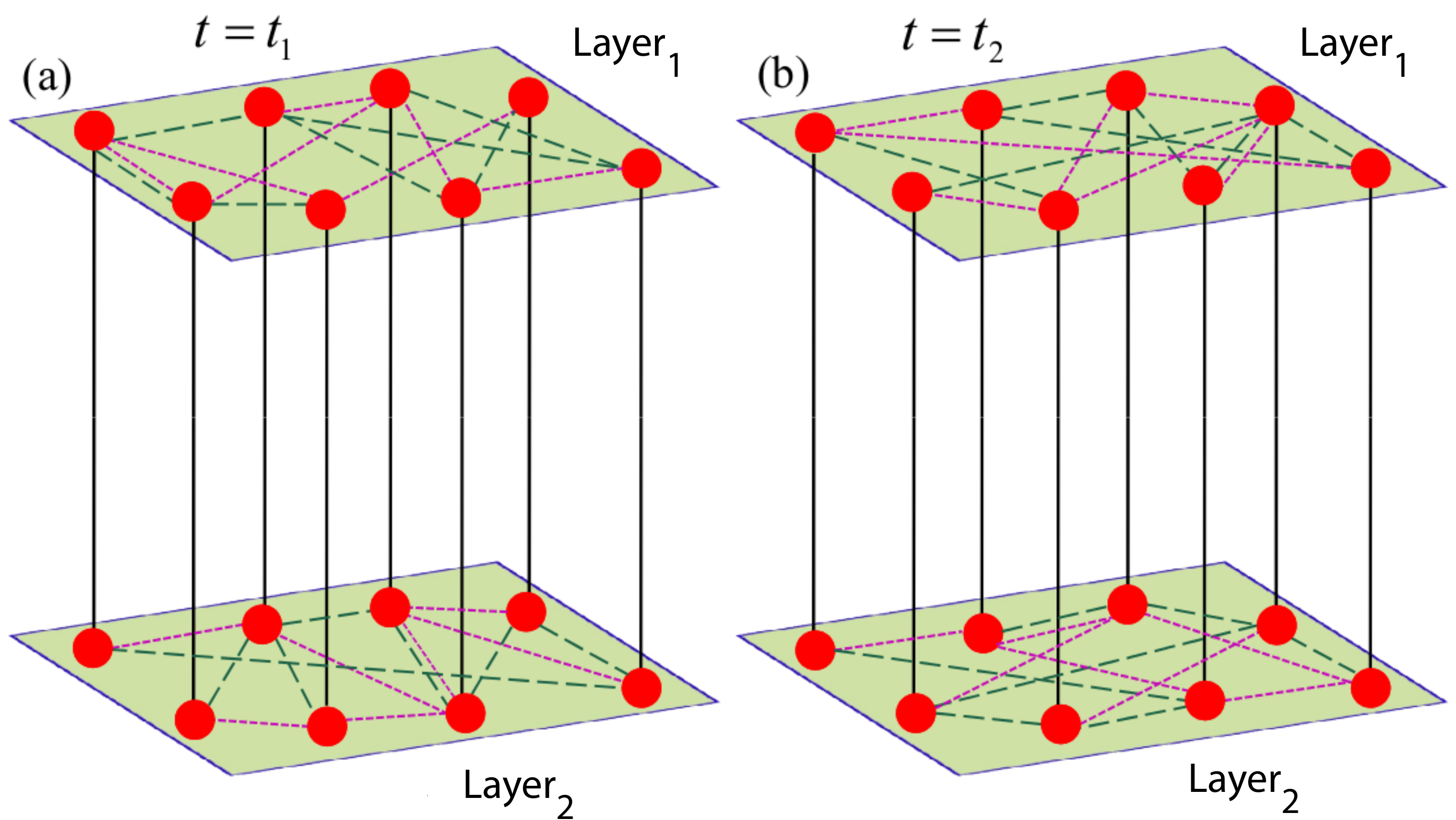}}
\caption{Schematic illustration of time-varying interactions in a multiplex hypernetwork at two different time instants: (a) $t_1$ and (b) $t_2$. The inter-layer connections are time-invariant and they are denoted by solid black line.
\\ {\it Source}: Reprinted with permission from Ref.~\cite{rakshit2020intralayer}.
}\label{Diagram}
\end{figure}
\par The schematic diagram in Fig.~\ref{Diagram} represents a time-varying hypernetwork with a multiplex structure of two layers consisting of $N=8$ nodes and $M=2$ interaction types in each layer. Two different types of interacting tiers are shown at two particular instances of times $t=t_1$ and $t=t_2$ in Figs.~\ref{Diagram}(a) and~\ref{Diagram}(b), respectively. The links of one tier are denoted by the green dashed lines while the links of the other tiers are depicted with magenta dotted lines, whereas the connections between the layers are time-independent and are represented by black solid lines.

\subsection{Switched systems}

\par Switched systems are a class of systems whose coefficients undergo abrupt changes. Consider the linear state equation
\begin{equation}\label{eq_2}
\dot{\bf x}(t)=A_{\rho(t)}{\bf x}(t),
\end{equation}
where $\rho:\mathbb{R}\to\mathbb{Z}^+$ is a switching sequence that selects elements from a family of matrix valued coefficients $\{A_1,A_2,\dots,A_n,\dots\}$ and dot stands for temporal derivative. When all of these elements are Hurwitz, stability of Eq.~\eqref{eq_2} is guaranteed if $\rho(t)$ switches at a sufficiently high rate~\cite{liberzon1999basic}. Further restrictions on these elements, such as the existence of a common Lyapunov function, can guarantee stability for arbitrary switching functions, even for slow switching.
\par When either the elements $A_i$'s are not all Hurwitz or none of them is Hurwitz, the stability of Eq.~\eqref{eq_2} is yet possible, although the class of switching functions is further restricted. For such a case, stability can be guaranteed if the switching sequence is sufficiently fast. Then it can be shown that Eq.~\eqref{eq_2} is asymptotically stable if there exists a constant $T$ such that the time-average matrix $\frac{1}{T}\int_{t}^{t+T}A_{\rho(\tau)}~d\tau$ is Hurwitz for all $t$ when $\epsilon$ is sufficiently small~\cite{aeyels1999exponential,aeyels1998new}. For each case, stability of the specific time-average system implies stability of the original system, and requires the existence of a time-independent Lyapunov function of the certain average system. 
\section{Static nodes}
\label{sec:monolayer}

This Chapter deals with temporal networks where links change as the result of diverse mechanisms such as deterministic or stochastic processes, external driving forces, or intrinsic adaptation capabilities. 

In the first part of the Chapter, we review a series of fundamental theoretical tools developed to study the onset and stability of synchronization in temporal networks. These methods are developed in the framework of continuous-time nonlinear oscillators (including periodic and chaotic systems), but their analysis provides general notions and concepts (such as the role of the different time scales at work in a temporal network) that are also useful to study the synchronous behavior of purely phase oscillators. For this reason these methods are reviewed in Section~\ref{sec:mathtools}, whereas examples of synchronization in temporal networks of phase oscillators are dealt with in Section~\ref{sec:KuramotoTN} and limit cycle and chaotic systems in Section~\ref{sec:ChaoticTN}.
At the end of the Chapter, we will also present examples of synchronous behavior emerging in multilayer networks and hypernetworks.

\subsection{Methods for the analysis of stability of the synchronous state}
\label{sec:mathtools}

Let us begin with briefly reviewing a few fundamental notions on synchronization in classical networks, i.e., with time-independent links, that prove to be particularly important also for the analysis of temporal networks. The general model to study synchronization in ensembles of $N$ identical units which are networking with a complex connectivity structure is described by the following equations:
\begin{equation}
\label{eq:eqStaticLapl}
\dot{\mathbf{x}}_i=\mathbf{f}(\mathbf{x}_i)-\epsilon\sum_{j=1}^N\mathcal{L}_{ij} \mathbf{h}(\mathbf{x}_j),
\end{equation}
 
\noindent with $i=1,\ldots,N$. Here, ${\mathbf{x}}_i(t)\in \mathbb{R}^n$ represents the state vector of the oscillator at node $i$, $\mathbf{f}:\mathbb{R}^n \rightarrow \mathbb{R}^n$ the vector field describing the uncoupled node dynamics, $\mathbf{h}:\mathbb{R}^n \rightarrow \mathbb{R}^n$ the coupling function, $\epsilon$ the coupling strength, and $\mathcal{L}=\{\mathcal{L}_{ij}\}$ the Laplacian matrix mapping the interactions among the network units.

As the Laplacian matrix is a zero-row sum matrix, then Eqs.~(\ref{eq:eqStaticLapl}) always admit a solution of the type $\mathbf{x}_1=\ldots=\mathbf{x}_N=\mathbf{x}_s$ with $\mathbf{x}_s$ such that $\dot{\mathbf{x}}_s=\mathbf{f}(\mathbf{x}_s)$. These conditions define the so-called \emph{synchronization manifold} and the \emph{synchronous solution} $\mathbf{x}_s$. However, the mere existence of this solution does not suffice to guarantee that the system will converge towards it: to observe synchronization, the solution needs to be stable.

A widely used technique to study the stability of the synchronous solution is based on the derivation of a \emph{Master Stability Function} (MSF) \cite{pecora1998master}. The technique consists in linearizing Eqs.~(\ref{eq:eqStaticLapl}) around $\mathbf{x}_s$, and then applying a proper transformation of variables that leads to a generic variational block of this type:
\begin{equation}
\label{eq:genericblockStatic}
    \dot{\mathbf{\xi}}=\left [ J\mathbf{f}-\alpha J\mathbf{h}\right ]{\mathbf{\xi}},
\end{equation}

\noindent where ${\mathbf{\xi}}\in \mathbb{R}^n$ and $J\mathbf{f}$ and $J\mathbf{h}$ represent the Jacobian matrices of $\mathbf{f}$ and $\mathbf{h}$, computed around $\mathbf{x}_s$ and $\alpha$ is a generic complex parameter. For simplicity, let us focus here on the case of undirected connected networks where $\alpha$ is a real parameter.

From Eqs.~(\ref{eq:genericblockStatic}) the maximum conditional Lyapunov exponent $\Lambda_{max}$ can be computed as a function of the independent parameter $\alpha$, thus obtaining the MSF $\Lambda_{max}=\Lambda_{max}(\alpha)$. This function fully characterizes the stability of the synchronous state, as this requires that the maximum conditional exponent is negative for $\alpha=\{\epsilon \lambda_2, \ldots, \epsilon \lambda_N\}$, where $0=\lambda_1 < \lambda_2 \leq \lambda_N$ are the eigenvalues of the Laplacian matrix of the network assumed to be undirected and connected (such that its eigenvalues are all positive, except the first which is zero).

The way in which the maximum conditional Lyapunov exponent $\Lambda_{max}$ depends on $\alpha$ yields three different classes of systems: i) type I systems where $\Lambda_{max}(\alpha)>0$ $\forall \alpha$, which includes dynamical systems that will never synchronize, no matter which network's structure of interactions is considered; ii) type II systems where $\Lambda_{max}(\alpha)<0$ for $\alpha \in [\alpha_1,+\infty)$, such that by tuning the coupling strength the synchronization condition $\epsilon \lambda_2 > \alpha_1$ can be met; iii) type III systems where $\Lambda_{max}(\alpha)<0$ for $\alpha \in [\alpha_1,\alpha_2]$, resulting in a non-trivial condition for synchronization requiring simultaneously that $\epsilon \lambda_2 > \alpha_1$ and $\epsilon \lambda_N < \alpha_2$ \cite{boccaletti2006complex}.

A few fundamental aspects are here worth to remark. First, the criterion only provides a necessary (and local) condition for synchronization, but in many works it proved to accurately predict the onset of it. Second, the MSF $\Lambda_{max}=\Lambda_{max}(\alpha)$ solely depends on the node dynamics and the coupling function and therefore can be calculated independently from the specific network's structure. Third, the topology of the network plays a fundamental role in determining the stability of synchronization as the condition $\Lambda_{max}(\alpha)<0$ is checked in $N-1$ points that depend on the eigenvalues of the Laplacian matrix. This matrix is, therefore, of crucial relevance for synchronization stability. Finally, we notice that Eqs.~(\ref{eq:genericblockStatic}) are derived through the transformation that uses the eigenbasis which diagonalizes the Laplacian matrix.

These observations are particularly important when we move to consider the case of temporal networks. In a time-varying structure, the Laplacian matrix changes with time, and therefore its eigenvalues and corresponding eigenvectors are functions of time as well. Consequently, the basis furnished by the eigenvectors also changes whenever the Laplacian matrix does. As we will see in detail in this section, this is a crucial issue as, in the general case, it hampers the derivation of an equation similar to Eqs.~(\ref{eq:genericblockStatic}). However, there are particular cases when the structure of the temporal network simplifies so that the eigenvector basis is constant in time and only the eigenvalues are time-varying. In other circumstances, i.e., when the time scale at which the temporal network changes significantly differs from that of the process taking place in the units of the system, the analysis of the temporal network can be reconducted to the static case, enabling the extension of techniques based on the MSF. This section is devoted to discuss the techniques for studying synchronization stability in temporal networks that can be developed starting from these considerations.

\subsubsection{The fast switching stability criterion}

Stillwell {\it et al.}~\cite{stilwell2006sufficient} introduced a fast switching stability criterion, in which the time-scale of the network evolution is faster than the time-scale of the coupled oscillators. Before describing this technique, we need the following preliminary lemma.

\newtheorem{theorem}{Theorem}
\newtheorem{lemma}[theorem]{Lemma}
\begin{lemma}
\label{Lemma_1} (Ref.~\cite{stilwell2006sufficient})
Suppose that there exists a time-average matrix $\bar{E}$ of the matrix valued function $E(t)$, such that for all $t\in\mathbb{R}^+$ and for some constant $T$, $\frac{1}{T}\bigintssss_{t}^{t+T}E(\tau)~d\tau=\bar{E}$. Then, for sufficiently fast switching, the following system
\begin{equation}\label{Eq7_1}
\begin{array}{lcl}
\dot{\bf z}(t)=\big[A(t)+E(t)\big]~{\bf z}(t),~~{\bf z}(t_0)={\bf z}_0,~t\ge t_0,
\end{array}
\end{equation}
will be uniformly asymptotically stable whenever the time-average system
\begin{equation}\label{Eq7_2}
\begin{array}{lcl}
\dot{\bf x}(t)=\big[A(t)+\bar{E}\big]~{\bf x}(t),~~{\bf x}(t_0)={\bf x}_0,~t\ge t_0,
\end{array}
\end{equation}
is also uniformly asymptotically stable.
\end{lemma}
Here ${\bf z}_0$ and ${\bf x}_0$ are two different initial conditions from the basin of attraction of the asymptotically stable state of systems~(\ref{Eq7_1}) and~(\ref{Eq7_2}) respectively. The lemma is valid for any constant time $t$, and for sufficiently large $T$, it depends on $A(t)$ and how fast $E(t)$ is switching. Stability of the frozen-time system does not guarantee the stability of the switched system, but this lemma shows that the switched time-varying system can be asymptotically stable if  the time-average system is asymptotically stable for sufficient fast switching.

\par Now consider a temporal network of $N$ identical coupled oscillators
\begin{equation}\label{rfssc_3}
\begin{array}{lcl}
\dot{\bf x}_i(t)=F({\bf x}_i(t))+\epsilon\sum\limits_{j=1}^{N}\mathcal{A}_{ij}(t)~B~[{\bf x}_j(t)-{\bf x}_i(t)],
\end{array}
\end{equation}

\noindent where $i=1,\ldots,N$, ${\bf x}_i\in\mathbb{R}^n$ is the state variable of the $i^{th}$ node and $B\in \mathbb{R}^{n\times n}$ is the inner coupling matrix. The scalar $\epsilon$ is a control parameter that sets the coupling strength between the oscillators. $\mathcal{A}(t)$ is the $N\times N$ time-varying graph adjacency matrix, which describes the interconnections between the oscillators.

\par When complete synchronization emerges in system~\eqref{rfssc_3}, all the oscillators evolve in unison. Then, there exists a trajectory ${\bf x}_s(t)\in\mathbb{R}^d$ such that ${\bf x}_1(t)={\bf x}_2(t)=\dots={\bf x}_N(t)={\bf x}_s(t)$. Consequently, the complete synchronization manifold can be defined as
\begin{equation}
\mathcal{S}_{CS}=\big\{ {\bf x}_0(t)\in\mathbb{R}^d~:~{\bf x}_1(t)={\bf x}_2(t)=\dots={\bf x}_N(t)={\bf x}_s(t) \big\}.
\end{equation}
Owing to the diffusive nature of the coupling, the complete synchronization solution ${\bf x}_s(t)$ is an invariant state for all the coupling strengths $\epsilon$ and all choices of the inner coupling matrix $B$.

\par For sufficiently fast switching, the time-average Laplacian matrix $\bar{\mathcal{L}}$ satisfies $\bar{\mathcal{L}}=\frac{1}{T}\int_{t}^{t+T}\mathcal{L}(\tau)~d\tau$, for some constant $T$. The matrix $\bar{\mathcal{L}}$ has the same inherent zero-row sum property as the parent Laplacian $\mathcal{L}(t)$. But $\bar{\mathcal{L}}$ is not actually describing a particular network, rather it is just the term by term time-average of the time-varying graph Laplacian $\mathcal{L}(t)$. However, the real square matrix $\bar{\mathcal{L}}$ can be unitarily triangularizable. Then, there exists a unitary matrix $P$ each column of which is made of the orthonormal eigenvectors of $\bar{\mathcal{L}}$, such that
\begin{equation*}
P^{-1}\bar{\mathcal{L}}P=\bar{U}=
\begin{bmatrix}
0 & \bar{U}_1 \\
O_{N-1 \times 1} & \bar{U}_2
\end{bmatrix},
\end{equation*}

\noindent is the Schur transformation of~$\bar{\mathcal{L}}$. Here $\bar{U}_2\in \mathbb{C}^{(N-1)\times (N-1)}$ is an upper triangular matrix containing in the main diagonal elements the $N-1$ eigenvalues of $\bar{\mathcal{L}}$ excluding $0$. The equations of motion of the coupled system incorporating the above average Laplacian are obtained from Eq.~(\ref{rfssc_3})  just by replacing $\mathcal{L}(t)$ by $\bar{\mathcal{L}}$.

Considering the Schur transformation and using the unitary matrix $P$, the equation for the error ${\bf \eta}$ transverse to the synchronization manifold can be written as
\begin{equation}\label{rfssc_5}
\begin{array}{lcl}
\dot{\bf \eta}(t)=[I_{N-1}\otimes F({\bf x_0}(t))-\epsilon\bar{U}_2\otimes B]~\eta(t).
\end{array}
\end{equation}

\par By considering the same Schur transformation applied to Eq.~(\ref{rfssc_3}), the equation of motion of the transverse error system becomes
\begin{equation}\label{rfssc_4}
\begin{array}{lcl}
\dot{\bf \xi}(t)=[I_{N-1}\otimes F({\bf x_0}(t))-\epsilon U_2(t)\otimes B]~\xi(t),
\end{array}
\end{equation}

\noindent where
\begin{equation*}
P^{-1}\mathcal{L}(t)P=
\begin{bmatrix}
0 & U_1(t) \\
O_{(N-1) \times 1} & U_2(t)
\end{bmatrix},
\end{equation*}

\noindent is the Schur transformation of $\mathcal{L}(t)$.

\par Now, it is easy to derive that $ \bar{U}_2=\frac{1}{T}\int_{t}^{t+T}U_2(\tau)~d\tau$. Thus, Lemma~\ref{Lemma_1} yields that, if the time-average system has an asymptotically stable synchronization manifold, then the time-varying network also has asymptotically stable synchronization for sufficient fast switching. This fast switching stability criterion is a fundamental tool to assess the local stability of several temporal networks.

Quite interestingly, using the MSF approach, one can derive the conditions for synchronization also when the network dynamics is much slower than that of the nodes, as discussed in detail in Ref.~\cite{zhou2016synchronization}. In Sec.~\ref{sec:syncADN} we will see an application of this case.

\subsubsection{Commutative graphs}

As opposed to the limit of fast switching discussed above, Ref.~\cite{boccaletti2006synchronization} considers explicitly the case where the time scale at which the
network changes is commensurable with that of the dynamics taking place in each unit. The same study has shown for the first time that the synchronizability of a network can be significantly improved by evolving the graph along a time-dependent connectivity matrix.
In Ref.~\cite{boccaletti2006synchronization}, the Authors consider a network of $N$ coupled identical systems, whose evolution is described by
\begin{equation}\label{eq_9}
\dot{\bf x}_i(t)=f({\bf x}_i)-\epsilon\sum\limits_{i=1}^{N}\mathcal{L}_{ij}(t)h({\bf x}_j),\hspace{50pt}i=1,2,\dots,N.
\end{equation}
Once again, ${\bf x}(t)\in\mathbb{R}^n$ is the $n$-dimensional vector describing the state of the $i$th node, $f:\mathbb{R}^n\to\mathbb{R}^n$ governs the local dynamics of the nodes, $h:\mathbb{R}^n\to\mathbb{R}^n$ is a vectorial output function, $\epsilon$ is the coupling strength, and $\mathcal{L}(t)$ is the time-varying zero row sum $N\times N$ symmetric Laplacian matrix. $\mathcal{L}(t)$ specifies the evolution in strength and topology of the underlying connection wiring. Being symmetric, $\mathcal{L}(t)$ admits at all times a set $(\lambda_i(t),v_i(t))$ of real eigenpairs such that $\mathcal{L}(t)v_i(t)=\lambda_i(t)v_i(t)$ and $v_j(t)^{T}v_i(t)=\delta_{ij}$.

\par It is worth noticing that the zero row sum condition imposed on $\mathcal{L}(t)$ ensures that the spectrum is (at each time) entirely non-negative, i.e., $\lambda_i(t)\ge0$ for all $i$ and $t$. Moreover, $\lambda_1(t)=0$ with associated eigenvector $v_1(t)=\big(\frac{1}{\sqrt{N}},\frac{1}{\sqrt{N}},\dots,\frac{1}{\sqrt{N}}\big)^{T}$ that defines the synchronization manifold ${\bf x}_i(t)={\bf x}_s(t)$ for $i=1,\ldots,N$.

\par One can consider, for instance, $\delta{\bf x}_i(t)={\bf x}_i(t)-{\bf x}_s(t)$ to be the deviation of the $i$th state vector from the synchronization manifold, and focus on the column vector $\delta{\bf x}(t)=\big( \delta{\bf x}_1(t),\delta{\bf x}_2(t),\dots,\delta{\bf x}_N(t) \big)^T$. Then in linear order of $\delta{\bf x}(t)$, the evolution equation reads as
\begin{equation}\label{eq_10}
\delta\dot{\bf x}(t)=\big[I_N\otimes Jf({\bf x}_s)-\mathcal{L}(t)\otimes Jh({\bf x}_s)\big]\delta{\bf x}(t),
\end{equation}
where $\otimes$ stands for the matrix direct product and $J$ denotes the Jacobian operator.

Now, the arbitrary state $\delta{\bf x}(t)$ can be written, at each time, as $\delta{\bf x}(t)=\sum_{j=1}^{N}v_j(t)\otimes\eta_j(t)$. Then applying $v_j(t)^{T}$ to the left side of each term in Eq.~\eqref{eq_10}, one finally obtains
\begin{equation}\label{eq_11}
\frac{d\eta_i(t)}{dt}=\big[ Jf({\bf x}_s)-\epsilon\lambda_i(t)Jh({\bf x}_s) \big]\delta_i(t)-\sum\limits_{j=1}^{N}v_i(t)^{T}\frac{dv_j(t)}{dt}\eta_j(t),\hspace{20pt}i=1,2,\dots,N.
\end{equation}

If compared with the classic Master Stability Function approach, Eq.~\eqref{eq_11} contains an extra term $\left(-\sum\limits_{j=1}^{N}v_i(t)^{T}\frac{dv_j(t)}{dt}\eta_j(t)\right)$
which accounts for the projection of the new basis of eigenvectors into the old one. Such a term is of paramount importance, as it can completely change the stability properties of the synchronous solution (as we will shortly see).

Now notice that the above Eq.~\eqref{eq_11} transforms into a set of $N$ variational equations of the form $\frac{d\eta_i(t)}{dt}=\big[ Jf({\bf x}_s)-\epsilon\lambda_i(t)Jh({\bf x}_s) \big]\delta_i(t)$, as soon as all eigenvectors are fixed in time, i.e., $\sum_{j=1}^{N}v_i(t)^{T}\frac{dv_j(t)}{dt}\eta_j(t)=0$.

The above condition can be realized in two different ways. Namely, either the coupling matrix $\mathcal{L}(t)$ is constant (and therefore one recovers the classical case of Master Stability Function), or when starting from an initial wiring condition $\mathcal{L}(t=t_0)$, the coupling matrix $\mathcal{L}(t)$ commutes at any time with $\mathcal{L}(t=t_0)$.

Reference~\cite{boccaletti2006synchronization} considers the case of an evolution along commutative graphs, and demonstrates that (also under such a rather restrictive hypothesis) synchronization can be greatly enhanced.
Notice, indeed, that the initial Laplacian can be written as $\mathcal{L}(t_0)=V\Lambda V^{T}$, where $V=[v_1,v_2,\dots,v_N]$ is an orthogonal matrix whose columns are the eigenvectors of $\mathcal{L}(t_0)$ and $\Lambda_0=\mathrm{diag}(0,\lambda_2(0),\lambda_3(0),\dots,\lambda_N(0))$ is the diagonal matrix consisting of the eigenvalues of $\mathcal{L}(t_0)$. At any time $t$, a commuting matrix $\mathcal{L}(t)$ can be constructed as $\mathcal{L}(t)=V\Lambda(t)V^{T}$.  Here, $\Lambda(t)=\mathrm{diag}(0,\lambda_2(t),\lambda_3(t),\dots,\lambda_N(t))$ and, for all $i>1$, $\lambda_i(t)$ are positive real numbers. Therefore, $\mathcal{L}(t)$ is positive semi-definite and zero row-sum. Since the standard orthogonal basis vectors are not collinear with the eigenvectors $v_1$, therefore $\mathcal{L}_{ii}(t)>0$ for all $i$.

Because of the commuting properties of $\mathcal{L}(t)$, Eq.~\eqref{eq_11} becomes $\dot{\eta}_i(t)={\bf K}_i\eta_i(t)$ for all $i=2,3,\dots,N$. Replacing $\epsilon\lambda_i(t)$ by $\nu$ in the kernel ${\bf K}_i$, the problem of stability of the synchronization manifold is tantamount to study the $n$-dimensional parametric variational equation $\dot{\eta}=[Jf({\bf x}_s)-\nu Jh({\bf x}_s)]\eta$. Then the stability region corresponds to the region where the curve of $\Lambda_{\max}$, the largest Lyapunov exponents with respect to $\nu$, is negative.

Later on, Ref.~\cite{del2015synchronization} provided a rigorous solution to the problem of constructing a structural evolution that switches between topologies without constraints on their commutativity, thus providing the most general framework for studies of synchronization of identical units under smooth changes in time of their connectivity, independently on the particular topologies visited, and also on the time scale of the evolution, which can be faster than, comparable to, or even secular with respect to the dynamics of the units.

\subsubsection{Techniques for global stability analysis}
\label{sec:secIIIBasin}
The methods discussed so far refer to linear stability analysis, and therefore their predictions hold only for small perturbations from a desired state. However, perturbations are not always infinitesimal and, in these latter cases, other techniques should be used. In order to understand the dynamical response to any kind of perturbation, one should have a picture of the complete landscape of the coupled system, that is, one should know the size of basin of attraction~\cite{wiley2006size,zou2014basin} for all attractors of the system. Recently, the concept of a measure of basin stability has been proposed to quantify how stable a synchronization state is against large perturbations~\cite{menck2013basin,menck2014dead,leng2016basin}. It is a nonlinear and nonlocal approach that relies on the volume of the basin of attraction rather than the traditional linearization based approach. It is applicable even to higher dimensional  systems and is a robust measure for characterizing multi-stable states.

\par The basin stability paradigm is particularly useful in case of time-varying networks, as it can be applied to a very large class of systems, whereas the linear stability analysis can be done exactly only in some specific cases. As we have seen, derivation of analytical or semi-analytical conditions for stability often needs to make special hypotheses on the coupling scheme, such as fast rewiring~\cite{amritkar2006synchronized,so2008synchronization}, on-off coupling~\cite{chen2009synchronization} or a particular class of local dynamics~\cite{chen2007synchronization}. The basin stability measure is a general numerical technique that can be used to analyze the stability of high-dimensional systems and to quantify different stable steady states in coupled delayed~\cite{leng2016basin} and non-delayed systems~\cite{rakshit2017basin}, synchronized states~\cite{kohar2014synchronization} and chimera states~\cite{rakshit2017basin2}.

\par The basin stability measure is defined as $\text{BS}=\int_{\mathscr{B}} \chi({\bf x}) \rho({\bf x}) d{\bf x}$, where $\mathscr{B}$ is the set of all possible perturbations ${\bf x}$ and $\chi({\bf x})$ is equal to $1$ if the system converges to synchronization after the perturbation ${\bf x}$, and zero otherwise. $\rho({\bf x})$ is the density of the perturbed states with $\int_{\mathscr{B}} \rho({\bf x}) d{\bf x}=1$. To compute BS, the coupled system is integrated for a number $Q$ (sufficiently large) of different states distributed randomly over a prescribed phase space volume, then the evolution of the system from these different initial states is computed. Let $M$ be the number of states that reach the synchronous state, then the BS for the synchronous state is estimated as $\frac{M}{Q}$. BS takes values in $[0,1]$, with BS~$=0$ implying that for all random initial conditions the synchronized state is unstable, and BS~$=1$ indicating that it is globally stable for any perturbation. Intermediate values of BS represents the probability to find a synchronous state starting from an initial condition that lies in the prescribed phase space volume. Very often this measure complements the information obtained through linear stability analysis. Examples of applications of the concept of BS to temporal networks are found in Ref.~\cite{kohar2014synchronization}.

Another very important class of methods for the analysis of synchronization stability is based on Lyapunov functions. These methods may provide local or global conditions for synchronization stability, but, especially for the case of temporal networks, often need to be tailored on the specific assumptions on the network evolution. They have been proved to be particularly effective for the study of adaptive networks of limit cycle and chaotic oscillators, where also the law of evolution for the network is deterministic. For this reason, they will be discussed later on in Sec.~\ref{sec:adaptivenetworksChaotic} with specific reference to this type of temporal networks.

\subsection{Monolayer networks: phase oscillators}
\label{sec:KuramotoTN}

\subsubsection{Blinking networks}

In his pioneer work, Kuramoto introduced a simple model to study synchronization in interacting systems. He considered the scenario where the coupling is weak compared to the attraction force towards the limit cycle that represents the natural tendency of the unit to oscillate at its own frequency of oscillation (i.e., the uncoupled dynamics). Under these circumstances, each oscillator dynamics can be fully described by a single state variable, $\phi_i$, representing the phase of the $i-$th oscillator. In his work, Kuramoto considered the case where each oscillator is coupled to all the others, a scenario which, in the framework of complex networks, corresponds to consider interactions that are fixed in time and global and that is referred to as global coupling, fully connectivity or all-to-all coupling. The model was then extended to account for general topologies of interactions~\cite{strogatz2000kuramoto,acebron2005kuramoto,rodrigues2016kuramoto}, as well as higher-order coupling mechanisms and time-varying intrinsic parameters (e.g., the natural frequency)~\cite{petkoski2012kuramoto,pietras2016ott,lu2018stability}. Here, we are focused on Kuramoto oscillators, also known as pure phase oscillators, interacting via time-varying edges. In particular, we start with blinking networks.

A first interesting study of blinking networks of Kuramoto oscillators concerns a rather special setting of the time-varying interactions~\cite{so2008synchronization}. The Authors have considered a system composed by two large heterogeneous populations of phase oscillators interacting according to two fixed arrangements switched at a given blinking frequency. In agreement with the key findings of the fast switching approach, they demonstrate that, at sufficiently high blinking frequencies, the two populations of interacting phase oscillators behave as if their connectivity were static and equal to the time-average of the temporal structure of coupling. This blinking mechanism was proved to be capable of inducing synchronization,
even when the switching occurs between topologies that individually do not support a coherent state.

To delve into the analysis of blinking networks of phase oscillators, let us illustrate in more detail the model and the results discussed in Ref.~\cite{faggian2019synchronization}, where, in particular, two interesting findings emerge. First, the Authors are able to show that, if the coupling is sufficiently strong and the rewiring sufficiently fast, then partial synchronization can be reached even in the presence of extremely low instantaneous connectivity. Second, the Authors provide approximate
analytical arguments that are able to predict the transition to synchronization beyond the limit conditions for fast switching.

To begin our discussion, let us describe the dynamical equations governing the model considered in Ref.~\cite{faggian2019synchronization}. The system consists of $N$ phase oscillators interacting through a temporal network with adjacency matrix $\mathcal{A}_{ij}(t)$:

\begin{equation}
\dot {\phi}_i=\omega_i+\frac{\epsilon}{d_i(t)}\sum\limits_{j=1}^{N}{\mathcal{A}_{ij}}(t)\sin(\phi_j-\phi_i),\\
\label{vc1}
\end{equation}

\noindent with $i=1,\ldots,N$. Here, ${\phi_i} \in [0,2\pi]$ represents the instantaneous phase of the $i$-th oscillator and $\omega_i$ is the natural frequency chosen from a zero-mean Gaussian distribution with standard deviation $\sigma$ (notice that, in line with the Kuramoto model, the system is heterogeneous, with each oscillator characterized by its own natural frequency, in general different from that of the other units). The parameter $\epsilon$ is the coupling strength and $d_i(t)$ represents the instantaneous degree of the $i-$th node at time $t$, that is, $d_i(t)=\max\{1,\sum\limits_{j=1}^{N}{\mathcal{A}_{ij}(t)} \}$.

The connectivity of the temporal network is encoded in the adjacency matrix $\mathcal{A}_{ij}(t)$: at each time step, links are generated according to the ER model~\cite{erdos1959random} with wiring probability indicated as $p$; in addition, a Poissonian process for random rewiring of the edges is employed, with each individual node rewiring synchronously all its incident edges with probability rate $\frac{1}{T}$, with $T$ being the rewiring time period. To implement the Poissonian process in numerical simulations, at each integration time step, each node goes through a rewiring event with Poissonian probability given by

\begin{equation}
r_P=1-e^{(-\frac{\Delta t}{T})},\\
\label{vc2}
\end{equation}
\noindent where $\Delta t$ is the integration step size considered. Following Ref.~\cite{faggian2019synchronization}, one also defines $q=pN$ as the mean degree connectivity.

To monitor the level of synchronization, one can consider the classical Kuramoto order parameter:

\begin{equation}
r(t)=\Big|\frac{1}{N}\sum\limits_{j=1}^{N}e^{\iota{\phi_j}}\Big|,
\label{ps3}
\end{equation}

\noindent with $\iota=\sqrt{-1}$, and $R$ is defined as the steady-state value taken by this parameter:

\begin{equation}
R=\lim\limits_{T\rightarrow +\infty}\frac{1}{T_w}\int\limits_0^{T_w} r(t) dt,
\label{eq:KuramotoR}
\end{equation}

\noindent where $T_w$ denotes a sufficiently large window of time over which the system dynamics is calculated. The order parameter $r$ (and consequently $R$) varies in the interval $[0,1]$, with $r=1$ indicating complete synchronization and $r\sim 0$ for the incoherent state.

\begin{figure}
	\centerline{
		\includegraphics[scale=0.500]{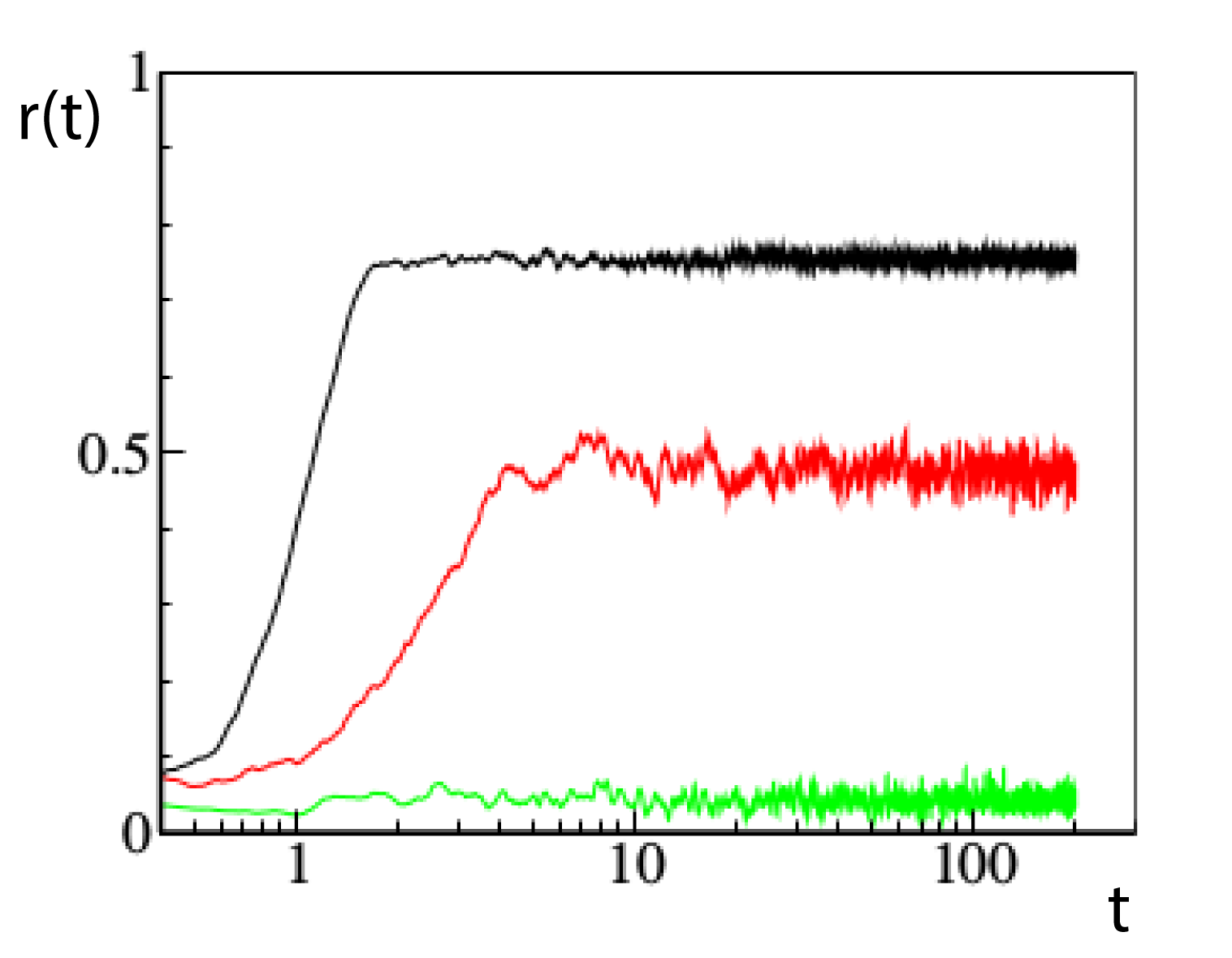}}
	\caption{Time evolution of the order parameter $r(t)$ for a blinking network of Kuramoto oscillators, for three different values of the blinking period $T=6.28$ (green), $T=0.63$ (red) and $T=0.31$ (black). The other parameters of the system are fixed as: network size $N=10^4$, connectivity $q=0.8$, coupling strength $\epsilon=8$, standard deviation of the natural frequency distribution $\sigma=1$.
	\\ {\it Source:} Reprinted figure with permission from Ref.~\cite{faggian2019synchronization}. }
	\label{fig16}
\end{figure}

In order to gain some insights on the system behavior, it is instructive to inspect the time evolution of the Kuramoto order parameter $r(t)$ for selected values of the rewiring period $T$. Fig.~\ref{fig16} illustrates it for a network with $N=10^4$ nodes, connectivity parameter $q=0.8$, coupling strength $\epsilon=8$, and standard deviation of the natural frequency distribution set to $\sigma=1$. We can notice that synchronization does not occur for high T, e.g. $T=6.28$, but, as the value of $T$ is lowered to $T=0.63$ or $T=0.31$, synchronization, although with different levels of coherence, emerges. The highest level of synchrony, in particular, is observed for the smallest value of the rewiring period $T$, indicating that not only fast switching enables synchronization, but also improves it.

\begin{figure}
	\centerline{
	\includegraphics[scale=0.5500]{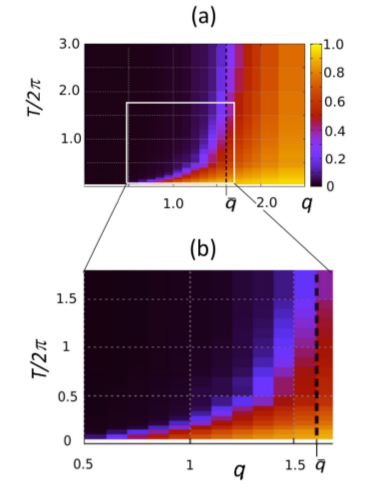}}
	\caption{Order parameter $R$ for a blinking network of Kuramoto oscillators as a function of the connectivity $q$ and blinking period $T$. The other parameters of the network are set to: $N=10^4$, $\epsilon=8$, and $\sigma=1$. Panel b) represents a magnified view of the same parameter in a region of interest.
	\\{\it Source:} Reprinted figure with permission from Ref.~\cite{faggian2019synchronization}.}
	\label{fig17}
\end{figure}

A more systematic analysis of the network dynamics can be carried out by considering how $R$ depends on two important control parameters, the connectivity $q$ and the blinking period $T$ (Fig.~\ref{fig17}(a)). For sufficiently large connectivity, i.e., $q>\bar{q}$, the system reaches synchronization irrespective of the value of $T$. Notice that the threshold $\bar{q}$ is higher than that for the emergence of a giant connected component in ER networks (i.e., $q=1$). This suggests that this structural condition promotes the observed dynamical regime. Instead, for small $q$, a sufficiently fast rewiring (associated to low values of $T$) is required to attain synchronization, at least for $q > 0.5$. It is also interesting to observe that, if we indicate as $T_c(q)$ the value of $T$ delimitating the boundary between partial synchrony and incoherence in the $q-T$ plane, it appears to be linear with $q$ (Fig.~\ref{fig17}(b)) before entering in the region where the giant connected component emerges and synchronization becomes independent of $T$.

The Authors of Ref.~\cite{faggian2019synchronization} propose an analytical argument to explain this behavior, based on the following considerations. There are three time scales at work in this system. The first one rules local synchronization and is measured by $\tau_{LS}$ defined as the time needed by a pair of coupled oscillators to synchronize. This parameter is particularly relevant in the case of low connectivity. The second time scale is the local desynchronization time $\tau_{LD}$, indicating the time to lose synchronization by a pair after being disconnected. The third time scale is the one ruling network blinking and is measured by the effective rewiring time $\tau_{ER}$. Now, suppose that $\tau_{LS}$ is much smaller than the other time scales, a regime where the coupled oscillators quickly synchronize. Then, if two coupled oscillators become disconnected, two events may occur: either the oscillators become linked to other units of the system, in a time shorter than $\tau_{LD}$, thus effectively propagating the information to synchronize the whole system, or they lose synchrony before being linked to other units, thus effectively stopping the propagation of the information for synchronization. The first event will likely occur when $\tau_{LD} < \tau_{ER}$, the second in the opposite scenario, i.e., then $\tau_{ER}<\tau_{LD}$, such that the boundary between synchrony and incoherence could be identified by the condition $\tau_{ER} \approx \tau_{LD}$. Taking into account that the relevant time scales can be approximated as $\tau_{LD} \approx \frac{\pi}{2\sqrt{2}\sigma}$ and $\tau_{ER}\approx \frac{T}{q}$ (see Ref.~\cite{faggian2019synchronization} for a detailed discussion), then one derives that:

\begin{equation}
T_c(q) \approx \frac{\pi}{2\sqrt{2}\sigma}q,
\end{equation}

\noindent that represents a linear relationship between $T_c$ and $q$ as expected.

Far from being limited to this specific example, the interplay between the diverse time scales that regulate the dynamics of coupled oscillators in temporal networks plays a crucial role in the emergence of synchronization. Along this review, we will see many other examples where the way in which these time scales interact is a fundamental determinant for the emerging dynamics.

Let us now continue the analysis of the model focusing on the limit of fast rewiring, i.e., $T\rightarrow 0$. Under this hypothesis the blinking network dynamics is equivalent to that of a system of oscillators coupled via an adjacency matrix fixed in time and obtained as the time-average of the temporal adjacency matrix. The generic $ij$ element of this matrix is computed as follows:
\begin{equation}
\frac{1}{T_{av}}\bigintss_{0}^{T_{av}}\frac{{\mathcal{A}_{ij}}^t(T)}{d_i(t)}dt=\sum\limits_{l=1}^{N-1}\frac{1}{l}
p^l{(1-p)}^{N-1-l}\frac{(N-2)!}{(l-1)!(N-1-l)!},
\label{vc3}
\end{equation}
\noindent where $T_{av}$ is the time window where the average is calculated. This has to be sufficiently large in order to include many rewiring events. Notice that the term $p^l{(1-p)}^{N-1-l}\frac{(N-2)!}{(l-1)!(N-1-l)!}$ represents  the probability that the $i$-th node is interacting with the $j$-th node having exactly $l-1$ other connections.

Plugging $p=\frac{q}{N}$ in Eq.~(\ref{vc3}) and considering the limit for large $N$, one can derive the following approximation:

\begin{equation}
\frac{1}{T_{av}}\bigintss_{0}^{T_{av}}\frac{{\mathcal{A}_{ij}}^t(T)}{d_i(t)}dt \approx \frac{1-e^{-q}}{N}.
\label{vc4}
\end{equation}

Consequently, under the hypothesis of fast switching, the blinking network in Eq.~(\ref{vc1}) will exhibit the same behavior of the time-invariant network described by the following dynamics:

\begin{equation}
\dot {\phi}_i=\omega_i+\frac{\epsilon(1-e^{-q})}{N}\sum\limits_{j=1}^{N}\sin(\phi_j-\phi_i),
\label{vc9}
\end{equation}
\noindent with $i=1,\ldots, N$. Here, it is very important to hallmark that the average network is characterized by an all-to-all topology. We will see the same results arising in several other setups of temporal networks under the fast switching approximation.

An interesting study has later pointed out that properly designing the temporal network can also favor synchronization without necessarily requiring to operate at a fast rewiring rate, see Ref.~\cite{leander2015controlling}. In particular, the results of this work leverage the technique of optimal control to consider either the case of minimizing the connectivity cost of a network of phase oscillators with prescribed synchrony or maximizing the synchrony of a network with bounded connectivity cost.
The snapshots of the temporal network obtained with this method will no more have random topologies, but will be given by the result of an optimization procedure. The physical mechanism induced by this technique is to preserve synchrony by shortening the duration of low synchrony states and lengthening the duration of high synchrony states.

\subsubsection{Adaptive networks}

As introduced in Sec.~\ref{subsec2:adaptativenetworks}, adaptive networks are time-varying graphs where the node dynamics and the network topology evolve together. In particular, here we consider the case where the collective dynamics influencing (and influenced by) the link evolution is synchronization, and the nodes are described by phase oscillators. Particularly relevant applications, where synchronization develops through links which continuously adapt in time, are  biological networks, and, more specifically, networks of neurons. Here, time-dependent synaptic plasticity and Hebbian learning  play a fundamental role in the emergence of high-level computational tasks such as learning and memory.

\begin{figure}[t]
\centerline{\includegraphics[scale=0.5]{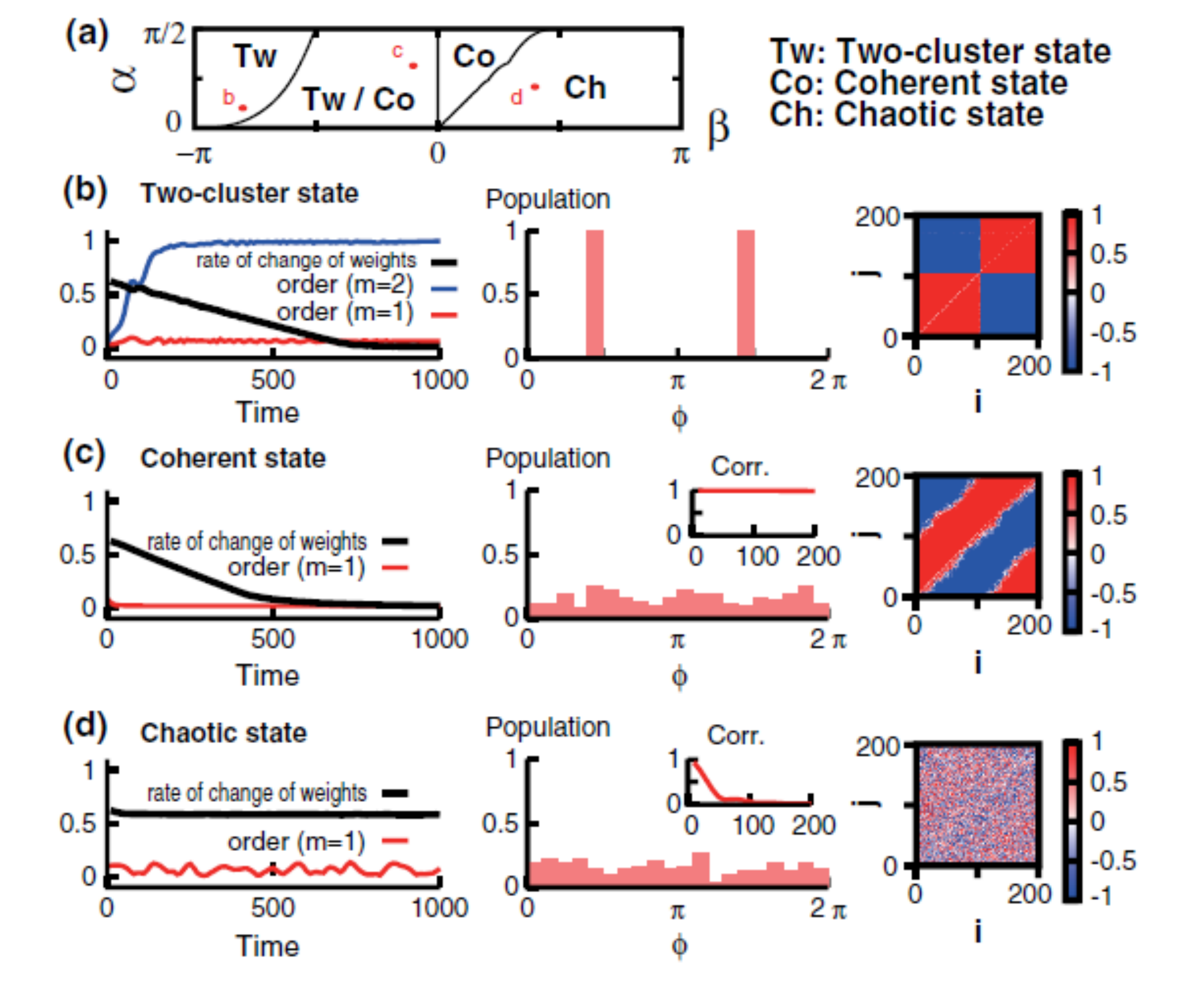}}
\caption{Synchronization in an adaptive network of Kuramoto oscillators ruled by Eqs.~(\ref{eq:adaptive1}) and~(\ref{eq:adaptive1weights}). (a) Phase diagram illustrating the emergence of three states: two-cluster state (Tw), coherent state (Co) and chaotic state (Ch). (b) Time evolution of the order parameters and of the normalized rate of change of the total weights (left panel), distribution of the phases at fixed time $t$ (middle panel), and color-coded representation of the weight matrix at the final time for the two-cluster state. (c) Same as in (b), for the coherent state. (d) Same as in (b), for the chaotic states. For all the parameters and settings of the simulation see Ref.~\cite{aoki2009co}.
\\{\it Source}: Reprinted figure with permission from Ref.~\cite{aoki2009co} \textcopyright~2009 by the American Physical Society.
\label{fig:adaptive1}
}
\end{figure}

As a first example of an adaptive network, let us consider the model introduced in Ref.~\cite{aoki2009co}, that consists of $N$ phase oscillators described by:

\begin{equation}
\dot {\phi}_i=1+\frac{1}{N}\sum\limits_{j=1}^{N}{w_{ij}}(t)\sin(\phi_j-\phi_i-\alpha),\\
\label{eq:adaptive1}
\end{equation}

\noindent where the coupling weights $w_{ij}(t)$ evolve in time with a dynamics given by:

\begin{equation}
\dot {w}_{ij}=-\epsilon \sin(\phi_i-\phi_j+\beta),\\
\label{eq:adaptive1weights}
\end{equation}

\noindent with the additional constraint that $|w_{ij}(t)|\leq 1$, so that to avoid an indefinite growth of the weights. In this model, all the oscillators have the same natural frequency, for convenience set to one, i.e., $\omega_i=1$. Furthermore, the time scale of the network evolution, which is given by $\epsilon^{-1}$, is set to be much larger than that of the oscillator dynamics, which implies $\epsilon \ll 1$. Here, $\alpha$ and $\beta$ are two control parameters of the model; by varying them as in the phase diagram shown in Fig.~\ref{fig:adaptive1}(a), the model displays three types of asymptotic behavior: a two cluster state, a coherent state, and a chaotic state.

In more details, for $\beta \in (-\pi,0)$ and $\alpha \in (0,\pi/2)$ a two cluster state may be observed. In this regime (Fig.~\ref{fig:adaptive1}(b)), the oscillators split in two groups with anti-phase synchronization, being the ratio of the two populations generally function of the initial conditions of the system (if the initial phases are chosen uniformly in the interval $[0,2\pi)$, then the two clusters have almost the same size, as in the middle panel of Fig.~\ref{fig:adaptive1}(b)). This is a stable state as the rate of change of the total weight converges to zero (left panel of Fig.~\ref{fig:adaptive1}(b)). The weights thus reach a stationary value such that if two nodes $i$ and $j$ belong to the same cluster then $w_{ij}=1$, and otherwise $w_{ij}=-1$ (right panel of Fig.~\ref{fig:adaptive1}(b)). For $\beta \in (-\pi/2,\pi/2)$ and $\alpha \in (0,\pi/2)$, the system also displays a coherent state with the phases of the oscillators almost uniformly distributed (Fig.~\ref{fig:adaptive1}(c)). Furthermore this state is stable as the rate of total weight change approaches zero. Finally, for $\beta \in (0,\pi)$ and $\alpha \in (0,\pi/2)$ a chaotic state with frustration is also found (Fig.~\ref{fig:adaptive1}(d)). In this case, the rate of total weight change does not vanish such that the network does not reach a stationary state. Instead, the weights continue to evolve in time. In this state, phases are not synchronized but evolve in a chaotic way. In all the scenarios illustrated in Fig.~\ref{fig:adaptive1}, the parameter $\epsilon$ has been set to $\epsilon=0.005$, and the initial values for the weights, i.e., $w_{ij}(0)$, have been set to random values selected from a uniform distribution in $[-1,1]$.

The effect of heterogeneous natural frequencies in this system has been studied in Ref.~\cite{aoki2011self}, where it is shown that the same qualitative states also emerge when the oscillators are non-identical. In particular, a larger heterogeneity is reflected into a slower convergence to the two-cluster state.

As we have already seen in our first example of adaptive network of Kuramoto oscillators, the coevolving dynamics that is obtained depends on the specific rule used to model weight adaptivity. In this example, in particular, changing a parameter ($\beta$) was sufficient to elicit different regimes both with respect to the collective behavior observed and to the structure of the emerging network. The rule to update the weights plays therefore a crucial role in shaping the behavior of an adaptive network. Other models for the weight dynamics, in fact, lead to networks exhibiting other self-organizing features. For instance, in Ref.~\cite{ren2007adaptive} the Authors study a network described by the following equations:

\begin{equation}
\dot {\phi}_i=\omega_i+\frac{1}{N}\sum\limits_{j=1}^{N}{w_{ij}}(t)\sin(\phi_j-\phi_i),\\
\label{eq:adaptive2}
\end{equation}

\noindent with weight dynamics given by:

\begin{equation}
\dot {w}_{ij}=\epsilon (\alpha |\sin(\beta(\phi_i-\phi_j))| -w_{ij}).
\label{eq:adaptive2weights}
\end{equation}

This rule consists of two factors. The first factor, $\alpha |\sin(\beta(\phi_i-\phi_j))|$, is a growth term that becomes stronger for oscillators having a larger phase difference, such that weights of links between oscillators with a large phase mismatch are updated by a larger quantity compared to oscillators with similar phases. The second factor, $-w_{ij}$, represents a forgetting/discharge term guaranteeing a continuous adaptation of the weights. With this choice of the coupling adaptation rule, the ability of the network to achieve synchronization is enhanced, since the weight of each link is adapted to the mismatch in the intrinsic frequencies of the oscillators at its extremes. In fact, the Authors of Ref.~\cite{ren2007adaptive} show that the steady-state distribution of the weights is $w_{ij} \propto |\omega_i-\omega_j|^\mu $ with $0.85 < \mu < 0.95$, which is very close to the criterion of optimal coupling law, in the sense of least average coupling cost, that corresponds to $\mu=1$. In addition, with such an adaptive coupling way, the system self-organizes such that larger coupling strengths are used when needed, i.e., during the transient, to quickly converge towards the synchronization regime, and then smaller weights are used to maintain the synchronization state.

Another very interesting model of adaptive network of Kuramoto oscillators investigates the structures that emerge in the presence of a constraint in the resources available to a node to establish links with the other units of the system~\cite{assenza2011emergence,gutierrez2011emerging}. The model incorporates two fundamental ingredients: homophily, taken into account by defining a weight update law that enhances connections between units with similar phases, and homeostasis, by incorporating a mechanism that preserves the sum of the weights of incoming connections at each node. In more detail, the dynamics of the adaptive network~\cite{gutierrez2011emerging} is described by:

\begin{equation}
\dot {\phi}_i=\omega_i+\epsilon\sum\limits_{j=1}^{N}{w_{ij}}(t)\sin(\phi_j-\phi_i),\\
\label{eq:adaptive3}
\end{equation}

\noindent and

\begin{equation}
\dot {w}_{ij}=p_{ij}-w_{ij}\sum\limits_{k \in \mathcal{N}_i}{p_{ik}},
\label{eq:adaptive3weights}
\end{equation}

\noindent where $p_{ij}$ is the average phase correlation between oscillators $i$ and $j$ in the time interval $T$:

\begin{equation}
p_{ij}=\frac{1}{T}\left | \,\, \int\limits_{-\infty}^t{e^{-(t-t')/T}e^{\iota (\phi_i(t') - \phi_j(t'))}dt'} \right |.
\label{eq:adaptive3p}
\end{equation}

\begin{figure}[t]
\centerline{\includegraphics[scale=0.38]{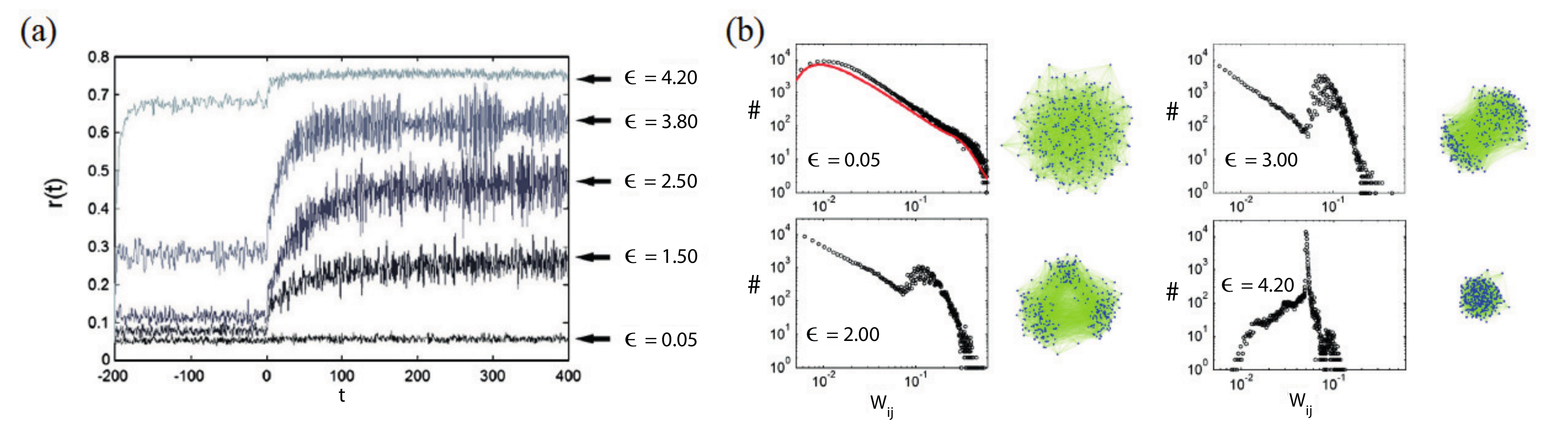}}
\caption{Synchronization in an adaptive network of Kuramoto oscillators ruled by Eqs.~(\ref{eq:adaptive3}) and~(\ref{eq:adaptive3weights}). The network has $N=300$ nodes. Other parameters are set as $K=20$ and $T=15$. (a) Time evolution of the Kuramoto order parameter $r(t)$ for different values of the coupling strength $\epsilon$. (b) Weight distribution and visual sketches of the corresponding network topology at different values of the coupling strength $\epsilon$.
\\{\it Source}: Reprinted figure with permission from Ref.~\cite{gutierrez2011emerging} \textcopyright~2011 by the American Physical Society.
\label{fig:adaptive3}
}
\end{figure}

From Eq.~(\ref{eq:adaptive3weights}) it follows that $\sum\limits_{j\in \mathcal{N}_i}w_{ij}=1$, such that the resources available for a node to establish/reinforce connections with other units are maintained constant in time. The competition between homophily (modeled by the first term in the right-hand side of Eq.~(\ref{eq:adaptive3weights})) and homeostasis (modeled by the second term in the right-hand side of the same equation) in this network leads to an enhancement in synchronization and to the emergence of a mesoscale of communities and a scale-free distribution in the connection weights. The enhancement in synchronization can be monitored by calculating the Kuramoto order parameter $r(t)$ at different values of the coupling strength $\epsilon$ (Fig.~\ref{fig:adaptive3}(a)). The network is initialized by randomly selecting a number $K$ of neighbors for each node and setting the initial value of each weight to $1/K$; the initial conditions for the phases are selected randomly from a uniform distribution in $[0,2\pi)$; the frequencies $\omega_i$ are also randomly assigned, from a uniform distribution in $(-\pi,\pi]$. Eqs.~(\ref{eq:adaptive3}) and~(\ref{eq:adaptive3weights}) are integrated without adaptation for 200 time units and, then, at $t=0$, the adaptation mechanism is activated, such that the adaptive case can be compared with the corresponding behavior in the non-adaptive counterpart of the temporal network. The time evolution of the Kuramoto order parameter $r(t)$ shows that synchronization is generally enhanced by the adaptive mechanism ruling weight evolution (Fig.~\ref{fig:adaptive3}(a)). This is particularly evident when the coupling strength is below the critical value to attain synchronization in a non-adaptive network, whereas for large $\epsilon$ the enhancement in synchronization is smaller.

However, the coupling strength influences not only the extent of enhancement in synchronization but also the  structures emerging during adaptation (Fig.~\ref{fig:adaptive3}(b)). For small $\epsilon$, the topology resulting from the competitive adaptive mechanism is highly heterogeneous, characterized by a weight distribution close to a power-law, indicating that in this case a complete redistribution of the weights occurred. As the coupling strength increases, the weight distribution exhibits a local maximum and the network is segregated into distinct modules, each characterized by its own frequency of oscillation. For instance, at $\epsilon=2$ three communities appear, whereas at $\epsilon=3$ the network splits into two communities. Finally, when the coupling strength is larger than the synchronization threshold in the non-adaptive counterpart of the temporal network, then the weight distribution becomes peaked at $1/K$ (this behavior is exemplified in Fig.~\ref{fig:adaptive3}(b), at $\epsilon=4.2$). In these conditions, as the network is able to reach synchronization even in the absence of adaptation, the weights do not significantly change from their initial value $1/K$.

In the last model we illustrate here, the rule of connectivity is stochastic and based on a fitness. At variance with what discussed so far, this means that the weights are all equal and constant in time, and only the topology is instantaneously updated following the so-called fitness or hidden variable network model~\cite{caldarelli2002scale,garlaschelli2004fitness}. According to this model, each pair of nodes is coupled with a probability given by a fitness function. In the adaptive network of Kuramoto oscillators, this fitness depends on the phases of the oscillators, i.e., $f(\phi_i,\phi_j)$, where $\phi_i$ and $\phi_j$ are the phases of the oscillators $i$ and $j$ that evolve as in Eqs.~(\ref{eq:adaptive3}). In Ref.~\cite{eom2016concurrent}, the Authors select the function $f(\phi_i,\phi_j)$ to embed a homophily mechanism in the adaptive network:

\begin{equation}
\begin{array}{lcl}
 f(\phi_i,\phi_j)=\frac{1}{N}[z\{1+\cos(\phi_i-\phi_j)\}],\\
\end{array}
\label{ps2}
\end{equation}

\noindent where $z$ is a positive parameter. In this way, the wiring probability for nodes with similar phases is higher than that for nodes having a larger phase difference. For oscillators with the same phase, i.e., $\phi_i = \phi_j$, the fitness becomes: $f(\phi_i,\phi_j)=\frac{2z}{N}$, whereas for oscillators with a phase lag equal to $\pi$, i.e., $| \phi_i-\phi_j|=\pi$, $f(\phi_i,\phi_j)=0$. Therefore, high values of $z$ yield a more connected structure. The parameter $\epsilon$ instead control the coherence of the oscillators, with higher values leading to a more coherent regime.

Numerical simulations of this model reveal interesting features, in particular how in this structure percolation and synchronization are simultaneously enhanced. Percolation, that is one of the most important phenomena occurring in networks~\cite{ho2007phase,dorogovtsev2008critical}, is the critical transition observed when the addition of a small number of links makes a substantial number of components of the network to become connected~\cite{albert2000error,cohen2000resilience,achlioptas2009explosive,buldyrev2010catastrophic,vespignani2010fragility,karrer2014percolation,radicchi2015percolation,riordan2011explosive}. Here, it is monitored by measuring the size of the largest connected component $s(t)$ and, in particular, its steady state value $s_\infty$. Synchronization is instead monitored by the classical Kuramoto order parameter as in Eq.~(\ref{ps3}) and its steady-state value $r_\infty$.

\begin{figure}
\centerline{\includegraphics[scale=0.700]{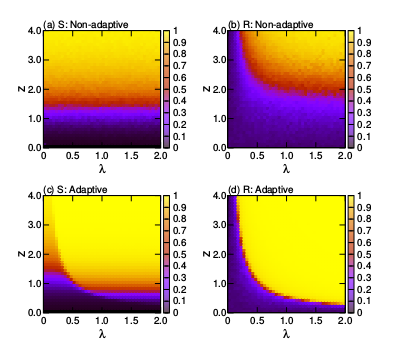}}
\caption{Concurrent enhancement of synchronization and percolation in an adaptive network of Kuramoto oscillators with fitness-based weight update. Phase diagrams of the non-adaptive (a,b) and adaptive (c,d) case for $N=150$. (a), (c) Percolation indicator $s_\infty(\epsilon,z)$. (b), (d) Steady-state Kuramoto order parameter $r_\infty(\epsilon,z)$.
\\{\it Source}: Reprinted with permission from Ref.~\cite{eom2016concurrent}.
\label{fig15}
}
\end{figure}

Following Ref.~\cite{eom2016concurrent}, let us consider a network with $N=150$ nodes and natural frequencies assigned from a uniform distribution in the range $[-1,1]$. Figure~\ref{fig15} shows how percolation and synchronization varies as a function of the two parameters $\epsilon$ and $z$, ruling the model, for both the non-adaptive (fixed weights) and the adaptive case (fitness-based weight update). The transition to percolation in the non-adaptive case only depends on $z$ (Fig.~\ref{fig15}(a)). In particular, the typical regimes are observed, with the subcritical regime ($s_\infty \sim 0.0$) appearing for $z < 1$, the critical regime for $z \sim 1$, the supercritical regime ($0 < s_\infty < 1$) for $1 < z < 3$, and the connected regime ($s_\infty \sim 1.0$) for $z > 3$. Correspondingly, diverse levels of synchronization are found, depending on the percolation state where the network is (Fig.~\ref{fig15}(b)): incoherent states $r_\infty < 0.05$ in the sub-critical and critical regimes ($z < 1$, $\forall \epsilon$); partial synchronization states ($0.1 < r_\infty < 0.9$) in the supercritical regime; and highly synchronized states in the connected regime $z > 3$. A different scenario emerges in the adaptive case, with a significant enhancement of both percolation and synchronization (Figs.~\ref{fig15}(c) and~\ref{fig15}(d)). Now, percolation becomes function of both $z$ and $\epsilon$, with larger values of $\epsilon$ favoring the emergence of a giant component at smaller values of $z$ and a significant enhancement of percolation observed in particular in the region $z<3$. Similarly, also synchronization benefits from the presence of the adaptation mechanism, and, again, this is particularly evident in the region $z<3$ where the connectivity is low. Overall, these results lead to the conclusion that the adaptivity instaurates a positive feedback loop between network structure and dynamics that enables the emergence of synchronization and connected components even in regions when the resources for interactions are limited.
Quite interestingly, recent results \cite{berner2020birth} have shown that adaptive mechanisms can also induce regime beyond that of global synchronization, leading, for instance, to the formation of partial synchronization patterns. To show this, the Authors of Ref.~\cite{berner2020birth} actually consider a network organized in more than one layer and demonstrate how multiplexing can induce various stable phase cluster states that are not stable or do not exist in the monolayer structure.

\subsection{Monolayer networks: limit cycle and chaotic oscillators}
\label{sec:ChaoticTN}

\subsubsection{Blinking networks}

Our first example of blinking networks of chaotic oscillators is devoted to illustrate the fast switching stability criterion. In particular, as in Ref.~\cite{stilwell2006sufficient}, we focus on a set of $N$ coupled identical R\"{o}ssler oscillators.

The dynamics of the system is described by the following equations:
\begin{equation}\label{eq_1}
\begin{array}{lcl}
\dot{x}_i(t)=-y_i(t)-z_i(t)-\epsilon\sum\limits_{j=1}^N\mathcal{L}_{ij}(t)x_j(t),\\
\dot{y}_i(t)=x_i(t)+ay_i(t),\vspace{7pt}\\

\dot{z}_i(t)=b+z_i(t)(x_i(t)-c),\\

\end{array}
\end{equation}

\noindent where $i=1,\ldots, N$ and $\mathcal{L}(t)$ is the time-varying Laplacian describing the blinking network. The system parameters are chosen as $a=0.2$, $b=0.2$ and $c=5.7$ so that the uncoupled dynamics is chaotic.

Here, we consider a blinking network with the following characteristics. At each snapshot the connectivity is given by an Erd\"{o}s–R\'{e}nyi network model with edge joining probability set to $p = 0.1$. At each time step the underlying network is rewired with probability $p_r$. The network comprises $N=200$ identical R\"ossler oscillators.

\begin{figure}
\centerline{\includegraphics[scale=0.5]{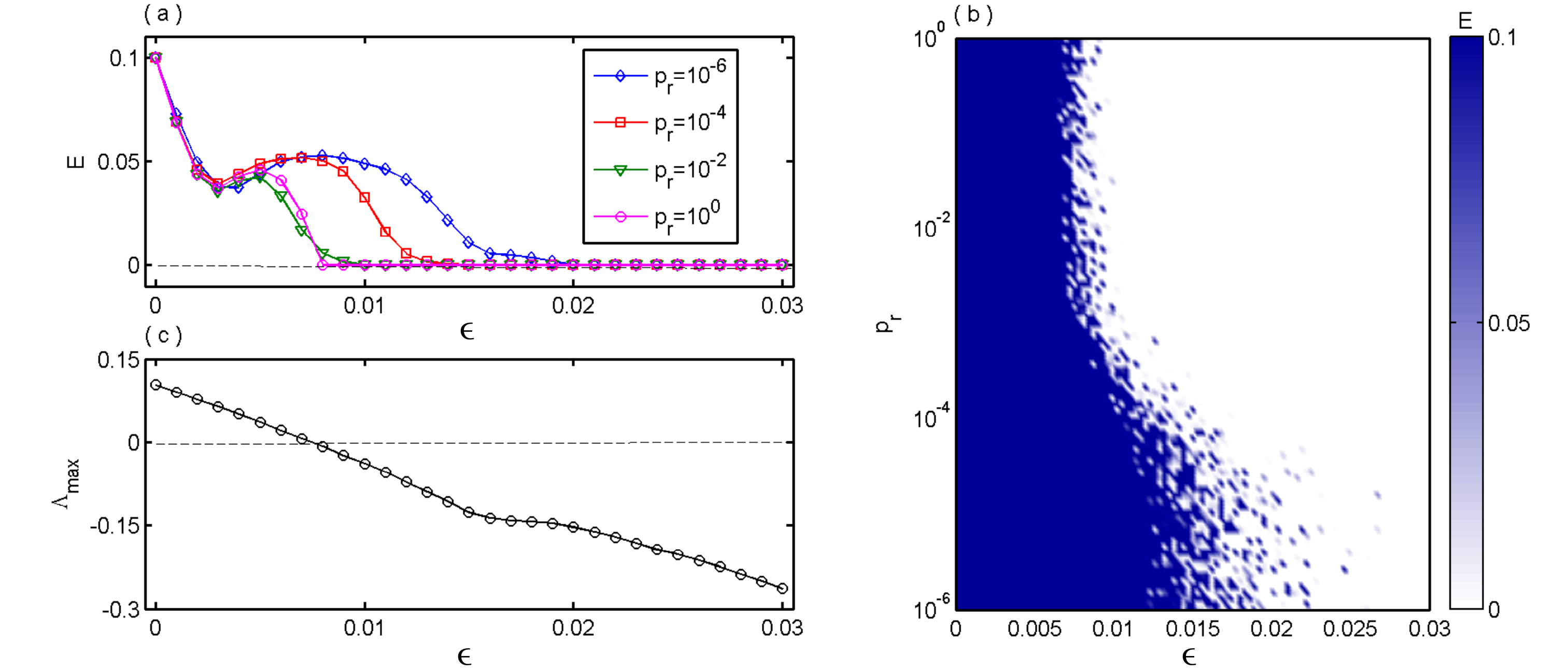}}
\caption{Synchronization in a blinking network of $N=200$ R\"{o}ssler oscillators. (a) Synchronization error $E$ vs. coupling strength $\epsilon$ for $p_r=10^{-6}$ (blue diamond line), $p_r=10^{-4}$ (red square line), $p_r=10^{-2}$ (green triangle line) and $p_r=10^{0}$ (magenta circle line). (b) Color coded two-dimensional plot of the synchronization error $E$ in the $(\epsilon,p_r)$ parameter plane. (c) Maximum transverse Lyapunov exponent $\Lambda_{max}$ of the time-average system vs. coupling strength $\epsilon$.
\label{fig_1}}
\end{figure}

To detect the emergence of complete synchronization, the synchronization error $E=\lim\limits_{T\to\infty}\frac{1}{T}\int_{t}^{t+T}\sum_{j=2}^{N}\frac{\norm{{\bf x}_j(t)-{\bf x}_1(t)}}{N-1}~d\tau$ is monitored.
In Fig.~\ref{fig_1}(a), this error $E$ is reported as a function of the coupling parameter $\epsilon$ for several values of the rewiring probability $p_r$. For small values of the rewiring probability, $p_r=10^{-6}$, the network is almost static as rewiring occurs very rarely, and global synchronization is observed for $\epsilon\geq 0.02$. Increasing the value of $p_r$ of two order of magnitude ($p_r=10^{-4}$) favors synchronization, that can be obtained for a lower value of the coupling $\epsilon$ (in particular, it is observed for  $\epsilon\ge0.015$). Finally, let us consider now the case of a very fast time scale in the temporal network evolution, obtained for  $p_r=10^{-2}$ or $p_r=10^{0}$. Synchronization at even lower values of the coupling, e.g., $\epsilon\ge0.012$ for $p_r=10^{-2}$ or $\epsilon\ge 0.008$ for $p_r=10^{0}$.

The results point out that a higher rewiring probability enhances synchrony, as it widens the interval of values of the coupling strength for which synchronization takes place, as compared to the case of a static network. To further illustrate the effect of the rewiring probability $p_r$ on synchronization, it is instructive to inspect the behavior of the synchronization error $E$ in the plane $(\epsilon,p_r)$ (Fig.~\ref{fig_1}(b)), where the enhancement in synchronization due to an increase in the rate of switching among the different link configurations is evident. Quite remarkably, this trend continues up to a certain limit, after which further increases of $p_r$ does not lead to changes in $E$.

The onset of synchronization, when $p_r$ is large, can be predicted by the fast switching stability criterion. In fact, one can calculate the time-average of the temporal network and use it to compute the maximum transverse Lyapunov exponent as a function of $\epsilon$, according to Eqs.~(\ref{rfssc_5}). The result is illustrated in Fig.~\ref{fig_1}(c) which clearly shows that the maximum transverse Lyapunov exponent becomes negative in correspondence of the values of $\epsilon$ for which the synchronization error $E$ for the blinking network with high $p_r$ becomes zero. Hence, we conclude that, as predicted by the fast switching stability criterion, if the static network of oscillators with topology given by the time-average of the temporal connectivity synchronizes, then the temporal network will also synchronize, provided that the switching occurs at a sufficiently enough high rate.

As a second example of synchronization in blinking networks, we now discuss oscillators interacting via on-off coupling strategy as in Ref.~\cite{chen2009synchronization}. In particular, the Authors of Ref.~\cite{chen2009synchronization} have found that switching the on-off coupling at a rate comparable to that of the node dynamics can be advantageous for synchronization, independently of the specific features of the network topology adopted to set the interactions during the `on' state. In more detail, the Authors have varied the on-off time scale considering a large interval of variations and studied the effects of this parameter on synchronization. When the on-off time scale is very small compared to that of the associated coupled node dynamics, the synchronization stability can be predicted by applying the fast switching stability criterion, and, hence, considering the static Laplacian matrix accounting for time-average couplings. As the on-off time scale is increased to become comparable to the time scale of node dynamics, several interesting features are identified, among which the most exciting finding is that one of the traditional bounds for synchronization, due to short-wavelength bifurcations (SWBs)~\cite{heagy1995short}, nearly disappears, yielding a neat advantage in allowing fast synchronization in large networks.

Taking into account that in on-off blinking networks a single, fixed set of interconnections is switched on and off, the dynamics of each node can be described as follows:
\begin{equation}\label{eq_5}
\dot{\bf x}_i=f({\bf x}_i)-\epsilon(t)\sum\limits_{i=1}^{N}\mathcal{L}_{ij}h({\bf x}_j),\hspace{20pt}i=1,2,\dots,N.
\end{equation}
\noindent where $\epsilon(t)$ is the on-off coupling strength. Two further parameters are needed to describe the network dynamics: $T$, representing the on-off period, and $\theta \in [0,1]$ representing the on-off rate. Connections among the oscillators are switched on for a fraction of the on-off period $T$ equal to $\theta$, such that for $hT<t<(h+\theta)T$ (with $h=0,1,2,\dots$) they are active and the coupling strength is set to $\epsilon(t)=\epsilon$, whereas for the remaining time $(n+\theta)T<t<(n+1)T$ they are turned off and $\epsilon(t)=0$. When $\theta=0$ the oscillators are isolated for all time, whereas for $\theta=1$ the connections are always active and the network is no more time-dependent.

For numerical illustration, we resort again to R\"{o}ssler oscillators, this time setting the parameters as $a=b=0.2$ and $c=7$, for which another chaotic attractor characterizes the uncoupled dynamics. In this case the time scale of the unit dynamics, estimated from the time evolution of the variable $x(t)$, is $T_{typical}=5.89$.

The specific way in which, in on-off blinking networks, the coupling depends on time prompts for a particularly convenient form of the linearized dynamics. To show this, let us consider again linearization around the synchronization manifold $\mathbf{x}_s$ and write the equations for the  deviation variables $\delta{\bf x}_i(t)={\bf x}_i(t)-{\bf x}_s(t)$ as:
\begin{equation}\label{eq_6}
\delta\dot{\bf x}_i=Jf({\bf x}_s)\delta{\bf x}_i-\epsilon(t)\sum\limits_{j=1}^{N}\mathcal{L}_{ij}Jh({\bf x}_s)\delta{\bf x}_j,
\end{equation}

At this point, one can apply the transformation $\mathbf{\eta}=\mathrm{T}^{-1}\otimes \mathrm{I}_n$ (with $\mathrm{T}$ such that $\mathrm{T}^{-1}\mathcal{L}\mathrm{T}$ is diagonal) to obtain $N$ blocks of the form
\begin{equation}\label{eq_7}
\dot{\bf \eta}_k=\big[Jf({\bf x}_s)-\epsilon(t)\lambda_kJh({\bf x}_s)\big]{\bf \eta}_k,
\end{equation}

\noindent and hence study the generic variational equation
\begin{equation}\label{eq_8}
\dot{\bf \xi}=\big[Jf({\bf x}_s)-\bar{\epsilon}(t)(\alpha+{\bf i}\beta)Jh({\bf x}_s)\big]{\bf \xi},
\end{equation}

\noindent where $\bar{\epsilon}(t)=\frac{\epsilon(t)}{\epsilon}$ is the normalized on-off coupling and equation (\ref{eq_8}) is called the coupling-dependent master stability equation (CMSE). The MSF then can be obtained by studying the largest Lyapunov exponent of the CMSE as a function of $\alpha$ and $\beta$, i.e., $\lambda_{\max}(\alpha+i\beta)$. At this point, the sign of $\lambda_{\max}$ at the points $\alpha+i\beta=\epsilon\gamma_k$ for the transverse modes $k=2,3,\dots,N$ is studied. It is here important to remark that $\gamma_k$ are the eigenvalues of the Laplacian matrix of the static configuration which is switched on and off in the blinking process. Only when all the transverse modes are located in the negative region of the MSF, the synchronous state is stable.

\begin{figure}
	\centerline{\includegraphics[width=0.5\textwidth]{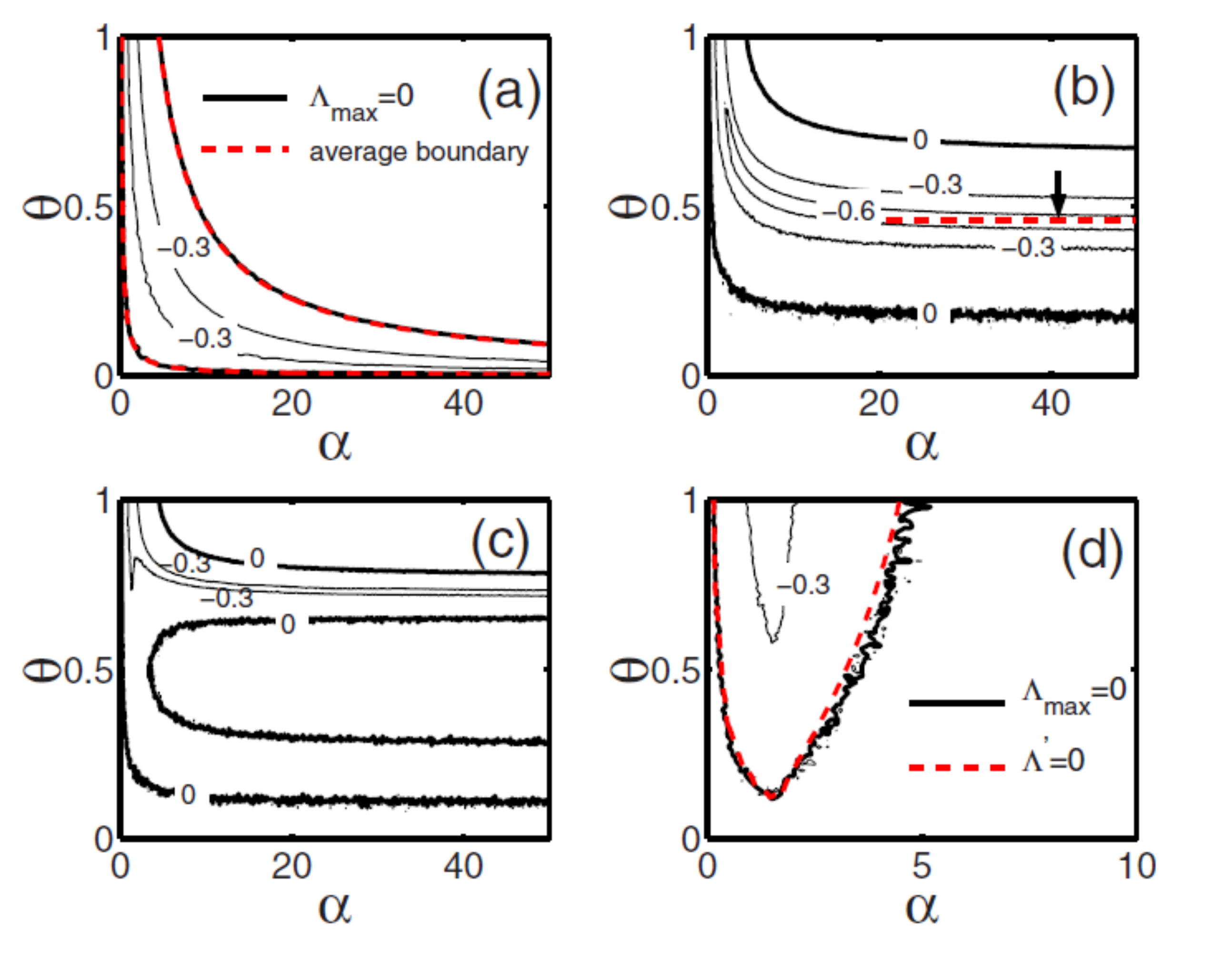}}
	\caption{Contour plot of the maximum transverse Lyapunov exponent $\Lambda_{\max}$ for an on-off blinking network of R\"{o}ssler oscillators in the plane $(\alpha,\theta)$ for different on-off period $T$: (a) $T=0.1$, (b) $T=3$, (c) $T=6$, (d) $T=60$. Black curves represent theoretical predictions, whereas the red ones the fitting from numerical simulations.
\\{\it Source:}	Reprinted figure with permission from Ref.~\cite{chen2009synchronization} \textcopyright~2009 by the American Physical Society. 
\label{fig_2}}
\end{figure}

To illustrate some numerical results, following Ref.~\cite{chen2009synchronization}, let us fix the coupling function $h({\bf x})$ as $h({\bf x})=[x~0~0]^T$, that produces in the case of time-independent links a type III MSF. We also consider that the interaction structure to switch is bidirectional, so that the Laplacian matrix $\mathcal{L}$ is symmetric and Eq.~(\ref{eq_8}) can be studied for real eigenvalues, i.e., $\beta=0$. The maximum Lyapunov exponent $\Lambda_{max}=\Lambda_{max}(\alpha,\theta)$ is shown for several values of the on-off period $T$ in Fig.~\ref{fig_2}. For $T=0.1 \ll T_{typical}$ (Fig.~\ref{fig_2}(a)) the fast switching stability criterion applies. Let $\alpha_1$ and $\alpha_2$ the thresholds for stability obtained for continuous coupling, i.e., for $\theta=1$ (for the given parameters, we have $\alpha_1 \simeq 0.14$ and $\alpha_2 \simeq 4.48$) such that $\Lambda_{max}(\alpha,1)<0$ if $\alpha \in [\alpha_1,\alpha_2]$. In addition, let
$\bar{\alpha}=\frac{1}{T}\int_{t}^{t+T}\bar{\epsilon}(\tau)\alpha~d\tau=\alpha\theta$. Then, the region of negative $\Lambda_{max}(\alpha,\theta)$, that is the region where the synchronization manifold is stable, is bounded by two hyperbola, given by $\epsilon\theta\gamma_2=\alpha_1$ and $\epsilon\theta\gamma_N=\alpha_2$. Notice that the prediction by the fast switching stability criterion is accurate as the theoretical boundaries of the region correspond with those derived from numerical simulations of the on-off blinking network (Fig.~\ref{fig_2}(a)).

Consider now the scenario where the time scale of network blinking is of the same order of that of the node dynamics. Two examples are shown in Fig.~\ref{fig_2}(b) for $T=3$ and in Fig.~\ref{fig_2}(c) for $T=6$. Here, the synchronization region is no more bounded by the two hyperbolic boundaries, that on the contrary gradually become parallel to each other for large $\alpha$. Noticeably, in Fig.~\ref{fig_2}(b) the stability region is larger than in the fast switching regime, whereas in Fig.~\ref{fig_2}(c) it appears intertwined by another region where synchronization is impossible to attain. In this intermediate time scale regime the striking feature that emerges is that in an interval of values of $\theta$ the second threshold disappears, denoting a transition from a type III to a type II MSF behavior.

The last scenario we study occurs for a time scale of the on-off blinking much larger than the time scale of the unit dynamics. The Authors of Ref.~\cite{chen2009synchronization} observe that, starting from the scenario of comparable time scales, further increasing the time scale of blinking reduce the region of synchronization stability up to the case illustrated in Fig.~\ref{fig_2}(d) where this region looks very different from that of Fig.~\ref{fig_2}(a)-(b). To account for the behavior in this regime, one can consider that during each period $T$, in the first fraction of time, $\theta T$, the coupling is turned on, and the divergence of nearby trajectory is dominated by the maximum Lyapunov exponent $\Lambda_{\perp}$ (which thus correspond to the continuous coupling case, when $\theta$ is always zero), whereas, in the second fraction of time $(1-\theta)T$, the coupling is off and the divergence is dominated by the maximum Lyapunov exponent of the uncoupled dynamics, here indicated as $\Lambda_{\max}^0$. Taking into account these considerations, the maximum Lyapunov exponent $\Lambda_{max}$ for a given $\theta$ can be approximated as

\begin{equation}
    \Lambda_{max} \approx \Lambda'=\Lambda_{\perp}\theta+\Lambda_{\max}^0(1-\theta).
\end{equation}

Figure~\ref{fig_2}(d) shows that this approximation proves to accurately predict the behavior when the blinking occurs at a rate very low compared to the evolution of the unit dynamics.

Reference~\cite{jeter2015synchronization} provides further examples of networks with on-off coupling (in particular, of R\"ossler and Duffing oscillators) that, for intermediate switching frequencies, exhibit windows (in the study called \emph{windows of opportunity}), where synchronization is stable in the temporal network, but unstable in the time-average structure, and so
in the fast switching regime.

Finally, it is interesting to note that the main features observed in the numerical simulations of the on-off blinking network illustrated above also appear in real experiments with physical systems, such as nonlinear electronic circuits~\cite{chen2009synchronization}. Earlier experiments conducted on two on-off coupled Chua's circuits~\cite{fortuna2003experimental} show the typical behavior arising for type II system, with synchronization achievable for switching frequencies greater than a certain threshold. A complex pattern of intertwined regions of synchronization or desynchronization in real systems can be also observed when nonlinear electronic oscillators are switched among two different configurations as in~\cite{buscarino2018synchronization}.

\subsubsection{Activity driven networks}
\label{sec:syncADN}

Let us now consider an activity driven network of chaotic oscillators as in Ref.~\cite{buscarino2018synchronization}. As discussed in Sec.~\ref{sec:adnmodel}, in this network links are set according to the node activity rate so that a sequence of snapshots switched with a time scale here fixed at $\tau \Delta t$ (where $\Delta t$ is the integration step size of the unit dynamics) is obtained. $\tau$ represents an important control parameter for the system as it modulates the rate of switching among the possible configurations. In particular, for small $\tau$ (e.g., $\tau=1$) the system is close to fast switching regime, whereas increasing $\tau$ the dynamics of the network evolution becomes first comparable and then slower than that of the units at the nodes.

We consider first the network behavior in the fast switching regime. Under this hypothesis, we can consider the time-average Laplacian matrix $\mathscr{\bar{L}}$ and rewrite it as $\mathscr{\bar{L}}=\frac{1}{N_c}\sum\limits_{k=1}^{N_c}\mathcal{L}{t_k}$ where $N_c$ is the number of possible instantaneous configurations of the network (each being equally probable). If the switching is sufficiently fast, the generic $ij$ element of this matrix is given by the probability $p_{ij}^{ADN}$ that nodes $i$ and $j$ are connected at a given time. As this probability~\cite{starnini2014temporal} can be written as:
\begin{equation}
    p_{ij}^{ADN}=1-\left [ \left ( 1-\frac{\chi_i}{N^2}\right )\left ( 1-\frac{\chi_j}{N^2}\right ) \right ]^{mN},
\end{equation}

\noindent so that for $i\neq j$ we can approximate $\mathscr{\bar{L}}_{ij}$ as
\begin{equation}
    \mathscr{\bar{L}}_{ij}\approx p_{ij}^{ADN},
\end{equation}

\noindent and finally $\mathscr{\bar{L}}_{ii}=-\sum\limits_{j}\mathscr{\bar{L}}_{ij}$. The availability of an approximated expression for $\mathscr{\bar{L}}$ makes possible, on the one hand, to predict the boundaries for the synchronization stability region and, on the other hand, to gain some insights on the effect of the parameters underlying the architecture of the temporal network such as $m$ and $\gamma$. This analysis carried out in Ref.~\cite{buscarino2018synchronization} yields the conclusion that in both type II and type III systems increasing $m$ and $\gamma$ generally favors synchronization, even if the eigenvalues of the time-average Laplacian, and so network synchronization, strongly depend on the specific instance of the power-law distribution of the node activity rates.

The other extreme scenario is represented by the regime when the network evolves with a time scale large compared to that of the system dynamics. In this case, the approach described in Ref.~\cite{zhou2016synchronization} can be applied. Given that there are $N_c$ possible configurations for the network, then one has to verify that the condition $\sum\limits_{k=1}^{N_c}\Lambda_{max}(\epsilon \lambda_k^{j_k}) < 0$ is true for each combination of the transverse modes of the $N_c$ configurations $(j_1,\ldots,j_{N_c})$. Notice, however, that, for activity driven networks with a power-law distribution of the node activity rates and small $m$, as it typically occurs, each snapshot will very likely include at least one node not connected to any other unit of the network, such that each configuration will include at least one zero eigenvalue. In this case, the condition for synchronization would require that $N_c \Lambda_{max}(0)<0$ which is impossible to obtain for chaotic dynamics where $\Lambda_{max}(0)$ (the maximum Lyapunov exponent of the isolated dynamics) is positive. This yields the conclusion that activity driven networks with a power-law distribution of the node activity rates and small $m$ are difficult to synchronize in the regime of slow switching. This analysis also inform about possible mechanisms to promote synchronization under this regime: they have to be directed towards reducing the likelihood of finding isolated nodes, for instance increasing $m$ or changing the shape of the activity rate distribution.

The numerical results discussed in Ref.~\cite{buscarino2018synchronization} confirm the suitability of the fast switching stability criterion to predict the region of synchronization stability as well as the analysis under the slow switching regime pointing out the difficulty to synchronize activity driven networks with power-law distribution of the activity rates and small $m$. The Authors of Ref.~\cite{buscarino2018synchronization} also investigate the behavior for intermediate values of $\tau$. Here, they find that increasing $\tau$ first the region of synchronization stability widens and then if becomes unbounded. We can conclude that in activity driven networks the same qualitative behavior observed in the examples of on-off blinking networks discussed above as well as in other temporal networks (see for instance Ref.~\cite{kohar2014synchronization}) arises, suggesting that this behavior does not depend on the specific features of how the time evolution shapes the link formation. The overall picture that emerges from these studies is that, in temporal networks, synchronization is generally promoted by fast network evolution, but more favourable conditions for synchronization can occur at switching rates slightly lower than the fast switching regime.

\subsubsection{Adaptive networks}
\label{sec:adaptivenetworksChaotic}

As the analysis based on the MSF formalism demonstrates, for unweighted networks of limit cycle and chaotic oscillators the problem of synchronization stability is not trivial. In particular, for systems with type III MSF the coupling strength needs to be properly selected and we can encounter situations where it is not possible to achieve synchronization. In this case, one can resort to weighted networks where links do not anymore have the same strength, but independent coupling coefficients. How can the values of such coefficients be found? Is it possible to use the self-organizing properties of temporal networks to let the network spontaneously evolve towards a set of suitable values? Adaptive networks provide an affirmative answer to these questions by furnishing a paradigm where the link weights change in time as a function of the network dynamical state. In addition, as we have already seen while illustrating the case of Kuramoto oscillators, in adaptive networks the structure of interconnections may be evolved as well, such that not only the values of the links, but also the presence or not of an interaction between two nodes is the result of the coevolution of network and dynamics.

There are several ways in which adaptation can be embedded in a network~\cite{delellis2010synchronization}. The adaptation strategy may be global (or centralized) if the same time-varying coupling $\epsilon=\epsilon(t)$ is used for all network links, or local (or decentralized), when an adaptive coupling coefficient is associated to each node, i.e., $w_{i}(t)$, or link, $w_{ij}(t)$. In addition, the network backbone can be kept constant, such that adaptation involves only a change of the weight of the interactions, or may be evolved, such that the adaptation mechanism is also able to cancel existing edges or form new links in the structure.

When a global adaptation strategy is used, the coupling strength is the same for all network nodes and is adapted taking into account the global status of the network, i.e., an information regarding all the oscillators of the system. An example of this strategy is provided in Ref.~\cite{de2008adaptive} where the following model is considered:
\begin{equation}\label{eq:adaptiveglobal}
\dot{\bf x}_i=f({\bf x}_i)-\epsilon(t)\sum\limits_{i=1}^{N}\mathcal{L}_{ij}h({\bf x}_j),
\end{equation}

\noindent with
\begin{equation}
\label{eq:adaptiveglobal_update}
\dot{\epsilon}=\frac{\mu}{N}\sum\limits_{j=1}^N \| h({\bf x}_j)- \frac{1}{N}\sum\limits_{j=1}^N h({\bf x}_j) \|,
\end{equation}

\noindent where the right hand side is function of the status of all network nodes. This strategy is able to drive the network towards synchronization. In correspondence of it, the coupling gain $\epsilon(t)$ reaches a stationary value. As further demonstrated in Ref.~\cite{chen2006network}, where, however, other update laws are used, the technique may be applied to diverse systems, such as networks of linearly or nonlinearly coupled oscillators, networks where the structure of interactions is not known, and networks where the backbone changes in time.

Notice that the adaptive mechanism of Eq.~(\ref{eq:adaptiveglobal_update}), as well as many others discussed in this section,  is based on increasing the coupling strength proportionally to a measure of the synchronization error. However, this adaptive mechanism fails when the units have a type III MSF and the initial weight is beyond the second threshold for synchronization. Similar cases require the use of a mechanism able to either increase or decrease the coupling strength. A suitable solution is the adaptive strategy introduced in Ref.~\cite{sorrentino2009adaptive} and successfully applied to synchronize a master-slave configuration of two Lorenz systems.

In local adaptive strategies, instead, each node makes use of information that can be gathered from its neighbors. To discuss these methods, let us begin with the so-called \emph{vertex-based approach} where an adaptive coupling coefficient is associated to each node. In particular, let us consider a system of coupled oscillators described by the following equations:
\begin{equation}\label{eq:adaptivelocalvertex}
\dot{\bf x}_i=f({\bf x}_i)-w_i(t)\sum\limits_{i=1}^{N}\mathcal{L}_{ij}h({\bf x}_j),
\end{equation}

\noindent where $w_i(t)$ is the time-varying coupling coefficient of node $i$ with $i=1,\ldots,N$. Several adaptive laws can be used for updating $w_i(t)$ to achieve synchronization. For instance, in Ref. \cite{zhou2006dynamical} the law is given by:
\begin{equation}\label{eq:adaptivelaw1}
\dot{w}_i=\frac{\mu \Delta_i}{1+\Delta_i},
\end{equation}

\noindent where $\mu>0$ and
\begin{equation}\label{eq:adaptivelaw2}
\dot{\Delta}_i=\left | h({\bf x}_i)-\frac{1}{k_i}\sum\limits_{i=1}^{N}\mathcal{A}_{ij}h({\bf x}_j) \right |.
\end{equation}

Another possibility is to select the update law as in Ref. \cite{de2008adaptive}:
\begin{equation}\label{eq:adaptivelaw3}
\dot{w}_i=\mu \left \| \sum\limits_{i=1}^{N}\mathcal{A}_{ij}\left ( h({\bf x}_j)-h({\bf x}_i) \right ) \right \|.
\end{equation}

In both cases, the network structure is kept constant, i.e., the terms $\mathcal{A}_{ij}$ are time-invariant. The two strategies also share the idea that the nodes negotiate the coupling strength with their neighbours by comparing their own output with the average output of their neighbors. Both strategies yield synchronization in many large networks where synchronization cannot be obtained if all the weights are equal~\cite{de2008adaptive,zhou2006dynamical}. This is a particularly relevant result, as these adaptive strategies not only demonstrate the possibility to achieve synchronization, but also provide a self-tuning approach to obtain the suitable values of the weights which, otherwise, would be difficult to obtain.
When synchronization is reached, the weights stop to update as the right-hand term of Eqs.~(\ref{eq:adaptivelaw2}) or~(\ref{eq:adaptivelaw3}) becomes zero. In correspondence of synchronization, therefore, the weights reach stationary values. Note, however, that, when the oscillators are non-identical, complete synchronization cannot be achieved, and the weights will increase indefinitely. For this reason, under these circumstances, the update laws have to be modified to incorporate some mechanism of weight saturation.

A related problem is that of using an adaptive strategy to track synchronization in a time-varying network. In particular, the Authors of Ref.~\cite{sorrentino2008adaptive} consider a slowly changing network and assume that each node only knows a single coupling signal received from the other nodes of the network, i.e., $\mathbf{s}_i(t)=\sum\limits_j \mathcal{A}_{ij}h(\mathbf{x}_j(t))$. The Authors show that, under these conditions, synchronization can be achieved and maintained by properly setting the coupling gains (one for each node) with an adaptive law. Crucially, this requires coupling coefficients that also evolve in time to track the characteristics of the network that changes in time.

At variance with the vertex-based approach, in the \emph{edge-based approach} the negotiation occurs between each pair of nodes. To implement this approach, the following model can be used:
\begin{equation}
\dot{\bf x}_i=f({\bf x}_i)+\sum\limits_{i=1}^{N}w_{ij}(t)(h({\bf x}_j)-h({\bf x}_i)),
\end{equation}

\noindent with
\begin{equation}\label{eq:adaptivelaw4}
\dot{w}_{ij}=\mu \left \| h({\bf x}_j)-h({\bf x}_i) \right \|,
\end{equation}

\noindent where $i$ and $j$ are such that $(i,j) \in \mathcal{E}$. In this way, the backbone of the network is kept constant in time, but its weights evolve in time.

For the special case where $h({\bf x}_i)={\bf x}_i$, that is, when coupling occurs through all the state variables, the global asymptotic stability of both vertex-based and edge-based adaptive strategies can be proved via Lyapunov-based techniques~\cite{de2008synchronization}. To illustrate this result, the notion of QUAD functions is required~\cite{chen2006network}. A function $f:\mathbb{R}^n\times \mathbb{R}^+ \rightarrow \mathbb{R}^n$ is QUAD if and only if for any $\mathbf{x},\mathbf{x}' \in \mathbb{R}^n$, then
\begin{equation}
(\mathbf{x}-\mathbf{x}')^T[f(\mathbf{x},t)-f(\mathbf{x}',t)]-(\mathbf{x}-\mathbf{x}')^T\Delta (\mathbf{x}-\mathbf{x}')\leq -\bar{\omega}(\mathbf{x}-\mathbf{x}')^T(\mathbf{x}-\mathbf{x}'),
\end{equation}

\noindent with $\Delta$ a diagonal matrix and $\bar{\omega}$ a positive constant.

Following Ref.~\cite{de2008synchronization}, we also need to introduce several matrices. Let $\mathcal{L}$ be the Laplacian matrix describing the network backbone (undirected and unweighted) and let $\mathbf{v}_1$ be the normalized left eigenvector of $\mathcal{L}$. Define $\mathrm{V}$ as the diagonal matrix that contains the elements of $\mathbf{v}_1$ and let $\mathrm{U}=\mathrm{V}-\mathbf{v}_1\mathbf{v}$. Finally, let $\mathcal{L}^w$ a matrix with coefficients
$\mathcal{L}^w_{ij}=w_{ij}(t)$ if $(i,j) \in \mathcal{E}$, $\mathcal{L}^w_{ii}=-\sum\limits_{j=1}^N w_{ij}(t)$, and $\mathcal{L}^w_{ij}=0$ otherwise.

At this point, we can summarize the sufficient conditions for the vertex-based and edge-based strategies to achieve synchronization (Theorem 1 of Ref.~\cite{de2008synchronization}). If $f$ is QUAD and the matrix $(\mathcal{L}\otimes \mathrm{I}_n)(\mathrm{I}_N\otimes \Delta)+(\mathcal{L}\otimes \mathrm{I}_n)(\mathcal{L}^w\otimes \mathrm{I}_n)$ is negative semi-definite $\forall t \geq 0$, then the vertex-based adaptive strategy of Eq.~(\ref{eq:adaptivelaw3}) guarantees synchronization. If $f$ is QUAD and the matrix $(\mathscr{V}\otimes \mathrm{I}_n)(\mathrm{I}_N\otimes \Delta)+(\mathscr{V}\otimes \mathrm{I}_n)(\mathcal{L}^w\otimes \mathrm{I}_n)$ is negative semi-definite $\forall t \geq 0$, then the edge-based adaptive strategy of Eq.~(\ref{eq:adaptivelaw4}) guarantees synchronization. These conditions take a particularly convenient form when $\Delta =0$ (Corollary 1 of Ref.~\cite{de2008synchronization}), in this case it suffices to check that $f$ is QUAD with $\Delta =0$ and that the network is connected, to conclude that the two strategies guarantee synchronization.

\begin{figure}[t]
	\centerline{
		\includegraphics[scale=0.35]{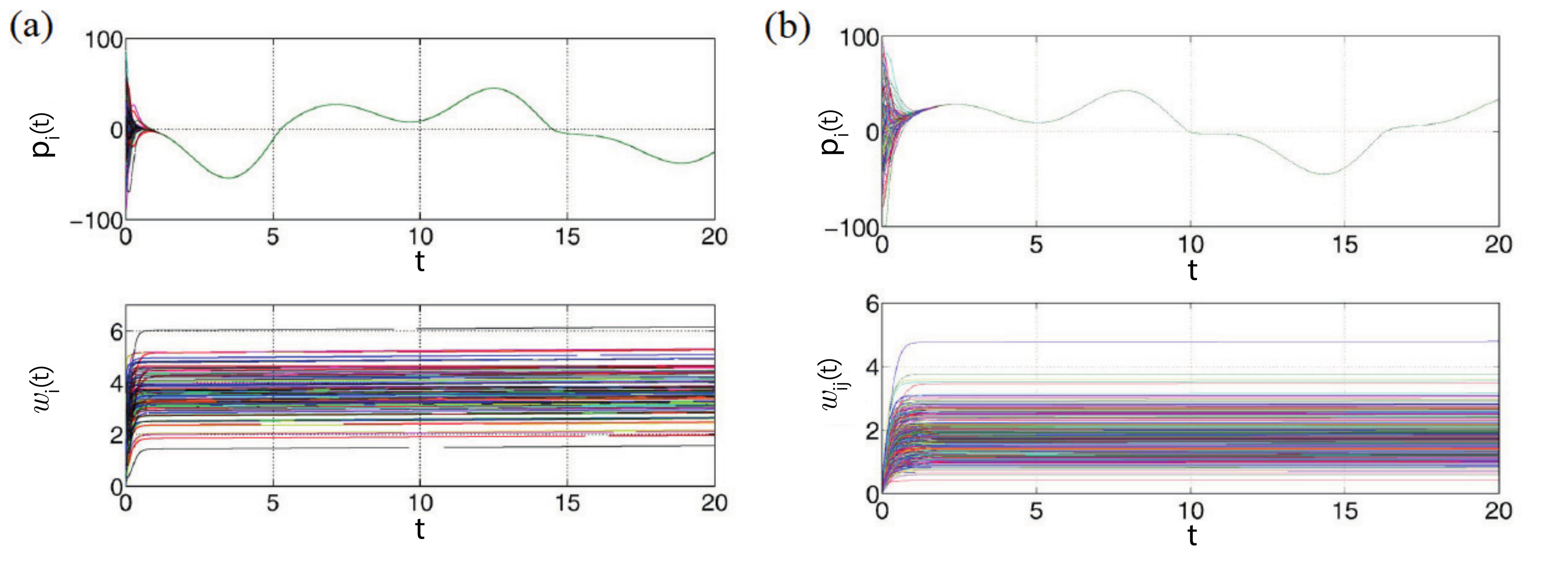}}
	\caption{Adaptive networks of $N=100$ chaotic oscillators (Chua's circuits): (a) vertex-based adaptive strategy; (b) edge-based vertex strategy. Time evolution of the first state variables for each node of the network (top) and of the link weights. The network topology is scale-free.
		\\{\it Source:} Reprinted figure from Ref.~\cite{de2008synchronization}, with the permission of AIP Publishing.
		\label{fig:adaptivenetworkchaotic}
}
\end{figure}

As a numerical example we refer to a network of Chua's circuits as in Ref.~\cite{de2008synchronization}. The Chua's circuit is QUAD with $\Delta=0$; in addition, the network considered is connected (in particular, its topology is scale-free), such that the conditions mentioned above (Corollary 1 of Ref.~\cite{de2008synchronization}) hold and the two strategies yield stable synchronization (Fig.~\ref{fig:adaptivenetworkchaotic}, top). As expected, the weights of the network links converge to stationary values as shown in Fig.~\ref{fig:adaptivenetworkchaotic}, bottom, where, in particular, the fast dynamics of the link evolution compared to that of the chaotic oscillators can be appreciated.

Interestingly, the conditions above mentioned for synchronization stability have been extended to a wider class of edge-based adaptive laws in Ref.~\cite{delellis2009novel}, where the Authors demonstrate that they also hold when a general formulation of the update law for the weights is adopted:
\begin{equation}
\label{eq:adaptivelaw5}
\dot{w}_{ij}=g({\bf e}_{ij}),
\end{equation}

\noindent with ${\bf e}_{ij}={\bf x}_j-{\bf x}_i$. In more detail, the Authors of Ref.~\cite{delellis2009novel} demonstrate that if $f$ is QUAD, the matrix $(\mathscr{V}\otimes \mathrm{I}_n)(\mathrm{I}_N\otimes \Delta)+(\mathscr{V}\otimes \mathrm{I}_n)(\mathcal{L}^w\otimes \mathrm{I}_n)$ is negative semi-definite $\forall t\geq 0$, and $g({\bf e}_{ij})$ is a monotonously increasing function of the error norm and is such that $g(0)=0$ and for some finite constant $m$, $0 \leq g({\bf e}_{ij}) \leq m$, or, alternatively, $g({\bf e}_{ij})=\mu \| {\bf e}_{ij} \|^p$ with $0 <p \leq 2$, then the edge-based strategy guarantees synchronization for any initial condition, i.e., $\lim\limits_{t\rightarrow +\infty}{\bf e}_{ij}=0$ $\forall {\bf e}_{ij}(0)$, and $\lim\limits_{t\rightarrow +\infty}{w}_{ij}(t)=\bar{w}_{ij}$ $\forall (i,j) \in \mathcal{E}$.

The conditions on the node dynamics can also be relaxed. The hypothesis we have discussed so far practically requires that the node vector field is contracting, but a larger class of systems, including unstable linear systems, systems in Lur’e form, Lipschitz vector fields, and systems with bounded Jacobians, can be considered as discussed in Ref.~\cite{yu2012distributed}.

Two other results of Ref.~\cite{yu2012distributed} are worth mentioning here. First, by strengthening the condition of the vector field $f$, the stability analysis can be also extended to the case where the oscillators are coupled through a subset of the state variables, i.e., $h((\mathbf{x}_j))=\mathrm{B}\mathbf{x}_j$ with $\mathrm{B}$ positive semi-definite. Second, synchronization can be adaptively achieved also by only acting on a reduced set of the network links. According to this strategy, named \emph{edge pinning synchronization}, the majority of the links are kept constant, whereas a fraction of them is updated according to:
\begin{equation}
\label{eq:edgepinningsync}
\dot{w}_{ij}=-\mu_{ij}(\mathbf{x}_i-\mathbf{x}_j)^T\mathrm{B}(\mathbf{x}_i-\mathbf{x}_j),
\end{equation}

\noindent for $(i,j) \in \mathcal{\bar{E}} \subset \mathcal{E}$, with $\mu_{ij}$ positive constants.

Here, the term `pinning' is borrowed from a more vast literature that deals with techniques aimed at controlling synchronization and, more in general, the collective behavior of a complex network, only acting on a subset of the nodes, and that also includes decentralized adaptive strategies. The problem tackled in pinning control~\cite{sorrentino2007controllability,porfiri2008criteria} is to steer the network towards a specific reference trajectory $\mathbf{s}(t)$ by controlling a fraction of its nodes, at variance with self-organized synchronization where the trajectory on which all the nodes converge is spontaneously chosen by the system itself. In the classical scheme several parameters, such as the control gain and the coupling coefficients, need to be tuned, such that the inclusion of adaptive mechanisms may be beneficial. This idea has been pursued in several works~\cite{chen2007pinning,wang2008adaptive,zhou2008pinning,turci2012hybrid,turci2012adaptive,turci2014adaptive}. To illustrate it, let us consider the following system of coupled oscillators:
\begin{equation}\label{eq:adaptivepinning}
\dot{\mathbf{x}}_i=f(\mathbf{x}_i)-\sum\limits_{i=1}^{N}w_{ij}(t)\mathrm{B}\mathbf{x}_j-\delta_i q_i(t)\mathrm{B}(\mathbf{x}_j-\mathbf{x}_i),
\end{equation}
\noindent where $\delta_i=1$ for the pinned nodes and $\delta_i=0$ otherwise, $q_i(t)$ is the adaptive control gain, and
$w_{ij}(t)$ (with $(i,j)\in \mathcal{E}$) the adaptive weights of the network links. Although also centralized strategies may be used to adjust the control gain~\cite{wang2008adaptive,zhou2008pinning} or both the control gain and the common coupling coefficient~\cite{chen2007pinning}, here we focus on the local, decentralized adaptive strategy studied in Ref.~\cite{turci2014adaptive}. To update the weights, the following law is adopted
\begin{equation}
\label{eq:edgelawpinned}
\dot{w}_{ij}=\mu_{ij}(\mathbf{x}_i-\mathbf{x}_j)^T\mathrm{B}(\mathbf{x}_i-\mathbf{x}_j),
\end{equation}

\noindent for $(i,j) \in \mathcal{E}$, while the control gain is updated as follows:
\begin{equation}
\label{eq:edgelawpinnedcontrolgain}
\dot{q}_{i}=\mu_{i}(\mathbf{x}_i(t)-\mathbf{s}(t))^T\mathrm{B}(\mathbf{x}_i(t)-\mathbf{s}(t)).
\end{equation}

It suffices that $f$ is QUAD to guarantee that, for any number of pinned nodes, the network reaches synchronization with all nodes variables asymptotically converging to the desired reference trajectory and the coupling weights and the control gains to finite steady-state values. Extensions derived from this technique are the non-identical reference hybrid pinning strategy~\cite{turci2012hybrid}, the network chaos control hybrid pinning strategy~\cite{turci2012hybrid}, and the node-to-node fully adaptive decentralized pinning strategy~\cite{turci2012adaptive}.

Let us now consider adaptation techniques where the network structure itself is evolved and, more specifically, the \emph{edge snapping adaptation strategy}~\cite{delellis2010evolution}. We refer to a network of oscillators coupled as follows:
\begin{equation}\label{eq:adaptivelocaledge}
\dot{\bf x}_i=f({\bf x}_i)+\epsilon\sum\limits_{i=1}^{N}w_{ij}(t)(h({\bf x}_j)-h({\bf x}_i)),
\end{equation}

\noindent which correspond to Eqs.~(\ref{eq:adaptivelocaledge}) with the inclusion of a further global (and constant in time) gain $\epsilon$. In addition, let us consider weights that, rather than obeying to an update law as in Eq.~(\ref{eq:adaptivelaw5}), undergo a second order dynamics
\begin{equation}
\label{eq:edgesnapping}
\ddot{w}_{ij} +\gamma \dot{w}_{ij}+\frac{\partial }{\partial w_{ij}}V(w_{ij})=g(\mathbf{e}_{ij}),
\end{equation}

\noindent where $\gamma$ is the damping coefficient, $V(w_{ij})$ a potential function, and $g(\mathbf{e}_{ij})$ a function of the error $\mathbf{e}_{ij}=\mathbf{x}_j-\mathbf{x}_i$. Suppose to select the potential function $V$ such that it has two local minima, one at $w_{ij}=1$ and one at $w_{ij}=0$. Correspondingly, the link $(i,j)$ will be considered as \emph{active}, when $w_{ij}=1$, or \emph{non-active} when $w_{ij}=0$. Now, let start the network from an initial condition $w_{ij}(0)=0$, i.e., where all link weights are zero. Then, due to the presence in Eq.~(\ref{eq:edgesnapping}) of the forcing term $g(\mathbf{e}_{ij})$, some of the variables $w_{ij}(t)$ can eventually leave the initial equilibrium point and reach the other equilibrium (at $w_{ij}=1$). In this way, the network will evolve towards a state where some of the links are adaptively \emph{activated} to support synchronization. As an example~\cite{delellis2010evolution}, let us consider a system of $N=100$ coupled Lorenz systems, with $g(\mathbf{e}_{ij})=\| \mathbf{x}_j-\mathbf{x}_i \|^2$, $V(w_{ij})=bw_{ij}^2(w_{ij}-1)^2$, $\epsilon=1$, $\gamma=5$, $b=180$, and
\begin{equation}
h(\mathbf{x}_j)=\left [ \begin{array}{lll} 1 & 0 & 0 \\ 0 & 0 & 0\\ 0 & 0 & 1 \\ \end{array} \right ]
\mathbf{x}_j.
\end{equation}


The time evolution of one representative state variable for each oscillator and the weights $w_{ij}(t)$ are illustrated in Fig.~9 of Ref.~\cite{delellis2010evolution}. There the authors notice that the network reaches synchronization in short time and, correspondingly, the weights asymptotically approach one of the two equilibrium points of the potential. In this way, the network is embedded with an adaptive mechanism that makes it able to self-determine its structure in order to support synchronization. The characteristics of the emerging topology depend on the dynamical evolution of the complex system and, thus, ultimately, by its initial conditions. Quite interestingly, the Authors of Ref.~\cite{delellis2010evolution} observe strong correlation between the initial conditions of the nodes and their degree in the final structure and between the initial conditions of the network and the maximum eigenvector of the topology obtained.

Adaptive mechanisms based on edge-snapping may be also applied to the problem of pinning control, in particular to select which nodes to pin~\cite{delellis2011pinning}. In this case, the values of $\delta_i$ are not decided a priori, but are the result of the co-evolution of the dynamics and the structure (that now includes the links from the reference system to the nodes that can be potentially pinned). To this purpose, the variables $\delta_i$ can be set as $\delta_i=b_i$ with $b_i$ obeying second-order dynamics similar to Eq.~(\ref{eq:edgesnapping}).

Finally, we mention that the weights of a network can be adapted not only to reach synchronization, but also to optimize other quantities such as: i) the network synchronizability as measured either by the smallest non-zero eigenvalue of the Laplacian matrix or the ratio between the largest and the smallest non-zero eigenvalue~\cite{kempton2017self}; ii) the robustness of consensus~\cite{kempton2017distributed}; iii) the selection of the control gains used in pinning control laws~\cite{di2019decentralized}. Remarkably, these objectives can be reached with a fully decentralized approach by designing a multi-layer structure where the layer devoted to weight optimization works in parallel with the layers estimating the actual value of the synchronizability measure.

To conclude this section on adaptive networks, we remark that most of the techniques illustrated also apply to consensus problems. Furthermore, with the use of memristors, electrical schemes realizing the key mechanisms underlying these techniques can be conceived~\cite{gambuzza2015memristor,gambuzza2017memristor}; the memristor technology is still at the early stage of its development, such that there is a considerable gap to be filled between the theoretical schemes designed and their practical counterparts, but the use of this technology is a promising approach to implement adaptive mechanisms in efficient ways requiring a small number of components.

\subsection{Multilayer networks}
\label{sec:secIV}
Up to now, we have discussed the time-varying effect on single layer networks. However, a large class of real-world and engineered systems are represented as networks whose architecture is multilayer~\cite{boccaletti2014structure,kivela2014multilayer}, i.e., it is made of two or more interaction layers. In these systems, one has to distinguish between  intra-layer and inter-layer interactions: all nodes within a layer interact among them via intra-layer links, whereas the interaction between nodes that belong to different layers take the name of inter-layer interactions (Fig.~\ref{Diagram}). There are many examples of such multilayer networks, like social networks~\cite{szell2010multirelational}, mobility transport networks~\cite{cardillo2013emergence,halu2014emergence},  air transportation networks~\cite{cardillo2013modeling}, subway networks~\cite{criado2007efficiency}, and neural networks~\cite{sartori1991simple}. Furthermore, the multilayered network structure greatly affects collective phenomena in such systems~\cite{radicchi2013abrupt,del2016synchronization}, such as the processes of epidemic spreading~\cite{saumell2012epidemic,granell2013dynamical,sanz2014dynamics}, diffusion~\cite{gomez2013diffusion}, controllability \cite{menichetti2016control}, percolation~\cite{buldyrev2010catastrophic,gao2012networks,bianconi2014multiple}, and evolutionary game dynamics \cite{wang2014rewarding}. On the other hand, due to the interplay between the local dynamics of the nodes and the network topology, different types of global states emerge, like intra-layer and inter-layer synchronization~\cite{gambuzza2015intra,sevilla2016inter,leyva2017inter}, explosive synchronization~\cite{zhang2015explosive}, cluster synchronization~\cite{jalan2016cluster} and chimera states~\cite{maksimenko2016excitation,majhi2016chimera}.

\par We here show an example of a multiplex network~\cite{rakshit2017time} where the intra-layer network architecture experiences processes of stochastic rewiring with a characteristic switching frequency, while the connections between the layers are static. In such evolving  network,  two types of synchronization patterns has been investigated, intra-layer synchronization and inter-layer synchronization. The effect of a time-varying layer in a chain of static layers gives an enhancing effect on inter-layer synchronization~\cite{rakshit2019enhancing}.

Then we will consider a temporal single layer network where more than one intra-layer connections exist between the nodes, and necessary and sufficient conditions are discussed for such a time-varying hypernetwork~\cite{rakshit2018emergence}. Finally, we review a series of works on the invariance and stability conditions for synchronization in time-varying multiplex hypernetworks~\cite{rakshit2020intralayer,rakshit2020invariance}.

\subsubsection{Time-varying multiplex network}
\par We begin with a bilayer multiplex network, where each layer contains $N$ nodes and the local dynamics in each node is a $d$-dimensional identical oscillator. The evolution of the $i^{th}$ node is described as
\begin{equation}
\begin{array}{lcl} \label{eq_14}
\dot{\bf x}_{i,1}=F({\bf x}_{i,1})-\epsilon\sum\limits_{j=1}^{N}\mathcal{L}^1_{ij}(t)G({\bf x}_{i,1})+\lambda[H({\bf x}_{i,2})-H({\bf x}_{i,1})],\hspace{30pt} \\
\dot{\bf x}_{i,2}=F({\bf x}_{i,2})-\epsilon\sum\limits_{j=1}^{N}\mathcal{L}^2_{ij}(t)G({\bf x}_{j,2})+\lambda[H({\bf x}_{i,1})-H({\bf x}_{i,2})],
\end{array}
\end{equation}
where ${\bf x}_{i,1}$ and ${\bf x}_{i,2}$ are the $i^{th}$ node's state variables in the two layers.  Here, $F:\mathbb{R}^d\rightarrow\mathbb{R}^d$ represents the vector field of the individual oscillators, $G:\mathbb{R}^d\rightarrow\mathbb{R}^d$ and $H:\mathbb{R}^d\rightarrow\mathbb{R}^d$ stand respectively for the output vectorial functions within the layers and between the layers. The parameters $\epsilon$ and $\lambda$ are the intra-layer and inter-layer coupling strengths, respectively.  
 Here, the architectures of the two layers are encoded by the Laplacian matrices $\mathcal{L}^1(t)$ and $\mathcal{L}^2(t)$. The intra-layer network topologies in both layers are small-world temporal networks, constructed by the method introduced by \textit{Watts \emph{and} Strogatz} in their seminal paper Ref.~\cite{watts1998collective}. These small-world networks evolve over time by rewiring every edges stochastically and independently with a characteristics frequency $f$. However, all the inter-layer links are fixed in time. In particular, each layer is rewired with probability $p_r=fdt$ by constructing a new small-world network, where $dt$ is a given time step. Due to the fixed choice of parameters $p_{\footnotesize{WS}}$ (small world probability) and $k$ (average degree of the nodes), these successively rewired small-world networks will be statistically equivalent throughout the procedure. Here, $p_r\sim 1$ implies that the two layers evolve rapidly due to a fast switching of links. However, when $p_r \sim 0$, the network edges have very small probability of change, hence the intra-layer networks are almost static.

 In these networks it is interesting to study \emph{intra-layer synchronization}, which occurs when all the units in each layer are synchronized each other, but not necessarily with their counterparts in other layers, i.e., $\mathbf{x}_{1,1}=\ldots=\mathbf{x}_{N,1}$, as well as \emph{inter-layer synchronization}, which occurs when all the nodes are synchronized with their counterparts in the other layers, but not necessarily with the other nodes in the same layers, i.e., $\mathbf{x}_{i,1}=\ldots=\mathbf{x}_{i,M}$, where $M$ is the number of layers (in this case $M=2$).

\par To numerically explore the emergent behavior of system~\eqref{eq_14}, the nodal dynamics is taken to be the three-dimensional R\"{o}ssler oscillators $\dot{x}=-y-z,\, \dot{y}=x+0.1y, \, \dot{z}=0.1+z(x-14)$. A diffusive coupling function is considered for both intra-layer and inter-layer couplings through the variable $y$, i.e., $G({\bf x})=[0~y~0]^{T}$ and $H({\bf x})=[0~y~0]^{T}$.
\par To explore the effect of the intra-layer rewiring frequency $f$ on the synchronization states, one defines the errors as
\begin{equation}
\begin{array}{lcl} \label{eqer1}
 E_{intra}=\lim_{T_t\to\infty} \frac{1}{T_t}\int_{0}^{T_t}\sum_{j=2}^{N} \frac{\|{\bf x}_{j,1}(t)-{\bf x}_{1,1}(t)\|+\|{\bf x}_{j,2}(t)-{\bf x}_{1,2}(t)\|}{2(N-1)}dt,
 \end{array}
 \end{equation}
and
\begin{equation}
\begin{array}{lcl} \label{eqer2}
E_{inter}=\lim_{T_t\to\infty} \frac{1}{T_t}\int_{0}^{T_t} \sum_{j=1}^{N} \frac{\|{\bf x}_{i,2}(t)-{\bf x}_{i,1}(t)\|}{N} dt,
\end{array}
 \end{equation}
where $T_t$ is a sufficiently large positive constant. The final state is considered as a stable synchronization state if the corresponding synchronization error is bounded by $10^{-4}$, otherwise the multiplex network Eq.~\eqref{eq_14} is said to be in an asynchronous state.

\par
\begin{figure}[t]
\centerline{~~\includegraphics[scale=0.42]{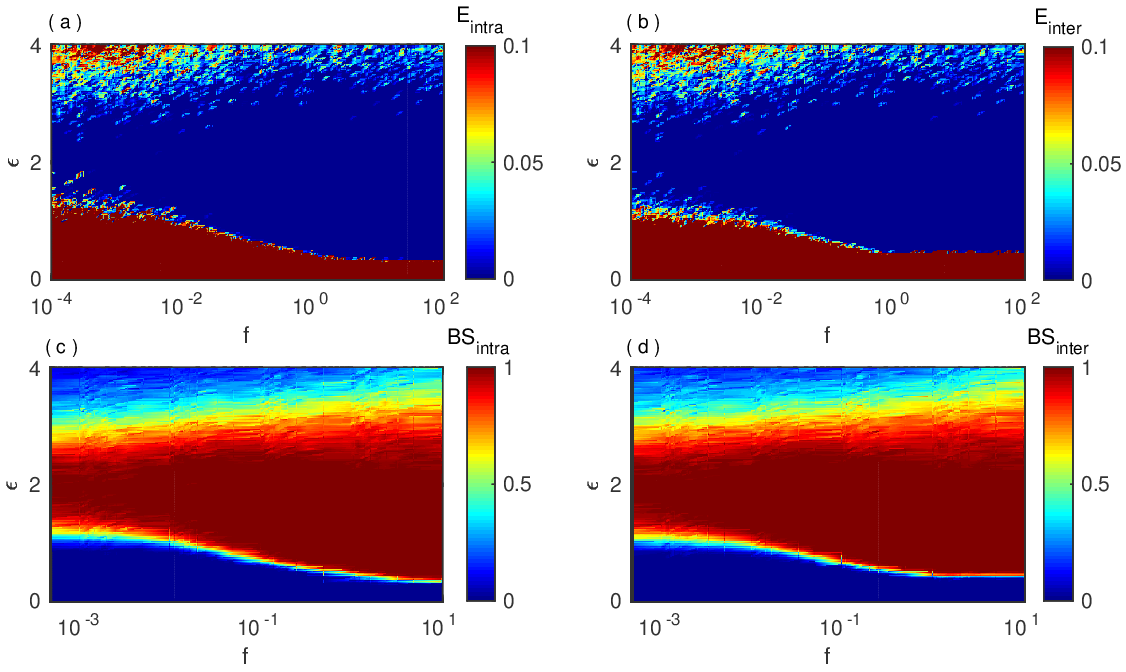}}
\caption{Synchronization error (upper panels) and basin stability (bottom panels) corresponding to the intra-layer (left panels) and inter-layer (right panels) synchronization states in $(\epsilon,f)$ parameter plane, where $\lambda=\frac{\epsilon}{6}$. Here, N = 200, the small world probability is $p_{SW} = 0.1$, and the average degree is $k = 3$.
}
\label{fig1}
\end{figure}

 Figure~\ref{fig1}(a) reports the variation of $E_{intra}$ in the $(\epsilon,f)$ parameter plane. From these results, it is evident that larger values of $f$ enhance intra-layer synchronization. Compared to the static multiplex networks, time-varying networks enlarge the range of coupling strengths yielding complete intra-layer synchronization. Similarly, the variation of $E_{inter}$ is drawn in Fig.~\ref{fig1}(b). Here, $E_{inter}$ first decreases as $\epsilon$ gradually increases, and becomes zero as it crosses a certain threshold value, indicating the emergence of the inter-layer synchronization state. Although, in this case, the critical values of $\epsilon$ only mildly depend on the rewiring frequency $f$. Further, increasing the intra-layer coupling strength beyond certain values, the multiplex network loses both types of synchronization patterns. Such a synchronization to desynchronization critical transition point of $\epsilon$ depends on $f$. Larger values of the switching frequency $f$ yield higher persistence of intra-layer and inter-layer synchronization states. Notice that both synchronization features occur for $\epsilon\ge 0.5$ irrespectively of the values of $f$, hence the whole network exhibits a global synchronization state.

\subsubsection{Linear stability analysis}
\par Next, we discuss the conditions for the appearance of  intra-layer and inter-layer synchronization states. For intra-layer synchrony, let the layer-$1$ and layer-$2$ evolve synchronously with the respective synchronization manifolds ${\bf x}_{0,1}$ and ${\bf x}_{0,2}$. Let $\delta{\bf x}_{i,1}$ and $\delta{\bf x}_{i,2}$ be the perturbations from such manifolds, i.e., ${\bf x}_{i,1}={\bf x}_{0,1}+\delta{\bf x}_{i,1}$ and ${\bf x}_{i,2}={\bf x}_{0,2}+\delta{\bf x}_{i,2}$. Then the linearized equations can be written as
\begin{equation}
\begin{array}{lcl} \label{eq_15}
\delta\dot{\bf x}_{i,1}=JF({\bf x}_{0,1})\delta{\bf x}_{i,1}-\epsilon\sum\limits_{j=1}^{N}\mathcal{L}^1_{ij}(t)JG({\bf x}_{0,1})\delta{\bf x}_{j,1}+\lambda[JH({\bf x}_{0,2})\delta{\bf x}_{i,2}-JH({\bf x}_{i,1})\delta{\bf x}_{i,1}],\\
\delta\dot{\bf x}_{i,2}=JF({\bf x}_{0,2})\delta{\bf x}_{i,2}-\epsilon\sum\limits_{j=1}^{N}\mathcal{L}^2_{ij}(t)JG({\bf x}_{0,2})\delta{\bf x}_{j,2}+\lambda[JH({\bf x}_{i,1})\delta{\bf x}_{i,1}-JH({\bf x}_{0,2})\delta{\bf x}_{i,2}],
\end{array}
\end{equation}
where $J$ represents the Jacobian operator and the intra-layer synchronization solution $({\bf x}_{0,1},{\bf x}_{0,2})$ satisfies the following evolution equation,
\begin{equation}
\begin{array}{lcl} \label{eq_16}
\dot{\bf x}_1=F({\bf x}_1)+\lambda[H({\bf x}_2)-H({\bf x}_1)],	\hspace{50pt} \\	\dot{\bf x}_2=F({\bf x}_2)+\lambda[H({\bf x}_1)-H({\bf x}_2)].
\end{array}
\end{equation}
If the largest Lyapunov exponent $\Lambda_{\max}^{intra}$ of the coupled linearized systems~\eqref{eq_15} is negative, then the intra-layer synchronization state is stable. In the linearized error system~\eqref{eq_15}, the dependency of the switching link frequency $f$ is incorporated through the intra-layer Laplacian matrices $\mathcal{L}^1(t)$ and $\mathcal{L}^2(t)$.

\par In Ref.~\cite{rakshit2017time} the Authors report that  the variation of $\Lambda_{max}^{intra}$ by changing the coupling strength $\epsilon$ and rewiring frequency $f$ becomes negative where $E_{intra}$ turns out to be zero (see Fig.~1(a) and 2(a) in Ref.~\cite{rakshit2017time}). Further increasing the values of $\epsilon$, $\Lambda_{max}^{intra}$ assumes positive values, as a result of a short-wavelength bifurcation~\cite{heagy1995short}. These transition points accurately match where $\Lambda_{max}^{intra}$ becomes again non-zero.
For larger values of $f$, $\Lambda_{max}^{intra}$ becomes negative faster, and persists to be negative at higher values of $\epsilon$.

\par As for inter-layer synchronization, let us consider a small difference $\delta {\bf z}_i={\bf x}_{i,2}-{\bf x}_{i,1}$ between the state of the $i^{th}$ replica node and that corresponding to complete synchronization. Then, around the inter-layer synchronization solution, the linearized  equation is
\begin{equation}
\begin{array}{lcl}\label{eq4}
\delta \dot{\bf z}_i=[JF({\bf \tilde{x}}_i)-2\lambda JH({\bf \tilde{x}}_i)]\delta{\bf z}_i-\epsilon\sum\limits_{j=1}^{N}\mathcal{L}^2_{ij}(t)JG({\bf \tilde{x}}_j)\delta{\bf z}_j+ \epsilon\sum\limits_{j=1}^{N} \Delta L_{ij}(t) G({\bf \tilde{x}}_j),
\end{array}
\end{equation}
where $\Delta L(t)$ is defined as $\Delta L(t)=\mathcal{L}^1(t)-\mathcal{L}^2(t)$. Here, ${\bf \tilde{x}}_i$ denotes the state variable of the $i^{th}$ node at the inter-layer synchronization solution, and its dynamics can be written as,
\begin{equation}
\begin{array}{lcl} \label{eq5}
\dot {\bf \tilde{x}}_i=f({\bf \tilde{x}}_i)-\epsilon\sum\limits_{k=1}^{N} \mathcal{L}^1_{ik}(t)G({\bf \tilde{x}}_k).
\end{array}
\end{equation}

The numerical results show that the quantity $\Lambda_{max}^{inter}$ experiences, as a function of $\epsilon$,  a first transition from the asynchronous to the synchronous state which is independent on the rewiring frequency $f$. This scenario changes for higher values of $\epsilon$, where inter-layer synchronization becomes unstable, and $\Lambda_{max}^{inter}$ becomes again positive, with a transition point which now strongly depend on $f$~\cite{rakshit2017time}.

\subsubsection{Basin stability measure}
\par In this subsection, we will discuss the effects of the rewiring frequency $f$ on the global stability of intra-layer and inter-layer synchronization states. The global stability of these states is assessed by means of the basin stability (BS) measurement (as introduced in Sec.~\ref{sec:secIIIBasin}). To calculate BS numerically, the entire system is integrated for $1,000$ distinct initial conditions chosen from the three-dimensional space $[-20,20]\times[-20,20]\times[0,35]$ in which the isolated attractor resides.
The bottom panel of Fig.~\ref{fig1}(c) and~\ref{fig1}(d) depicts the variation of basin stability for intra-layer ($BS_{intra}$) and inter-layer ($BS_{inter}$) synchronization states in the $(f, \epsilon)$ parameter plane, respectively.
As $\epsilon$ increases, the values of BS$_{intra}$ gradually increases from zero, and approaches the maximum value one. As the values of $f$ increases, BS$_{intra}$ varies more sharply. Irrespective of the intra-layer rewiring frequencies and up to a certain value of $\epsilon$, unit values of BS$_{intra}$ persists. Moreover a transition toward the intra-layer desynchronization is observed, no matter how promptly the intra-layer networks vary. Here, BS$_{intra}$  progressively goes to zero, as $\epsilon$ systematically increases. Different $BS_{intra}$ values for distinct values of $f$ demonstrate the influence of temporal intra-layer network on this synchronization pattern. As the intra-layer coupling strength increases, the value of $BS_{inter}$ grows abruptly, and do not significantly rely on the intra-layer rewiring frequencies. This result is in agreement with the aptitude of the variation of $E_{inter}$ already observed in Fig.~\ref{fig1}(b). Similar to the variation of BS$_{intra}$, unit value of BS$_{inter}$ persists up to a certain value of $\epsilon$. Then, the inter-layer synchronization state gradually disappears, with a tendency that significantly depends on $f$. At the critical transition point of $\epsilon$ for which synchronization patterns emerge, the corresponding basin stability values stay in the open interval $(0,1)$. This implies that a fraction of the initial conditions supports the respective synchronization patterns.

\subsubsection{Effect of demultiplexing}
\begin{figure}[t]
\centerline{\includegraphics[scale=0.38]{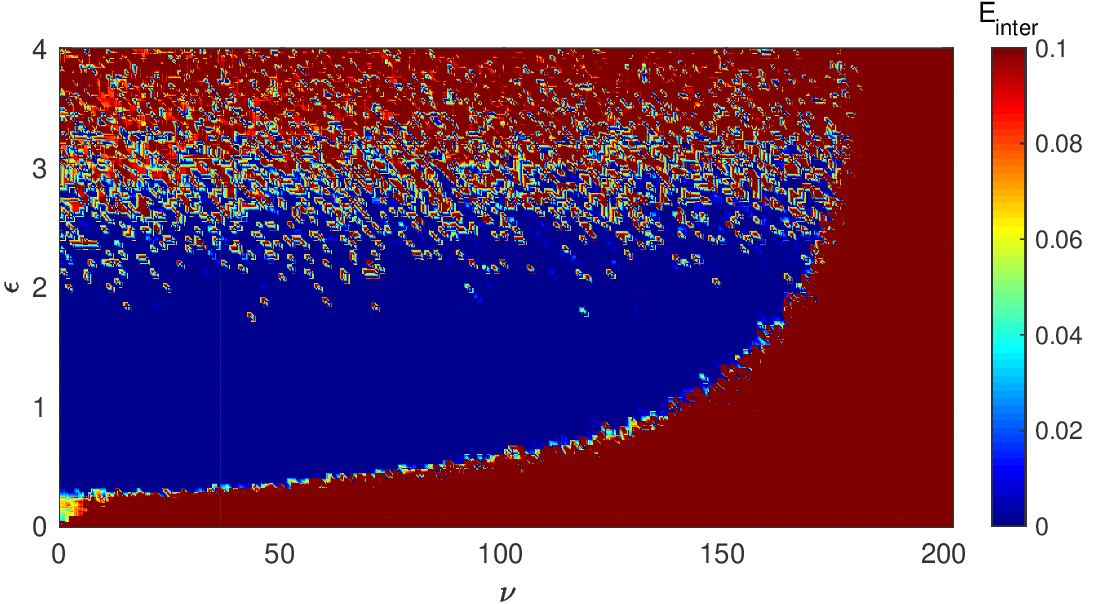}~~~~~}
\caption{Continuous variation of $E_{inter}$ in the $(\nu,\epsilon)$ parameter plane with $f=0.1$ and $\lambda=2.0$. The number of demultiplexed links increases by increasing the value of $\epsilon$ and maximum 170 inter-layer links can be removed to get inter-layer synchronization state. 
}
\label{fig5}
\end{figure}

In multiplex networks, one of the issues that needs to be addressed is the robustness of inter-layer synchronization against a progressive demultiplexing of the replica connections: from a given multiplex structure, one has to progressively take off inter-layer links  until both layers become isolated, and monitor the resilience of inter-layer synchronization. The number of demultiplexed replicas is denoted by $\nu$ and varied between $0$ and  $N=200$. Here, $\nu=0$ means a complete multiplex network, i.e., all the replicas are present. When $\nu = N$, the two layers become completely decoupled. The influence of $\epsilon$ on the resilience of  inter-layer synchronization  is illustrated in Fig.~\ref{fig5}. Here,  E$_{inter}$ continuously change in the $(\nu,\epsilon)$ parameter plane, and inter-layer synchronization persists up to a certain value of $\nu$. By increasing the values of $\epsilon$, the number of demultiplexed replicas that preserve the inter-layer synchrony grows as well. When $\epsilon=2$, the inter-layer synchronization can be observed even when nearly $170$ replica connections are removed. However, this tendency holds up to a given value of $\epsilon$: increasing the intra-layer coupling strength above this threshold reduces the critical value of demultiplexed replicas. 

The study has been later extended in Ref.~\cite{rakshit2019enhancing}, which explored intra-layer synchronization in presence of static as well as of multiple time-varying layers, and reveals how already a single time-varying layer is able to enhance intra-layer synchronization.

\subsection{Hypernetworks}

\par Synchronization of coupled Hindmarsh-Rose neurons with hypernetwork architecture has been studied in Ref.~\cite{rakshit2018emergence}, with two different kinds of neuronal communications, namely electric gap junctional  and chemical synaptic interactions. The simultaneous occurrence of these two kinds of network architectures in a single network gives rise to a hypernetwork architecture. For inter-neuronal communication, both  kinds of synaptic interactions coexist and also work independently over time in several neuronal network. Mimicking a  realistic neurobiological scenario, every links of this hypernetwork are permitted to switch over time, and are varied with a switching frequency $f$. Various kinds of spatiotemporal dynamics are explored by tuning the comparative gap junctional strength $\epsilon$, the synaptic interaction strength $g_c$, the characteristics rewiring frequency $f$, and the small-world probability $p_{SW}$, which accounts for the randomness of the network. For instance, if $p_{SW}=0$, the network is fully non-local, while $p_{SW}=1$ indicates it is completely a random network.

\par Consider $N=200$ Hindmarsh-Rose (HR) model neurons, coupled concurrently through linear electric gap junction  and non-linear chemical synaptic interactions. Then, the evolution equation of such neuronal hypernetwork is described by
\begin{equation}\label{eq:system}
\begin{array}{lcl}
	\vspace{5pt}
\dot{x_i}(t)=y_i-ax_i^3+bx_i^2-z_i+I+\frac{g_c}{k_c}(v_s-x_i)\sum\limits_{j=1}^{N}\mathcal{A}^{(c)}_{ij}(t)\frac{1}{1+\text{exp}[-\lambda({\bf x_j}-\Theta_s)]}- \epsilon\sum\limits_{j=1}^{N}\mathcal{L}^{(e)}_{ij}(t)x_j,\\
\vspace{10pt}
\dot{y_i}(t)=c-dx_i^2-y_i,\\
\dot{z_i}(t)=r((x_i-x_0)s-z_i),
\end{array}
\end{equation}
where $i=1, 2, \dots, N$ is the oscillator index, the membrane potential of neuron $i$ is represented by the state variable $x_i(t)$, $y_i(t)$ and corresponds to the ions transportation of the $i^{th}$ neuron across the membrane through the fast current associated with $Na^+$ and $K^+$ ions. The variable $z_i(t)$ is the transportation of ions for the $i^{th}$ neuron across the membrane potential through the slow current associated with $Ca^{2+}$ ions, and the speed of such current is controlled by the system parameter $r$. $g_c$ represents the chemical synaptic strength, while $\epsilon$ is the electrical gap junctional strength. These two synaptic strengths regulate how information will be dispensed among the neurons mediated by the interaction channels. The value of the system parameters are chosen as $a=1$, $b=3$, $c=1$, $d=5$, $r=0.005$, $s=4$, $x_0=-1.6$, and $I=3.25$. For this set of parameter values, the membrane potential of the individual HR system demonstrates multi-time scale chaotic dynamics, known as spiking bursting pattern. Moreover, the chemical coupling parameters are fixed at $\Theta_s=-0.25,~v_s=2$ and $\lambda=10$.

\par In neuronal system, the electrical interaction is bidirectional in nature, however, the chemical coupling is unidirectional type~\cite{pereda2014electrical}. The former type of coupling has been considered to display a small-world network architecture~\cite{bassett2006small,liao2017small,bassett2006adaptive}, formulated through the construction algorithm proposed by Watts-Strogatz~\cite{watts1998collective}. The corresponding Laplacian is  the square matrix $\mathcal{L}^{(e)}$ of order $N$, with average degree $\langle k_e \rangle$ and edge rewiring probability $p_{SW}$. However, the second type of interaction is represented by a unidirectional random network, encoded by the adjacency $\mathcal{A}^{(c)}$ and the matrix $\mathcal{L}^{(c)}$ denotes the associated Laplacian. The in-degrees of all the vertices are considered here as identical and equal to $k_c$. Complete neuronal synchrony indicates that all the neurons evolve in unison with identical trajectory in the phase-space. For such case, the associated synchronization error becomes zero, i.e., $E=0$. Here, stability of the neuronal synchronization solution against any arbitrary perturbations is characterized through the basin stability measurement. The phase-space volume has been sampled from $[-1.5,2.0]\times[-7.0,1.0]\times[2.9,3.4]$, and if $E <10^{-5}$, the entire coupled system is assumed to be synchronized.
\par The synchronization error and the corresponding basin stability the $(\epsilon,f)$ parameter plane are reported in Fig.~4 of Ref~\cite{rakshit2018emergence}, respectively in the upper and lower panels. Color bar in the upper panel depicts the variation of $E$. On the other hand, the color bar in lower panel represents the variation of basin stability and hence characterize global robustness of neuronal synchronization. The left, middle and right panels are respectively drawn for $k_e=4$, $k_e=6$ and $k_e=8$. Here, the chemical synaptic interaction strength systematically varies as $g_c=\frac{\epsilon}{2}$, and the in-degree of the underlying network is fixed at $k_c=5$. The upper panel shows that the region of neuronal synchrony monotonically enlarges as the average degree of the SW network gradually increases. Likewise, a similar tendency occurs for the corresponding BS diagram in lower panel.

\par To gather more information on neuronal synchrony, the synchronization time, i.e., the mean time to attain that solution is plotted in Fig.~8 of Ref~\cite{rakshit2018emergence} for five exemplary values of switching frequency $f$. Here, left and right panels depict the variation of the synchronization time with respect to $\epsilon$ for $k_e=6$, and $k_e=8$ respectively. The synchronization time gradually reduces as the electrical interaction strength increases.

\par Let $\bar{\mathcal{L}}^{(e)}$ and $\bar{\mathcal{L}}^{(c)}$ denote the time-average Laplacians for the electrical gap junctional network and chemical synaptic network, respectively.  Additionally, $\{0,\gamma_2^{(e)},\gamma_3^{(e)},\dots,\gamma_{N}^{(e)}\}$ and $\{0,\gamma_2^{(c)},\gamma_3^{(c)},\dots,\gamma_{N}^{(c)}\}$ are the set of eigenvalues of the zero-row sum matrices $\bar{\mathcal{L}}^{(e)}$ and $\bar{\mathcal{L}}^{(c)}$, respectively. For sufficiently rapid rewiring, the time-average weighted network warrants a synchronization transition to occur~\cite{stilwell2006sufficient}. Then Master Stability Equation (MSE) transverse to the complete synchronization manifold reads as follows,
\begin{equation}\label{eq:mse}
\begin{array}{lcl}
\dot{\xi_i}^{(x)}=(-3ax^2+2bx)\xi_i^{(x)}+\xi_i^{(y)}-\xi_i^{(z)}-\frac{g_c\xi_i^{(x)}}{1+exp(\lambda(\Theta_s-x))}-\epsilon\gamma_i^{(e)}\xi_i^{(x)}+\frac{g_c}{k_c}(v_s-x)\frac{\lambda exp(\lambda(\Theta_s-x))}{[1+exp(\lambda(\Theta_s-x))]^2}[k_c\xi_i^{(x)}-\gamma_i^{(c)}\xi_i^{(x)}], \\
\dot{\xi_i}^{(y)}=-2dx\xi_i^{(x)}-\xi_i^{(y)}, \\
\dot{\xi_i}^{(z)}=r(s\xi_i^{(x)}-\xi_i^{(z)}),
\end{array}
\end{equation}
where $i=2, 3, \dots, N$, and $(x(t), y(t), z(t))$ is the state vector for the neuronal synchronization manifold, which dominates the following evolution equation,
\begin{equation}
\begin{array}{lll}
\dot{x}(t)=y-ax^3+bx^2-z+I+g_c(v_s-x)\Gamma(x),\\
\dot{y}(t)=c-dx^2-y,\\
\dot{z}(t)=r(s(x-x_0)-z).
\end{array}
\end{equation}
The variation of $\Lambda_{max}$ of Eq.~(\ref{eq:mse}) is drawn in Fig.~11 of Ref.~\cite{rakshit2018emergence} by simultaneously varying the coupling strengths $\epsilon$ and $g_c$, where the color bar depicts the variation of $\Lambda_{max}$. Here, the largest transverse Lyapunov exponent turn out to be negative value exactly where $E$ becomes zero for $f\in[10,100]$, i.e., for enough rapid switching.

\par According to the Wu-Chua conjecture \cite{wu1996conjecture}, the critical coupling threshold for $\epsilon$  can be derived as \begin{equation}
		\epsilon^*_N(g_c)=\frac{2~\epsilon^*_2(g_c)}{\gamma^{(e)}_2(N)}.
\end{equation}		
For a given values of chemical synaptic strength $g_c$, $\epsilon^*_2(g_c)$ and $\epsilon^*_N(g_c)$ are respectively the electrical coupling threshold for network size $2$ and $N$. For several chemical synaptic strength $g_c$, the threshold value of electrical interaction strength $\epsilon_N^*(g_c)$ has been obtained. In that figure, the dashed black curve represents the analytically obtained critical coupling strength. In the $(\epsilon,g_c)$ parameter plane, the regions below and above the critical curve respectively correspond to the desynchronization and synchronization solution. From this figure, it is evident that the analytical curve obtained from the Wu-Chua conjecture agrees very well with the transition of maximum transverse Lyapunov exponent.

Recently, studies on intra-layer synchronization have been extended to time-varying multiplex hypernetworks~\cite{rakshit2020intralayer}, and inter-layer synchronization has also been described in stochastic multiplex hypernetworks~\cite{rakshit2020invariance}.

\subsubsection{Global stability analysis: Single-node basin stability approach}

\par Global stability analysis of synchronization in time-varying multiplex networks will be discussed in this section via basin stability measurement (discussed in Sec.~\ref{sec:secIIIBasin})~\cite{menck2013basin,leng2016basin}.  Recently, the authors of Ref.~\cite{majhi2019emergence} went through such a study for a multiplex time-varying network of mobile oscillators based on the idea of targeted dynamical attacks of a specific node in the network, termed as ``single-node basin stability (SNBS)". This SNBS scheme is mainly used to understand the probability of return to the synchronization manifold when a particular node experiences an arbitrary nonlocal perturbation. Notice that the problem of global stability and resilience of networked dynamical systems in response to small perturbations simultaneously affecting multiple nodes can be studied using the multiple-node basin stability approach introduced in Ref.~\cite{mitra2017multiple}.

\par Reference~\cite{majhi2019emergence} considered a bilayer multiplex network with $N$ mobile multi-agent systems, which are performing a 2D-lattice random walk  in each layer. The local dynamics of each node $i~ (i=1,2,...,N)$ is associated with a dynamical system $\dot{X_i}=F(X_i)$, where $X_i$ is a $d$-dimensional vector of dynamical variables and $F(X_i)$ is a vector field characterizing the dynamical units. Mathematically the entire dynamical network is described as

\begin{equation}
	\begin{array}{lcl} \label{mnbs1}
		\dot X_{1,i}=F(X_{1,i})-k_1\sum\limits_{j=1}^N g^1_{ij}(t)E_1 (X_{1,j})+\epsilon[H (X_{2,i})-H (X_{1,i})],\\
		\dot X_{2,i}=F(X_{2,i})-k_2\sum\limits_{j=1}^N g^2_{ij}(t)E_2 (X_{2,j})+\epsilon[H (X_{1,i})-H (X_{2,i})],
	\end{array}
\end{equation}
where $k_1$ and $k_2$ are the intra-layer coupling strengths among the mobile multi-agent in layer-$1$ and layer-$2$, respectively, and  $\epsilon$ is the inter-layer coupling strength. The time-varying intra-layer connections are governed by the zero-row sum Laplacian matrices of order $N$ as $G_1(t)=[g^1_{ij}(t)]_{N\times N}$ and $G_2(t)=[g^2_{ij}(t)]_{N\times N}$. The matrix element $g^k_{ij}(t)=-1$, if the  $j$-th agent lies in the vision size of the $i$-th agent and zero otherwise. Also, $E_k:\mathbb{R}^d\rightarrow\mathbb{R}^d$ ($k=1, 2$) and $H:\mathbb{R}^d\rightarrow\mathbb{R}^d$ are the intra-layer and inter-layer output vectorial functions, respectively.

\par Paradigmatic chaotic R\"{o}ssler oscillators~\cite{rossler1976equation} are then chosen to represent the dynamics of the nodes,  with  $F(X_{k,i})$ equal to
\begin{equation}
	F(X_{k,i})=\left(
	\begin{array}{c}
		-y_{k,i}-z_{k,i}\\
		x_{k,i}+0.2 y_{k,i}\\
		0.2+z_{k,i}(x_{k,i}-5.7)\\
	\end{array}
	\right). \\
\end{equation}
The intra-layer and inter-layer coupling functions $E_k(X_k)=(0,y_k,0)^T$ and $H(X_k)=(0,y_k,0)^T$ with $k=1,2$ are taken to act at the $y$ variable. Further, the movement algorithm of the multi-agent systems is on 2D-lattice of $M_1 \times M_2$ mesh points.

During movement, each agent creates a much smaller square shape region of $\phi^2$ area (calling them {\it vision size}) than the physical space $M_1 \times M_2$. The $i$-th agent interact with other agents if they belong to the $i$-th agent's vision range. The vision range of a particular agent creates in the direction of the agent's motion and so the multi-agent generate a random geometric graph~\cite{dall2002random,penrose2003random} which is a simple and interesting model in spatial networks~\cite{barthelemy2011spatial}.

\begin{figure}
	\centerline{
		\includegraphics[scale=0.5700]{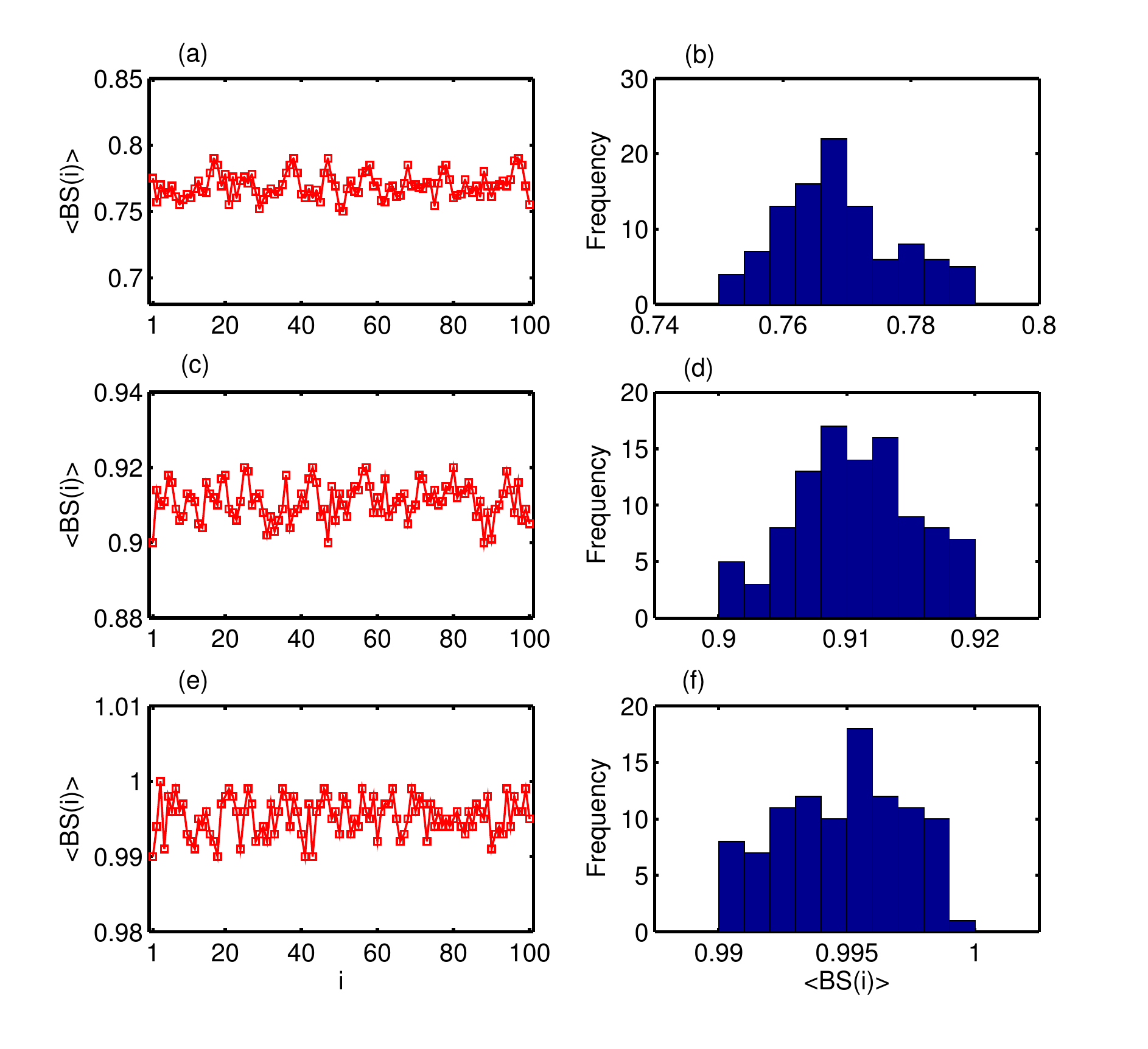}}
	\caption{ Left panel: Value of the single-node basin stability $ \langle BS(i)\rangle $ of the inter-layer synchronization state of the replica nodes for (a) $\phi=10$, (c) $\phi=20$ and (e) $\phi=30$. Right panel: The corresponding histograms are shown in (b), (d) and (f), respectively. Here, $I=1000$ random initial conditions are chosen from the phase space volume $[-15, 15]\times [-15, 15]\times [0, 35]$ of the individual dynamical units. Other parameters are $k_1=k_2=3.0$, $\epsilon=0.20,$ $u=6$, $M_1=M_2=300$, $N=100$ and $M=10$.
	}
	\label{fig18}
\end{figure}

\par Now, one can calculate the SNBS Eqs.~(\ref{mnbs1}), and to this purpose one takes $M$ number of different points from the attractor far away from the synchronization manifold  $S(t)=(X_1,X_2,\dots, X_N)$. The mean SNBS of the $i$-th agent is written as,
\begin{equation}
	\langle BS(i)\rangle=\frac{1}{M}\sum\limits_{m=1}^M BS(i,m),
\end{equation}\\
where $BS(i,m)=\frac{J}{I}$ is the SNBS of the $i$-th agent starting from the $m$-th point on the attractor, and $J$ is the number of initial conditions that return to the synchronized manifold among $I$ number of initial conditions.

\par For inter-layer synchronization, one calculates $ \langle BS(i)\rangle $ of all the pairs of replica nodes $(i=1,2, ..., N)$. The value of the vision range $\phi$ of each agents play an important role for the time-varying nature of the network. Increasing $\phi$ indicates the possibility of more interaction among mobile agents. For this reason, one changes the vision range $\phi$ to study the global stability.  Figure~\ref{fig18} shows the results on SNBS of each replica nodes and their corresponding histograms. For the vision range $\phi=10$, the values of SNBS of the replica nodes $i=1, 2, ..., N$ are shown in Fig.~\ref{fig18}(a). Here, the SNBS lies in the range $[0.75, 0.79]$. With increase of the vision range $\phi=20,$ the values of SNBS  lie in the range $[0.90, 0.92]$ as evident from Fig.~\ref{fig18}(c).  With further increase $\phi=30$, almost all the initial conditions eventually lead to the synchronized state with $ \langle BS(i)\rangle \in [0.99, 1.0]~~(i=1,2,..., N) $ and no matter which pairs of nodes are perturbed (see  Fig.~\ref{fig18}(e)) The corresponding frequencies of $ \langle BS(i)\rangle $ for all these three values of $\phi \in \{10,~20,~30\}$ are respectively shown in Figs.~\ref{fig18}(b,d,f). Interestingly, one sees note that $ \langle BS(i)\rangle $ increases monotonically with  $\phi$, while its dispersion has the opposite trend. Thus, a sufficient vision range can substantiate an optimal response  of the nodes to perturbations.

\section{Mobile agents}
\label{sec:secV}
Multi-agent systems are composed of many agents interconnected by a communication network and capable to deal with problems that are difficult (or even impossible) to solve by a single agent~\cite{wooldridge2009introduction}. Such agents are autonomous and can in fact take decisions by their own, but it is only due to their interactions that they (as a whole system) can perform a task in an efficient way. Agents are usually equipped with limited knowledge, such that their control occurs in a decentralized and distributed way.

Synchronization and consensus are two central mechanisms in such systems~\cite{li2009consensus,sun2018synchronization} and several works have dealt with issues like robustness~\cite{trentelman2013robust,seyboth2015robust}, resilience~\cite{chen2020adaptive}, observer-based methods~\cite{jiang2006state,selvaraj2018observer}, time-delayed communications~\cite{liu2011synchronization,jia2019synchronization}, event-triggered control~\cite{liuzza2016distributed}, adaptive coupling~\cite{gambuzza2015memristor,ma2018adaptive,liu2018state}, and many other aspects (a comprehensive list of all of them is beyond the scope of this work).

In this Chapter we review the relevant literature on synchronization in systems of mobile agents. These systems form, indeed, intrinsically time-dependent networks, as the communication network depends on the agent position and, hence, is not fixed in time. A typical example is given by a temporal proximity graph, where agents are assumed to be equipped with a communication system with limited range and to be able to communicate only with those agents that are at a distance shorter than the so-called interacting or sensing radius. Therefore, the network of interactions is determined by the way in which connections are established between agents and by the characteristics of the agent motion. The third key ingredient of the models discussed in this section is the dynamics associated to the agents. We start with what is probably the most simple representation of oscillator dynamics, i.e., phase oscillators (Sec.~\ref{sec:phaseoscillators}), then move to pulse-coupled oscillators (Sec.~\ref{sec:pulsecoupledmovingagents}), and, finally, investigate the case of limit cycle and chaotic systems (Sec.~\ref{sec:chaoticoscimovingagents}).

\subsection{Phase oscillators}
\label{sec:phaseoscillators}

\subsubsection{Phase oscillators moving on lattices}

We start to elucidate a simple case of agents' motion and dynamics, namely the motion over a lattice and phase oscillators in continuous time~\cite{uriu2013dynamics}.

We then consider $N$ phase oscillators, indexed by $i=1,\ldots,N$, each occupying the position of a node in a linear lattice of size $N$. Nodes can interact with neighbors within a discrete radius $R$, and mobility is dictated by position exchanges with only one of the node's immediate neighbors at random times, generated as identical and independently distributed random variables with Poisson distribution of rate $\lambda$. The system can be described in a rotating reference frame in terms of locally coupled oscillating dynamics, as

\begin{equation}
	\frac{d \phi_i}{dt} = \frac{\epsilon}{n_i} \sum_{|x_{i^\prime} - x_i| \leq R} \sin [\phi_i(t)-\phi_{i^\prime}(t)],
	\label{eq:coupled_uriu}
\end{equation}

\noindent where $\phi_i$ is the phase of the $i$th oscillator, $R$ is the interaction radius, $n_i$ is the size of the interaction neighborhood, and $\epsilon$ is the coupling strength, considered as uniform along the whole lattice. Initial phases $\phi_i(0)$ are uniformly distributed at random in the interval $[0, 2 \pi]$. Boundary conditions are non-periodic, whereby an oscillator at the border can interact and exchange positions only with neighbors on one side.

The model comprises two time scales: $1/\epsilon$ accounts for the coupled phase dynamics, and $1/\lambda$ is representative of the motion dynamics. The parameter $\lambda/\epsilon$ is therefore representative of the interplay between the two time scales~\cite{uriu2013dynamics} and, without loss of generality, we can set $\epsilon=1$ to elucidate salient properties of system~\eqref{eq:coupled_uriu}.

\begin{figure}
	\centering
	\includegraphics[scale=0.4]{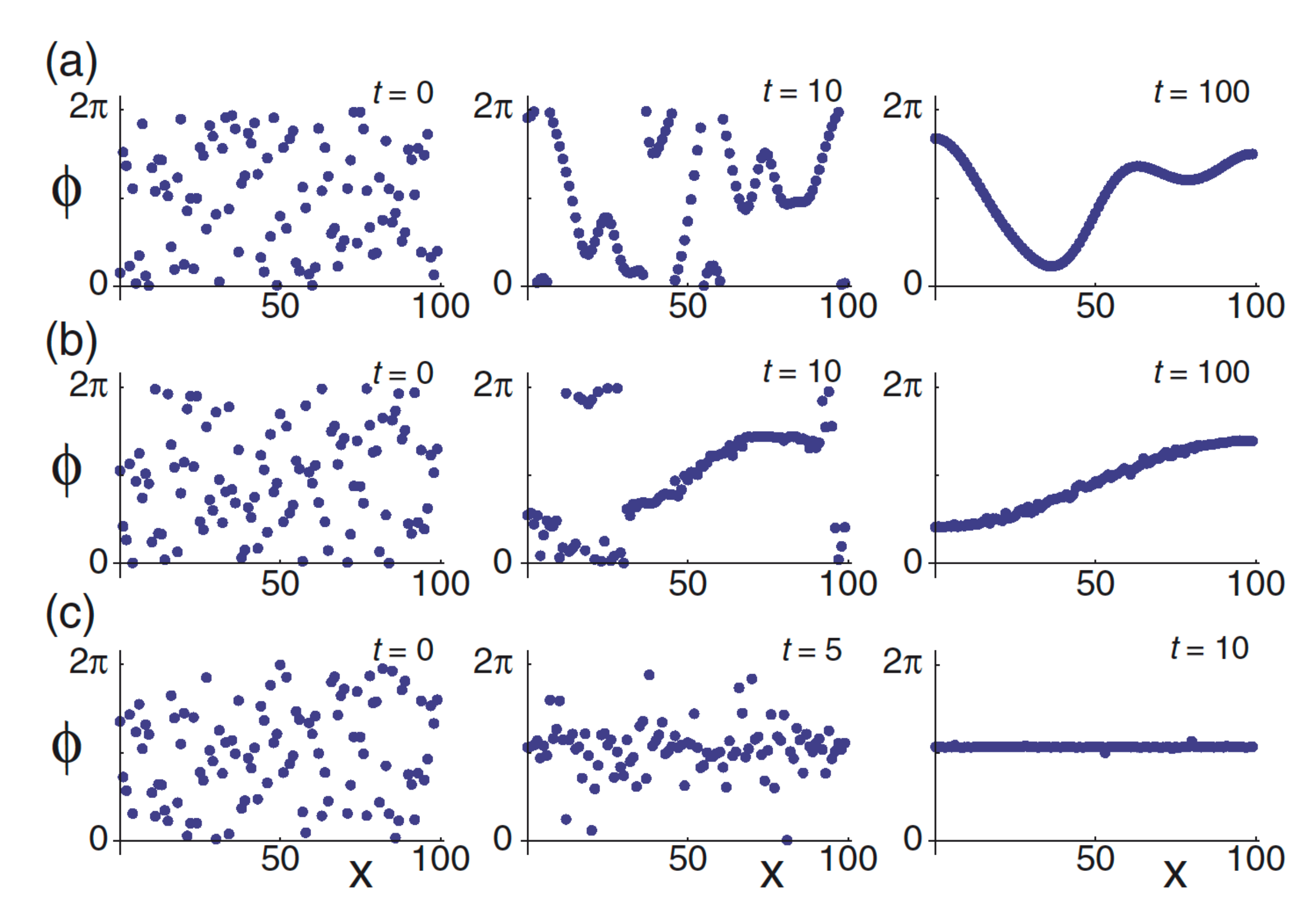}
	\caption{\label{fig:UruiFig2} Snapshots of spatial phase profiles. (a) $\lambda/\epsilon=0$, (b) $\lambda/\epsilon=10$, (c) $\lambda/\epsilon=10^4$. In all panels, $\epsilon=1$, $R=1$, and $N=100$.
	\\ {\it Source}: Reprinted figure with permission from Ref.~\cite{uriu2013dynamics} \textcopyright~2013 by the American Physical Society}
\end{figure}

The first set of experiments of Ref.~\cite{uriu2013dynamics} is conducted with unitary mobility range and connection radius, that is, $R=1$. Due also to the selected boundary condition, the oscillators tend toward complete phase synchronization, passing from the initial random distribution of the phases, through a snaky phase pattern that comprises several spatial modes with heterogeneous wavelengths, eventually attaining complete synchronization. For non-mobile nodes, that is, $\lambda/\epsilon=0$, the transition to complete synchronization is extremely slow. When dealing with mobile oscillators, increasing the value of $\lambda/\epsilon$, the pattern is dominated by longer wavelength modes. For high values of the coefficient $\lambda/\epsilon$, the snaky pattern does not appear and the oscillators rapidly converge toward complete synchronization. Intuitively, as long as the value of the parameter $\lambda/\epsilon$ increases, neighboring oscillators spend a shorter and shorter time coupled together before exchanging places. On the one hand, this prevents neighboring oscillators to settle  toward a common phase before moving elsewhere; however, mobility globally speeds up complete synchronization, making all the phases rapidly converge to the mean phase of the population. Results are summarized in Fig.~\ref{fig:UruiFig2}.

To characterize the transient dynamics, one can introduce a parameter quantifying the correlation between two lattice sites
\begin{equation}
	\rho(d,t)= \langle \cos [\phi_{k+d}(t) - \phi_k (t)] \rangle_k,
	\label{eq:rhocorrelation}
\end{equation}
where $\phi_{k}(t)$ and $\phi_{k+d}(t)$ are the phases at time $t$ in nodes $k$ and $k+d$, respectively, $d$ is the distance between two nodes and the average $\langle \cdot \rangle_k$ is executed for variable $k=1,\ldots,N-d$. The parameter quantifies the average instantaneous phase correlation at a distance $d$ in the lattice. Namely, if $\rho \approx 0$, phases are uncorrelated at the given distance, whereas if $\rho \approx \pm 1$, phases are totally correlated at the given distance. Value $-1$ denotes correlation with opposite sign. The time evolution of $\rho(t,d)$ elucidates the convergence dynamics of the oscillator network toward the common phase.

The dynamics of parameter $\rho$ follows an exponential relaxation, as $1-\rho(d) \approx e^{-t/T_c}$,
where $T_c$  is a relaxation time that depends on the interplay between mobility and dynamics. It has been shown that, in a wide mobility range between $\lambda/\epsilon=1$ and the onset of the mean field condition (very high $\lambda/\epsilon$), the following relationship holds:
\begin{equation}
	T_c \approx \frac{N^2}{\pi^2 \epsilon} \frac{1}{1+ \lambda/\epsilon}.
\end{equation}
At the extremes of the mobility range, the relaxation time $T_c$ does not depend on the ratio $\lambda/\epsilon$. For very low mobility, $T_c$ strongly depends on the system size. For very high mobility, the collective behavior can be interpreted through a mean-field analysis, yielding $T_c \approx 0.5$. In such a mean-field regime, mobility has an effect equivalent to extending the coupling range to an effective value $R_e$, which is estimated as $R_e = (1/4)(-3+\sqrt{49+48\lambda/\epsilon})$, in the regime where the longest spatial mode keeps its sinusoidal shape, that is, for $(\lambda/\epsilon)/N^2 << 1$.

Considering a long-range coupling ($R>1$) yields a qualitatively similar, yet richer, spectrum of phenomena. A mean-field behavior, similar to that observed for $R=1$, emerges also in this case, and for $R>>1$ and $r/N<<1$, becomes apparent for $\lambda/\epsilon > R^2 / 3$. Casting the latter inequality as $(R^2/\lambda)/(1/\epsilon)<3$, we can observe that mobility plays an active role in affecting synchronization when the time scale for each oscillator to explore the coupling range $R$ is comparable to that of the phase dynamics. Also in the case of long-range coupling, a widening of the effective coupling range $R$ is observed under mean-field condition, indicating an effective role of mobility on the overall system dynamics.

\subsubsection{Metapopulations of phase oscillators}

In the previous section we have discussed a model of phase oscillators moving on a lattice structure where, at each time, each node is occupied by a single agent and interactions occur among adjacent nodes. We now move to consider a more general, \emph{metapopulation} model where i) motion occurs along an arbitrary network structure; ii) each node can host a population of agents; and iii) interactions occur within each node of the network (i.e., among the agents that occupy the same node). The model, first introduced in Ref.~\cite{gomez2013motion}, consists of a set of $W$ agents distributed along the $N$ nodes of a network with adjacency matrix $\mathcal{A}$. The agents of the system are labeled with the subscript $i$ ($i=1,\ldots,W$), while the nodes of the network with subscript $I$ ($I=1,\ldots,N$).

At any time $t$, each agent is located in one of the nodes of the network where it interacts with all the other agents located in that node. Then, after a time interval equal to $\tau_M$, the agent moves to one of the adjacent nodes where it will interact with other agents of the system and so on. Therefore, at each node of the network an all-to-all interaction among the agents of the node takes place, such that, assuming that agent $i$ at time $t$ is in node $I$, the dynamics of the phase variables is given by:

\begin{equation}
	\label{eq:coupled_jesus}
	\dot{\phi_i} = \omega_i+\epsilon \sum_{j\in I} \sin [\phi_j(t)-\phi_{i}(t)],
\end{equation}

\noindent where $\omega_i$ are the natural frequencies of the oscillators, drawn from a distribution $g(\omega)$, and $\epsilon$ is the coupling strength.

Mobility of the agents, which corresponds to migration from one population to the other of the metapopulation model, consists of a degree-biased random walk on the network \cite{gomez2008entropy,sinatra2011maximal,masuda2017random}. According to this mobility rule, the probability $\prod_{I\rightarrow J}$ that an agent at node $I$ moves to an adjacent node $J$ depends on the degree $k_J$ of the destination node as follows:

\begin{equation}
	\label{eq:probBiasedRandomWalk}
	\prod_{I\rightarrow J}=\frac{k_{J}^{\alpha}}{\sum\limits_{l=1}^N k_l^{\alpha}},
\end{equation}

\noindent where $\alpha$ represents a parameter tuning the bias of agent motion towards low-degree nodes (for negative $\alpha$) or high-degree ones (for positive $\alpha$), whereas $\alpha=0$ implements the unbiased random walk. The model hence has three control parameters: the coupling strength $\epsilon$, acting on the dynamical interactions among the phase variables; the bias parameter $\alpha$, determining the type of motion; and the interval $\tau_M$ between subsequent steps of motion, which fixes the
ratio between the time scales of interaction and motion.

\begin{figure}
	\centering
	\includegraphics[scale=0.45]{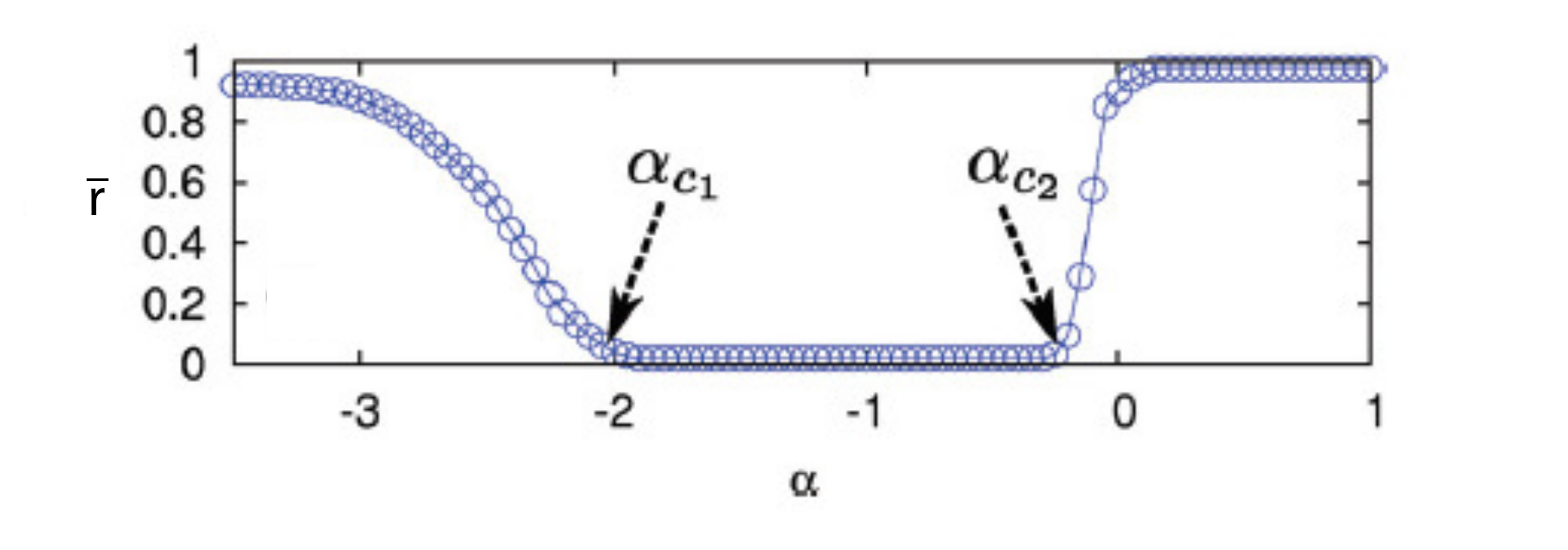}
	\caption{\label{fig:rJesus} Order parameter $r$ for a metapopulation of phase oscillators as function of the motion parameter $\alpha$ (see Eq.~\eqref{eq:probBiasedRandomWalk}), showing how the type of motion can dramatically change the macroscopic behavior of the system in terms of synchronization.
	\\ {\it Source}: Reprinted figure with permission from Ref.~\cite{gomez2013motion} \textcopyright~2013 by the American Physical Society.}
\end{figure}

It is instructive to show how the order parameter $\bar{r}=\lim\limits_{T\rightarrow \infty}\frac{1}{T}\left |\frac{1}{N} \sum\limits_{j=1}^N e^{\iota \phi_j(t)} \right |$ varies with $\alpha$. An example, obtained for a fixed value of $\epsilon$ ($\epsilon=0.08$) in a metapopulation model with $W=5000$ agents moving on a scale-free network with $N=500$ nodes, is shown in Fig.~\ref{fig:rJesus}. Suppose to start from a value of $\alpha=-1$, where oscillations are incoherent as $\bar{r} \simeq 0$, then, a fully coherent state can be reached either increasing or decreasing $\alpha$. This clearly demonstrates that it is possible to tune the level of synchronization by acting on the motion parameter, \emph{i.e.,} changing the type of motion of the walkers on the network. However, the microscopic mechanisms underlying synchronization in the case of large negative $\alpha$ and of small positive $\alpha$ are quite different. The Authors of Ref.~\cite{gomez2013motion}, in fact, observe two microscopic paths to synchronization: one, arising for $\alpha < -1$, driven by low-degree nodes, and one, arising for $\alpha > -1$, by the hubs.

Another interesting result illustrated in Ref.~\cite{gomez2013motion} is the behavior with $\tau_M$, the time constant of motion. In the limit case that $\tau_M$ is infinite, the metapopulation model becomes equivalent to a set of $N$ independent globally coupled populations of Kuramoto oscillators. Under these conditions, synchronization can be attained locally, but not globally, as there are no interactions among the different populations. This consideration also explains the behavior for finite, but large $\tau_M$, where rare motion events yield to poor mixing, preventing the emergence of global order. On the contrary, for small $\tau_M$, the time spent by agents at each node is so small to hinder local synchronization, but the fast motion favors the onset of global synchronization.

\subsubsection{Phase oscillators coupled through temporal proximity graphs}
\label{sec:secDiazGuilera}

To illustrate the case where the oscillators are carried  by agents moving in a continuous space, we begin with the model investigated in Ref.~\cite{fujiwara2011synchronization}. According to this study, agents are assumed to perform random walks in a two-dimensional plane, in particular a square of size $L$ and periodic boundary conditions. In addition, agents are assumed to communicate with each other through short-range devices, such that interactions among them can be modeled by a proximity graph that changes in time as the result of the agent motion.

The $i$-th agent moves with velocity $\mathbf{v}_i(t)$, having constant modulus $v$ and variable heading $\theta_i$, such that $\mathbf{v}_i(t)=ve^{\iota\theta_i(t)}$. The headings are updated randomly at discrete time steps $t_k$, with $t_k-t_{k-1}=\tau_M$, such that, indicating with $\mathbf{y}_i(t)\in \mathbb{R}^2$ the position of the $i$-th agent in the plane at time $t$, we have that motion is governed by Eqs.~(\ref{eq:randomwalkers}), here repeated for convenience

\begin{equation}
	\left. \begin{array}{l} \mathbf{y}_i(t_k+\tau_M)=\mathbf{y}_i(t_k)+\tau_M\mathbf{v}_i(t_k),
		\\
		\theta_i(t_k)=\eta_i(t_k)m
	\end{array}
	\right. 
\end{equation}

\noindent where $\eta_i(t_k)$ ($i=1,\ldots,N$) are $N$ independent random variables chosen at each time $t_k$ with uniform probability in the interval $[-\pi,\pi]$.

The Authors of Ref.~\cite{fujiwara2011synchronization} consider that the oscillator associated with each agent is described by a single phase variable, $\phi_i(t)$ with $i=1,\ldots,N$, as in the Kuramoto model~\cite{acebron2005kuramoto,rodrigues2016kuramoto}. Here, however, unlike the classical Kuramoto model, all the agents have the same natural frequency, that without lack of generality is set to zero, \emph{i.e.,} $\omega_i=0$, and interact according to the temporal network induced by the motion scheme discussed above. Accordingly, the dynamics of the oscillators are described by:

\begin{equation}
	\label{eq:phaseoscillators}
	\phi_i(t+\tau_P)=\phi_i(t)+\epsilon \sum_{j=1}^N\mathcal{A}_{ij}(t)\sin[\phi_j(t)-\phi_i(t)],
\end{equation}

\noindent where the phases are updated at discrete time intervals, here indicated with $\tau_P$.

In Eq.~(\ref{eq:phaseoscillators}), the time-varying pattern of interactions among agents is encoded in the coefficients $\mathcal{A}_{ij}(t)$ of the adjacency matrix $\mathcal{A}(t)$ that is function of time $t$. Interactions are mutual, so that at each time instant the adjacency matrix is symmetric. The coefficients are, therefore, defined as $\mathcal{A}_{ij}(t)=\mathcal{A}_{ji}(t)= 1$ if $i$ and $j$ are connected by a link at time $t$, while $\mathcal{A}_{ij}(t)=\mathcal{A}_{ji}(t)=0$ otherwise. In addition, $\mathcal{A}_{ii}(t)=0$, as there are no self-loops. At each time instant, links are established based on the mutual distance between agents, such that $\mathcal{A}(t)$ is the adjacency matrix of the proximity graph induced by the positions $\mathbf{y}_i(t)$, i.e., $\mathcal{A}_{ij}(t)=1$ if $\|\mathbf{y}_j(t)-\mathbf{y}_i(t)\|<R$, where $R$ is the sensing radius.

Before discussing how synchronization arises in this system, it is important to note that, similar to continuum percolation, when $R$ grows the interaction topology has a transition from a regime characterized by small-size, disconnected components to a single giant component including all the agents. The transition occurs at $(N-1)\pi R_c^2/L^2 \approx 4.51$, and, accordingly, different synchronization behaviors are observed for $R<R_c$ or $R>R_c$.

To characterize the system behavior, the Authors of Ref.~\cite{fujiwara2011synchronization} introduce the average phase difference

\begin{equation}
	E_\phi=\sqrt{\frac{1}{N(N-1)}\sum\limits_{i,j=1}^N(\phi_j-\phi_k)^2}
\end{equation}

\noindent and observe that, after an initial transient, it decays exponentially such that it is possible to define the synchronization characteristic time $T_s$ by the relationship $E_\phi \propto e^{-t/T_s}$. In turns, this allows the definition of the number of phase updates needed by the system to reach synchronization as $n_T=T_s/\tau_P$. The smaller is this parameter the more efficient is the system, as it can reach synchronization in a shorter time.

To illustrate the system behavior, let us start considering the scenario where agent motion is fast compared to the oscillator dynamics. Under this assumption, the topology changes fast enough such that the system behavior can be described by taking into account the average interaction matrix from which one can provide the following fast-switching estimate $T_{FS}$ of the synchronization characteristic time $T_s$:

\begin{equation}
	T_{FS}=-\tau_P/\log (1-\epsilon (N-1)p)
\end{equation}

\noindent where $p=\pi R^2/L^2$ represents the probability that two agents are interacting each other under the fast-switching hypothesis. In fact, in this scenario the updated position becomes uncorrelated with the previous one, and the probability that two agents interact becomes equal to the probability that an agent lies in the sensing area of another agent. The latter is given by the ratio between the sensing area $\pi R^2$ and the overall area where agents move, \emph{i.e.,} $L^2$.

\begin{figure}
	\centering
	\includegraphics[scale=0.4]{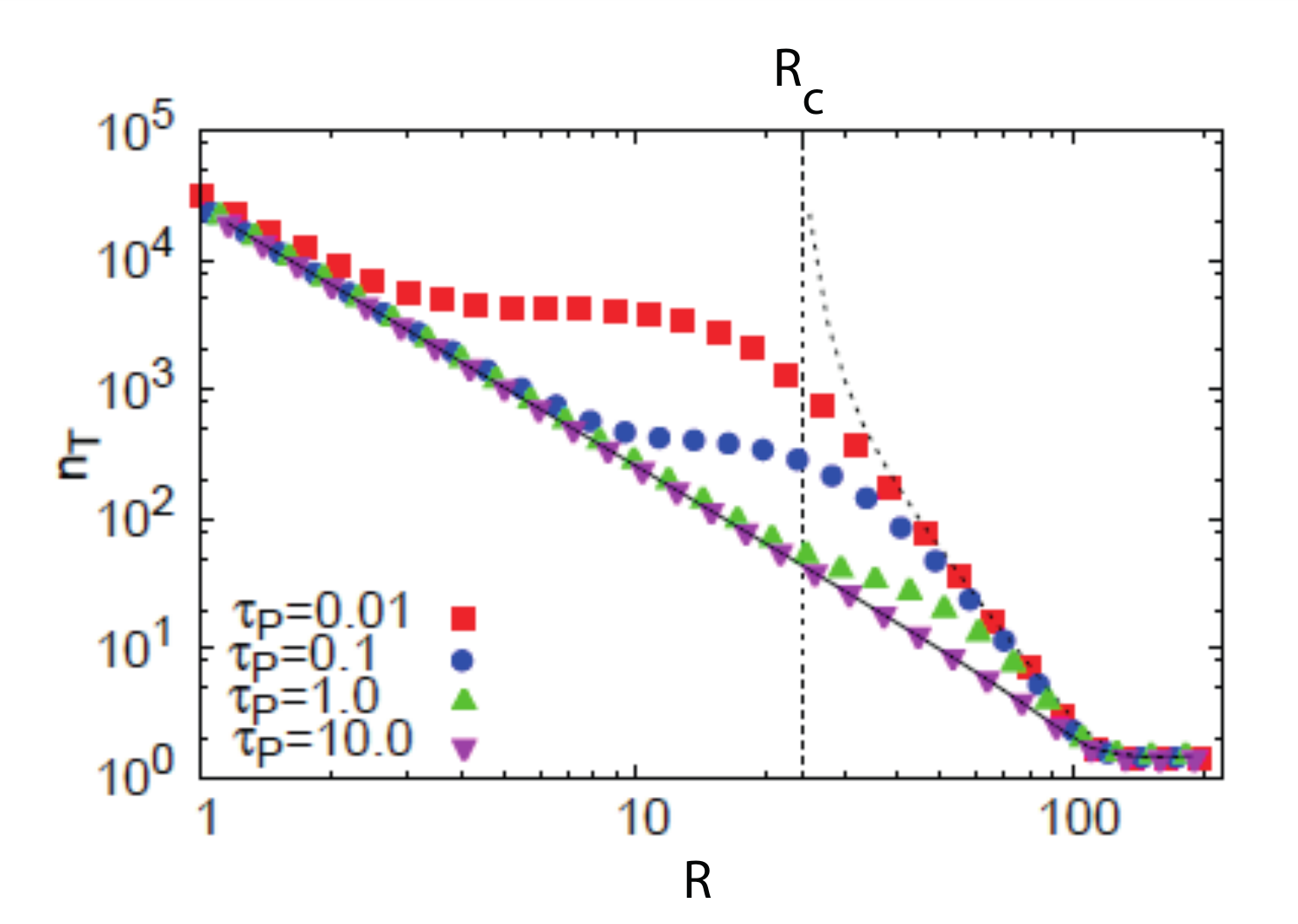}
	\caption{\label{fig:nTvsR} Number of updates required to reach synchronization, $n_T$, vs. the interaction radius $R$ in a system of phase oscillators interacting through a temporal proximity graph.
		\\ {\it Source}: Reprinted figure with permission from Ref.~\cite{fujiwara2011synchronization} \textcopyright~2011 by the American Physical Society.}
\end{figure}

The behavior of $n_T$ as a function of $R$ is illustrated in Fig.~\ref{fig:nTvsR} for several settings of $\tau_P$ along with the prediction of the fast-switching assumption (represented by the continuous line). We notice that $n_T$ decreases with increasing $R$, as a larger interaction radius favors synchronization. In addition, for large $\tau_P$ the fast-switching prediction is accurate in the whole interval of $R$, whereas for small $\tau_P$ the behavior of $n_T$ deviates from the prediction, specially close to the percolation transition $R \simeq R_c$. In fact, a large $\tau_P$ indicates that the oscillator dynamics is slow, a scenario where the fast-switching approximation well reproduces the system behavior.

Altogether, the analysis of the model reveals three diverse asymptotic behaviors that can be explained taking into account that there are two characteristic time scales, one that accounts for the typical time for clusters to synchronize and the other one for leaving the cluster. The first regime takes place when $R$ is small and $\tau_P$ is large. In this case, the network connectivity changes very fast, before the time required by the agents to synchronize with their neighbors, such that all units reach synchronization at approximately the same rate. In Ref.~\cite{fujiwara2011synchronization} this regime is called  global synchronization. The fast-switching assumption accurately describes this mechanism of synchronization. The second regimes occurs when $\tau_P$ is decreased or $R$ is increased, still remaining below the threshold for percolation, \emph{i.e.,} $R<R_c$. In this scenario, the formation of clusters of synchronous nodes is favoured by the fact that the topology changes slowly compared to the time scale of the synchronization process. This mechanism is called multiple cluster local synchronization and is not well reproduced by the fast switching approximation that neglects the correlation between consecutive positions of agents and system dynamics. Finally, for $R>R_c$ a third regime is observed. Here, all the units are connected into a single giant component and motion becomes unnecessary for synchronization, but still contributes to the evolution towards the final state of the system. When $R$ is large enough that the connectivity becomes that of a complete graph, the fast switching assumption again predicts the system behavior.

In the model discussed so far, agents follow Brownian motion. The case of superdiffusive motion has been, instead, considered in Ref.~\cite{grossmann2016superdiffusion}, using the following equations to describe the motion dynamics:

\begin{equation}
	\label{eq:superdiffusive}
	\dot{\mathbf{y}}_i(t)=\xi_i(t)
\end{equation}

\noindent and phase dynamics:

\begin{equation}
	\dot{\phi}_i(t)=\frac{1}{N_i(t)}\sum\limits_{j=1}^N\mathcal{A}_{ij}(t)\sin[\phi_j(t)-\phi_i(t)]+\sqrt{2D_{\phi}}\eta_i(t)
\end{equation}

\noindent where $\xi_i(t)$ indicate L\'evy noise, $N_i(t)$ represents the number of neighbors of agent $i$ at time $t$ and the term $\sqrt{2D_{\phi}}\eta_i(t)$ an additive, Gaussian white noise with intensity $D_\phi$. The Authors of Ref.~\cite{grossmann2016superdiffusion} define a parameter, $\alpha$ in their study, that characterize the distribution of the random displacements and tune the characteristics of the motion, from Brownian (obtained for $\alpha=2$) to L\'evy flights (for $\alpha \in (0,2)$). A remarkable feature of the interplay between motion and dynamics that is unveiled by this model is that, while for Brownian motion, the order parameter $\Phi=\lvert\frac{1}{N} \sum\limits_{j=1}^N e^{\iota \phi_j(t)} \rvert^2$ decreases according to a power law with the number of units, for L\'evy flights the order parameter tends towards a nonzero constant value for large system sizes, indicating  a robust synchronized state.

An important generalization of the model of mobile phase oscillators is dealt with in Ref.~\cite{levis2017synchronization}, where the hypothesis of point-like particles is removed to consider active disks with a non-zero volume. In this study, the dynamics of the oscillators is again described by phase oscillators as in Eq.~(\ref{eq:phaseoscillators}), whereas the kinetic Monte Carlo model of Ref.~\cite{levis2014clustering} is used to represent particle motion. According to this model, agents are self-propelled hard disks with diameter $\sigma$. Their positions are updated in time as follows:

\begin{equation}
	\mathbf{y}_i(t_k+\tau_M)=\mathbf{y}_i(t_k)+\tau_M\mathbf{\delta}_i(t)P_{acc}
\end{equation}

\noindent where $P_{acc}$ represents the acceptance probability of the update, allowing to model how interactions among particles take place. In particular, $P_{acc}=1$ if the update does not yield any overlap among particles, whereas $P_{acc}=0$ in the opposite case. $\mathbf{\delta}_i(t)$ represents the displacement with dynamics given by:

\begin{equation}
	\mathbf{\delta}_i(t_k)=\mathbf{\delta}_i(t_k-\tau_M)+v_1\mathbf{\eta}_i(t)
\end{equation}

\noindent with $\mathbf{\delta}_i(0)=v_0\mathbf{\eta}_i(0)$. Here, $\mathbf{\eta}_i(t)$ is a vector of independent random components drawn at each time step from a uniform distribution in $[-1,1]$. In addition, $\mathbf{\delta}_i(t_k)$ is constrained to lie in a square box of size $v_0$. In the extreme case that $v_1 \ll v_0$, the random shift $v_1\mathbf{\eta}_i(t)$ becomes negligible and particle motion is ballistic with velocity $v_0$. On the other extreme, when $v_1 \gg v_0$, one finds that the displacements are no more correlated in time and the particles move as Brownian disks. Finally, between these two extremes, the model describes an overdamped persistent random walk with persistence time $\tau=\frac{v_0^2}{v_1^2}\tau_M$. The persistence time can be considered a control parameter of the model, along with the packing fraction $\phi=\frac{\pi \sigma^2 N}{4V}$. Varying the two control parameters, one has the possibility to model different phases such as fluid, cluster and gel-like ones~\cite{levis2014clustering}. Considering excluded volume interactions yields $P_{acc}=1$ and $\phi=0$ in the model such that it is possible to recover the case of point-like particles. For other values of the parameters, the model offers a general framework to explore the interplay between self-propulsion, steric repulsion, and phase coupling.

From the analysis of the model carried out in Ref.~\cite{levis2017synchronization}, several interesting conclusions may be drawn. It seems particularly relevant the fact that while, as we have seen, in the absence of particle-particle interactions self-propulsion promotes synchronization of locally coupled oscillators, the same is not true for disks of a finite volume. In fact, in the presence of repulsive interactions, one finds that the behavior of the synchronization time with the persistence time  is not monotonic, and synchronization can be optimized by choosing a precise value of this parameter as a function of the density. This optimum represents a tradeoff between the enhancement of particle motion and the tendency to form clusters as the persistence of the particles is increased.

Another relevant scenario to consider is represented by the presence of non-identical natural frequencies in the oscillators~\cite{ma2015shuttle,ling2020explosive}. Also in this case, agent mobility plays an important
role in achieving synchronization, which in general is favored by an increasing velocity~\cite{ma2015shuttle}. Quite interestingly, in this scenario explosive synchronization may also appear, as we now briefly discuss referring to the model presented in Ref.~\cite{ling2020explosive}.

In classical complex networks of phase oscillators, although in most situations the transition to synchronization is second-order, there also exist settings where the transition is first-order. This phenomenon is known as explosive synchronization (ES) and has been observed in structures with a correlation between the node degree and its natural frequency~\cite{gomez2011explosive}, between the  strength of the coupling and the node natural frequency~\cite{zhang2013explosive}, or between the strength of the coupling and the frequency mismatch with the neighboring oscillators~\cite{leyva2013explosive}, as well as in structures with an adaptive coupling modulated by a local order parameter~\cite{zhang2015explosive,avalos2018emergent} and several other settings (for a recent review see Ref.~\cite{d2019explosive}).
The Authors of Ref.~\cite{ling2020explosive} shows that explosive synchronization may also be observed in systems of mobile phase oscillators. In particular, they model mobility as in Eq.~(\ref{eq:randomwalkers}) with agents interacting according to a temporal proximity graph, and assume the following dynamics for the phase of the oscillators:

\begin{equation}
	\label{eq:phaseforexplosive}
	\dot{\phi}_i(t)=\omega_i+\epsilon\alpha_i\sum\limits_{j=1}^N\mathcal{A}_{ij}(t)\sin[\phi_j(t)-\phi_i(t)]
\end{equation}

\noindent where $\alpha_i$ is a parameter tuning the coupling strength and $k_i(t)$ the degree of node $i$ at time $t$. The natural frequencies $\omega_i$ of the oscillators are assumed to be uniformly distributed in the interval $[-\pi,\pi]$. For a fraction $f$ of randomly chosen units, $\alpha_i=r_i$ where $r_i=\left | \frac{1}{k_i(t)}\sum\limits_{j=1}^N\mathcal{A}_{ij}(t)e^{\iota \phi_i(t)}\right |$ is the local order parameter, whereas for the remaining fraction of units, i.e., $1-f$, $\alpha_i=1$. In this way, similarly to Ref.~\cite{zhang2015explosive}, the effective strength of the coupling is correlated with the local level of synchrony. However, at variance with the model investigated in Ref.~\cite{zhang2015explosive}, here the neighborhood of each unit changes in time as the result of agent motion.

\begin{figure}
	\centering
	\includegraphics[scale=0.5]{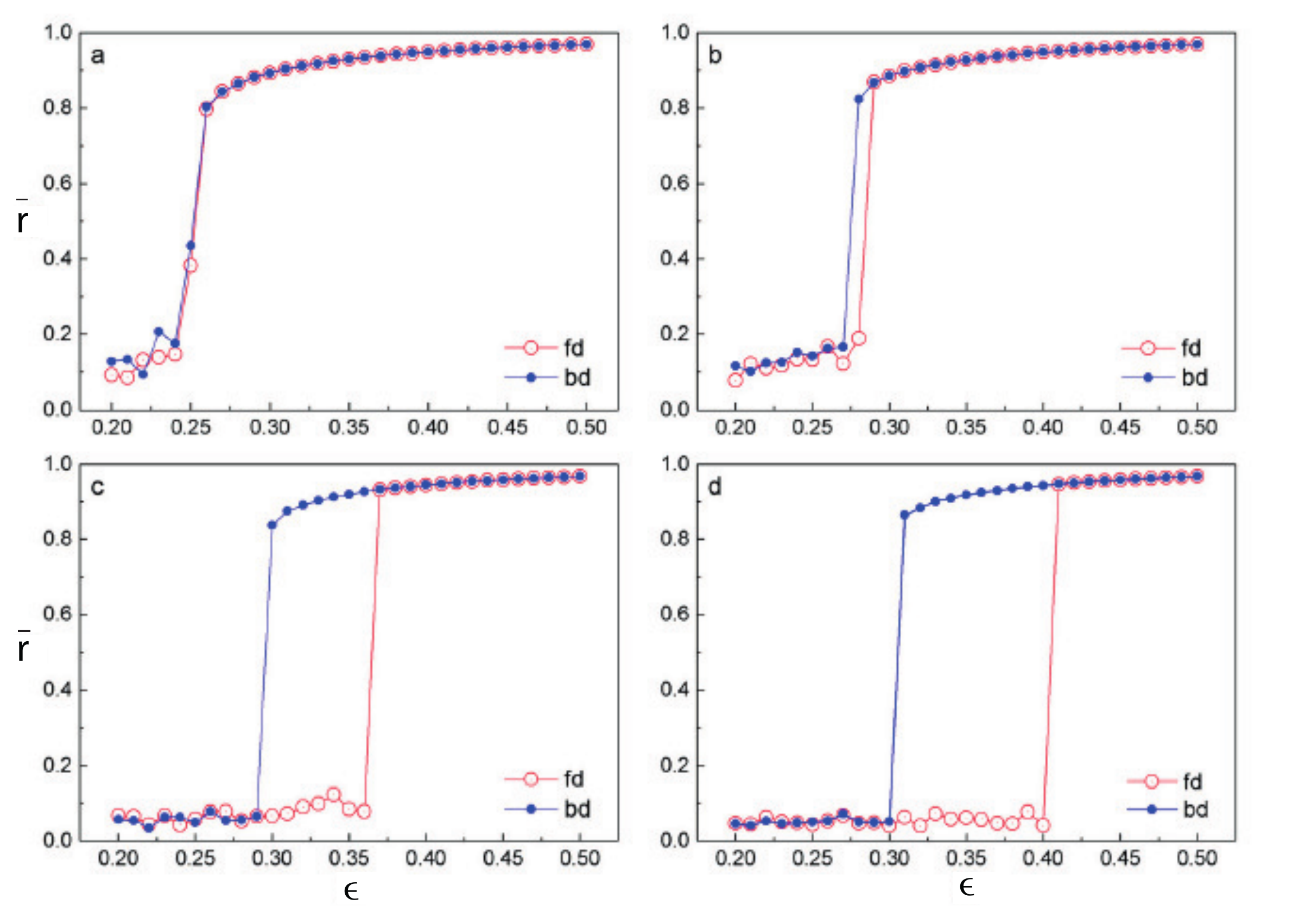}
	\caption{\label{fig:explosive} Order parameter
		$\bar{r}=\lim\limits_{T\rightarrow \infty}\frac{1}{T}\left |\frac{1}{N}\sum\limits_{j=1}^N e^{\iota \phi_j{t}}\right |$ with respect to coupling strength $\epsilon$ for a model of mobile phase oscillators as in Eqs.~(\ref{eq:phaseforexplosive}) for different values of the tuning parameter $f$: (a) $f=0$; (b) $f=0.2$; (c) $f=0.8$, and (d) $f=1$. The red open circles and blue solid circles indicate represent the forward and backward continuation diagrams, respectively.
		\\ {\it Source}: Reprinted figure from Ref.~\cite{ling2020explosive}, \textcopyright~2020, with permission from Elsevier.}
\end{figure}

The peculiar feature of this model is that for $f=0$ the transition to synchronization is second-order, whereas for larger $f$ it becomes first-order with a hysteresis window of increasing width. Figure~\ref{fig:explosive} illustrate this result, showing forward and backward continuations with the coupling strength $\epsilon$ for the order parameter $\bar{r}=\lim\limits_{T\rightarrow \infty}\frac{1}{T}\left |\frac{1}{N}\sum\limits_{j=1}^N e^{\iota \phi_j{t}}\right |$.

\subsubsection{Application to flocking control}

An important application of the model of phase oscillators coupled according to a proximity graph is in multi-agent systems, where it is used to design a protocol for addressing the flocking problem~\cite{zhu2012flocking}. The flocking problem consists of determining a distributed control law ensuring that the velocities of all agents asymptotically converge to the same value, while avoiding collisions among the agents. In recent years, diverse solutions to the problem have been proposed, including leaderless or leader-follower strategies, methods based on artificial potential fields, or techniques relying on the knowledge of the relative distances among the agents~\cite{mesbahi2010graph,olfati2006flocking,zavlanos2007potential}.

The model of coupled phase oscillators provides a solution to the flocking problem when the neighbors' heading is the only information available to agents. The main idea is to reinterpret the phase of the oscillator as the heading of the agent, such that $\phi_i=\theta_i$. The dynamics of the system can be hence rewritten as:

\begin{equation}
	\label{eq:coupledphaseoscToFlockingNL}
	\begin{array}{l}
		\left [ \begin{array}{c}
			\dot{y}_{i,1}  \\
			\dot{y}_{i,2}
		\end{array}\right ] = v  \left [ \begin{array}{c}
			\cos \theta_i  \\
			\sin \theta_i
		\end{array}\right ], \\
		\dot{\theta}_i=\epsilon\sum\limits_{j=1}^N \mathcal{A}_{ij}\sin(\theta_j-\theta_i),
	\end{array}
\end{equation}

\noindent or, in the case that a linear protocol is adopted, as:

\begin{equation}
	\label{eq:coupledphaseoscToFlockingLinear}
	\begin{array}{l}
		\left [ \begin{array}{c}
			\dot{y}_{i,1}  \\
			\dot{y}_{i,2}
		\end{array}\right ] = v  \left [ \begin{array}{c}
			\cos \theta_i  \\
			\sin \theta_i
		\end{array}\right ], \\
		\dot{\theta}_i=\epsilon \sum\limits_{j=1}^N \mathcal{A}_{ij}(\theta_j-\theta_i).
	\end{array}
\end{equation}

There is a profound difference between system~(\ref{eq:coupledphaseoscToFlockingNL}) and the model discussed in Sec.~\ref{sec:secDiazGuilera}. Here, in fact, the dynamics of synchronization influences the motion of the agents, such that the system becomes a closed-loop one, where motion affects synchronization and synchronization, in turn, impacts motion. Consequently, the structure of interactions is no more exclusively determined by the type of motion, but also by how the synchronization dynamics develops. In particular, it may happen that, because of how their headings evolve, some agents become disconnected from the network and are no more able to re-join the team. For this reason, flocking control requires algorithms that explicitly consider the problem of preserving network connectedness. The Authors of Ref.~\cite{zhu2012flocking} demonstrate that both models~(\ref{eq:coupledphaseoscToFlockingNL}) and~(\ref{eq:coupledphaseoscToFlockingLinear}) may solve flocking while preserving network connectedness and avoiding collisions among agents. To discuss in more detail the conditions guaranteeing this, let us start from the linear case.

Consider system~(\ref{eq:coupledphaseoscToFlockingLinear}). If the initial proximity graph is connected and the initial positions of the agents are distinct, then there exists a critical value $\epsilon_c$ of the control gain, such that if $\epsilon>\epsilon_c$ the proximity graph preserves connectedness for any time $t>0$, collisions are avoided and the flocking problem is solved. This result shows that, provided that the control gain is sufficiently large, flocking can be achieved for any velocity $v$. However, the value of $\epsilon_c$ increases with $v$, meaning that faster agents require a higher control gain. The dependence of $k_c$ on the algebraic connectivity of the initial proximity graph, here indicated as $\mu$, is also interesting: $\epsilon_c$ scales as $\mu^{-1}$, such that a larger connectivity yields a smaller threshold for flocking control.

A similar result holds for the nonlinear case~(\ref{eq:coupledphaseoscToFlockingNL}). However, in this case a further condition is required: for any pair of agents, $i$ and $j$, the difference of their initial headings need to be bounded, namely $|\theta_j(0)-\theta_i(0)|< \pi$. The expression for $k_c$ in this case is different, but the functional dependence on $v$ and $\mu$ is the same.

Notice that the two results constitute sufficient, but not necessary conditions, and yet they provide an analytic proof of the existence of a solution to the problem based on the information on the headings on neighbors in the temporal proximity graph.

An interesting generalization of this approach to flocking control is the inclusion of both metric and topological interactions~\cite{wang2016synchronization}. The idea is that each agent interacts with all the other units within a circle of a fixed radius, and, in addition, with a number of other agents determined according to the topological criterion of the smallest distance. In more detail, the Authors of Ref.~\cite{wang2016synchronization} set a minimum number of interacting agents, $M$, and, whenever an agent finds $m<M$ neighbors within its sensing radius, it also interacts with $M-m$ further units, selected as those at the smallest distance from its position. In this way, in any condition each agent will interact at least with $M$ agents, a configuration which proves to be particularly useful when the agent density is low and protocols based only on metric interactions may show poor performance. Such an \emph{hybrid} protocol also solves the problem of preserving network connectedness, but requires a communication system that is able to reach any other agent, independently from its effective distance. Here, a large communication radius $R$ and number $M$ both favor synchronization, as they generate a configuration with a larger number of neighbors. We will see that, while this principle holds true in different models of coupled phase oscillators, the same does not apply when the unit dynamics is more complex, as, for instance, it happens in the case of chaotic oscillators discussed in Sec.~\ref{sec:chaoticoscimovingagents}. After deriving the conditions guaranteeing convergence of the protocol, the Authors of Ref.~\cite{wang2016synchronization} discuss a series of numerical results showing that the metric-topological model conjugates the advantages
of the metric and topological interaction models and has the fastest convergent rate and smallest heading difference for different interaction ranges.

We conclude this section remarking that there is a huge literature on flocking control, which goes beyond the purpose of this section whose aim was instead to hallmark the relationship between it and the model of coupled phase oscillators in temporal proximity graphs.

\subsubsection{Interplay between mobility and synchronization}
\label{sec:swarmalators}

Several complex phenomena in nature triggered the definition of models in which mobile phase oscillators have a two-way interaction. Most of the examples in the literature, in fact, deal with the influence of motion on the oscillator's phases, whereas a limited effort has been devoted to the converse phenomenon, that is, how the oscillator phase can influence motion. For example, it is well known that various animal species, such as frogs, crickets, and katydids, tend to synchronize their sound emissions~\cite{aihara2008mathematical,walker1969acoustic,greenfield1994synchronous,aihara2014spatio}. However, whether the characteristics of such sound collective phenomena may drive collective or individual motion phenomena is still an open question. Motivating examples toward a mathematical modeling of such phenomena come from the physics of magnetic colloids~\cite{yan2012linking,martin2013driving,snezhko2011magnetic} and the microfluidic mixtures of active spinners~\cite{nguyen2014emergent,van2016spatiotemporal}, where  particles or spinners tend to attract or repel each other, based on their orientation. In the biological realm, a population of myxobacteria that exhibit a bidirectional coupling between spatial and phase dynamics has been observed and modeled by Igoshin and colleagues in Ref.~\cite{igoshin2001pattern}. Also Tanaka et al., in Ref.~\cite{tanaka2007general} laid the foundations of such investigations, considering chemotactic oscillators whose movements in space are mediated by the diffusion of a background chemical. These observations call for a new modeling paradigm, introduced by O'Keeffe and colleagues, called \em swarmalators, \em i.e., oscillators that synchronize and swarm~\cite{o2017oscillators}. Despite a simple mathematical formulation, the model exhibits an ample range of complex phenomena that can parallel relevant natural and physical phenomena.

The model comprises $N$ oscillators, indexed with $i=1, 2,\ldots, N$, endowed with  phase $\phi_i$ and natural frequency $\omega_i$. Each oscillator occupies position $\mathbf{x}_i=(x_i,y_i)$  and moves with velocity $\mathbf{v}_i$.

The system dynamics is modeled as

\begin{equation}
	\label{eq:swarmaspace}
	\mathbf{\dot{x}_i} = \mathbf{v}_i + \frac{1}{N} \sum_{j=1}^N \left[ \mathbf{I}_\mathrm{att} (\mathbf{x}_j - \mathbf{x}_i) F(\phi_j - \phi_i) - \mathbf{I}_\mathrm{rep}  (\mathbf{x}_j - \mathbf{x}_i) \right],
\end{equation}

\begin{equation}
	\label{eq:swarmaphase}
	\dot{\phi}_i = \omega_i + \frac{K}{N} \sum_{j=1}^N H_{\mathrm{att}} (\phi_j - \phi_i) G(\mathbf{x}_j - \mathbf{x}_i).
\end{equation}
Functions $\mathbf{I}_\mathrm{att}$ and $\mathbf{I}_\mathrm{rep}$ model spatial attraction and repulsion between swarmalators, respectively. Function $H_{\mathrm{att}}$, on the other hand, models the phase interaction. Finally, function $F$ in Eq.~\eqref{eq:swarmaspace} quantifies the influence of phase similarity on spatial attraction and function $G$ in Eq.~\eqref{eq:swarmaphase} quantifies the influence of spatial proximity on phase attraction.

Even a simplified version of the model can lead to a reach portfolio of complex phenomena. Let us consider for example
\begin{equation}
	\label{eq:simplswarmspace}
	\mathbf{\dot{x}_i} = \mathbf{v}_i + \frac{1}{N} \left[ \sum_{j=1, j\neq i}^N \frac{\mathbf{x}_j - \mathbf{x}_i}{|\mathbf{x}_j - \mathbf{x}_i|} \left( A + J \cos(\phi_j - \phi_i) \right) -B \frac{\mathbf{x}_j - \mathbf{x}_i}{|\mathbf{x}_j - \mathbf{x}_i|^2} \right],
\end{equation}

\begin{equation}
	\label{eq:simplswarmphase}
	\dot{\phi}_i = \omega_i + \frac{K}{N} \sum_{j=1, j\neq i}^N  \frac{\sin \left( \phi_j -\phi_i \right)}{|\mathbf{x}_j - \mathbf{x}_i|}.
\end{equation}
Formulating the additional assumptions of  identical swarmalators ($\omega_i=\omega$ and $\mathbf{v}_i = \mathbf{v})$, a common propulsion velocity with constant magnitude and direction $\mathbf{v}=v_0 \hat{n}$, where $\hat{n}$ is a constant vector, choosing a reference frame such that $\omega = v_0 = 0$, and rescaling time and space such that $A=B=1$, the system behavior can be parameterized according to the values of $J$ and $K$. These parameters regulate the interplay between motion and synchronization. In particular, parameter $K$ regulates the strength of the phase difference, and parameter $J$ regulates the relationship between phase similarity and spatial attraction. Positive values of $J$ induces behaviors in which swarmalators tend to come closer in space to those with similar phases, whereas the opposite phenomenon occurs for negative values of $J$.

\begin{figure}
	\centering
	\includegraphics[scale=0.3]{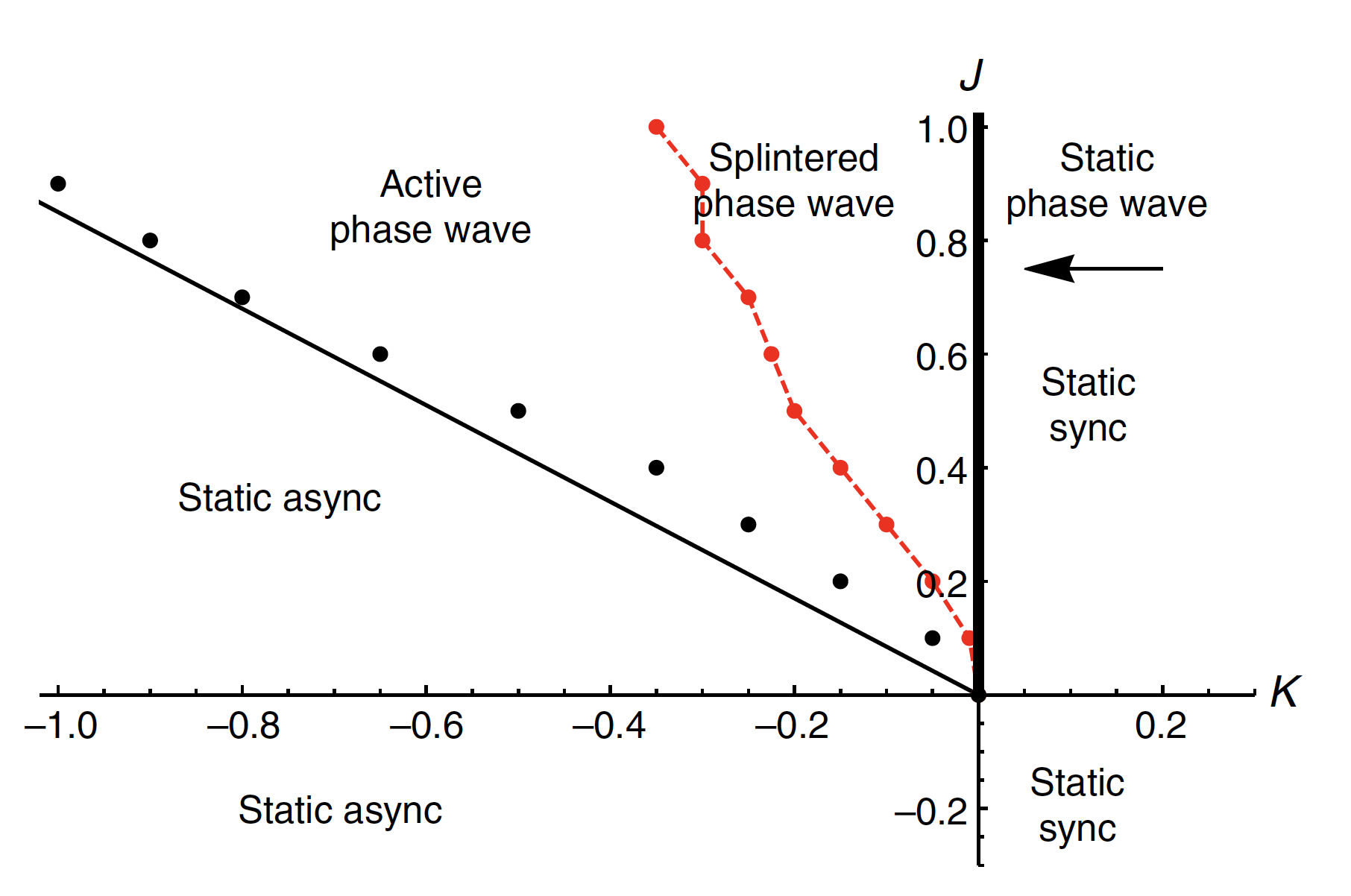}
	\caption{\label{fig:swarmaparameterspace} Location of the states of the model defined by Eqs.\eqref{eq:simplswarmspace}-\eqref{eq:simplswarmphase}, in the $(K,J)$ parameter space.
	\\ {\it Source}: Reprinted with permission from Ref. \cite{o2017oscillators}.}
\end{figure}

Even in this simplified setting, the swarmalator system described by Eqs.\eqref{eq:simplswarmspace}-\eqref{eq:simplswarmphase} exhibit five macroscopic behaviors, or states, corresponding to different areas of the $(K,J)$ parameter space, illustrated in Fig.~\ref{fig:swarmaparameterspace}. Among these five states, three are ultimately static in space and phase, whereas in the remaining two, the swarmalators move.

The five states (where the first three states are static, and the last two comprise moving swarmalators) can be summarized as follows:
\begin{itemize}
	\item \textbf{Static synchrony:} the swarmalators are distributed in space according to a circularly symmetric, cristal-like formation, and are fully synchronized in phase. This state occurs for $K>0$ and all $J$.
	\item \textbf{Static asynchrony:} the swarmalators are uniformly distributed in space, and can assume any phase everywhere in the spaces (indeed, the phase distribution results uniform over the space). This state occurs for $J<0$ and $K<0$, and also for $J>0$, provided that it lies in the wedge $J<|K_c|$, where $K_c$ is derived in Ref.~\cite{o2017oscillators} through a semianalytical approximation.
	\item  \textbf{Static phase wave:} In the last stationary state, the swarmalators' phases are frozen at their initial values. This state occurs for $K=0$ and $J>0$, implying that swarmalators tend to settle closer to others with similar phases. This, in turn, results in an annular structure where the spatial angle of each swarmalator is perfectly correlated with its phase $\phi$.
	\item \textbf{Splintered phase wave:} the transition to $K<0$ leads to the regimes where swarmalators move in space. A static phase wave is observed along the space, which splinters into disconnected clusters, each characterized by a distinct phase. Within each cluster, swarmalators execute rapid motions oscillating both in position and in phase about their mean values. The role of the parameters in shaping this regime is still unclear and deserves further investigations, see Ref.~\cite{o2017oscillators}.
	\item \textbf{Active phase wave:} with further decreases in $K<0$, swarmalators tend to execute regular cycles both in spatial angles and in phase. Further oscillations are revealed along the radial position, where each swarmalator travels back and forth from the inside to the outside of the global annular configuration, while orbiting around the annulus. Natural phenomena exhibiting states similar to the active phase state are double milling states in biological swarms~\cite{carrillo2009double} and vortex arrays observed in groups of sperm~\cite{riedel2005self}.
\end{itemize}

The swarmalator model keeps its characteristics also when different, and more generic functions $\mathbf{I}_\mathrm{att}$, $\mathbf{I}_\mathrm{rep}$, and $G$ are selected, thus highlighting the genericity of the model in describing models of dynamical systems that may synchronize and swarm at the same time, and where these two phenomena are intertwined. Qualitatively similar results are also found when some degree of heterogeneity among the swarmalator population is introduced in Ref.~\cite{o2017oscillators}.

\subsection{Pulse-coupled oscillators}
\label{sec:pulsecoupledmovingagents}

In many natural and artificial systems, interactions do not occur continuously, but only at discrete time instants. Consider, for instance, an ensemble of neurons in the human brain or a swarm of flashing fireflies; they are both examples of biological systems where the units communicate through short pulses to synchronize their activity~\cite{tateno2007phase,buck1988synchronous}. This type of coupling may also be found in engineered systems. An example is synchronization of the clocks embedded in a sensor network that may be efficiently achieved operating exclusively
at the physical layer by transmitting pulses rather than packet messages~\cite{wang2015synchronization,hong2005scalable,an2009nonidentical}. All these systems may be conveniently modeled by pulse-coupled oscillators~\cite{mirollo1990synchronization}.

Pulse-coupled oscillators are limit cycle oscillators
coupled through the links of a network through which they exchange pulses at discrete time instants. In this section we review the relevant literature concerning synchronization in these systems in the case that the network is time-varying, and is obtained as the result of agent motion.

Let us begin with the model discussed in Ref.~\cite{wang2015synchronization}, where the agents are described by a phase variable $\phi_i(t) \in [0,1]$, $i=1,2,\ldots,N$, whose dynamics is given by

\begin{equation}
	\label{eq:phasevariablePCO}
	\dot{\phi}_i=\frac{1}{T}
\end{equation}

\noindent where $T$ is the pulse period. In addition, a state variable $x_i(t) \in [0, 1]$, function of the oscillator phase, is associated with each oscillator:

\begin{equation}
	\label{eq:statevariablePCO}
	{x}_i=f(\phi_i)
\end{equation}

\noindent where $f:[0,1] \rightarrow [0,1]$ is a smooth, monotonically increasing and concave down function, such that $x_i=0$ when $\phi_i=0$ and $x_i=1$ when $\phi_i=1$.

Agents move as random walkers (see Eqs.~(\ref{eq:randomwalkers})) and interact according to the temporal proximity graph. In particular, suppose that at time $t$ the phase and state variables of oscillator $i$ become equal to 1 and indicate with $\mathcal{N}_i(t)$ the set of agents within distance $R$ from $i$ at this time. Then, the agent $i$ will send to all its neighbors ($j\in \mathcal{N}_i(t)$) a signal pulse that will immediately increase by a quantity $\epsilon$ (equal to the coupling strength) the variable $x_j$. At the same time the pulse firing will reset the phase and state variables of oscillator $i$. Hence, we have:

\begin{equation}
	x_i(t^-)=1 \Rightarrow \left \{ \begin{array}{ll} x_i(t^+)=\phi_i(t^+)=0 & \\
		x_j(t^+)=\min \{ x_j(t^-)+\epsilon,1\}, & j\in \mathcal{N}_i(t)\\
		\phi_j(t^+)=f^{-1}(x_j(t^+)), & j\in \mathcal{N}_i(t)
	\end{array} \right.
\end{equation}

For this model, there is a very interesting result that parallels a fundamental theorem derived for pulse-coupled oscillators in the case of static, full (i.e., all-to-all) connectivity~\cite{mirollo1990synchronization}. Mirollo and Strogatz have, in fact, found that if a system of statically pulse-coupled oscillators has identical, smooth, monotonically increasing and concave down state functions, identical and instantaneous coupling, and full connectivity, then for any coupling strength $\epsilon >0$, the set of initial states that do not lead to synchronization has zero measure~\cite{mirollo1990synchronization}. Noticeably, the conditions on the state functions are very mild, and make the result applicable to many scenarios.

The Authors of Ref.~\cite{wang2015synchronization} have discovered that a very similar result holds for moving pulse-coupled oscillators. In fact, if the pulse-coupled oscillators move according to the random walker model~(\ref{eq:randomwalkers}), are characterized by identical, smooth, monotonically increasing and concave down state functions, identical and instantaneous coupling, and are linked each other according to a temporal proximity graph that is connected at any time, then the set of initial states that do not lead to synchronization has zero measure. Again the hypotheses on the state function are very mild and, hence, the result is quite general. It is also interesting to observe that for fully connected temporal proximity graphs one recovers the result valid for static networks. Finally, we notice that the conditions are sufficient, but not necessary for synchronization, as shown by the numerical examples given in Ref.~\cite{wang2015synchronization}, which prove that synchronization in a system of moving pulse-coupled oscillators can be achieved even if the network is not connected at any time.

The numerical simulations discussed in Ref.~\cite{wang2015synchronization} reveal another important feature of the system: the synchronization time is not monotonic with respect to the coupling strength. In fact, while it first decreases with $\epsilon$, then it increases, before finally decreasing with further increasing coupling strengths. A similar non-monotonic, and somewhat unexpected, dependence of the synchronization time on some of the system parameters, in particular agent speed, is observed in other setups of moving pulse-coupled oscillators~\cite{prignano2012synchronization,prignano2013tuning,perez2017control,beardo2017influence} and worth to be discussed here in more detail.

To elucidate this aspect, let us consider the model with a minimal interaction rule discussed in Ref.~\cite{prignano2013tuning}. The model consists of $N$ integrate-and-fire oscillators characterized by a phase $\phi_i(t) \in [0,1]$ with dynamics as in Eq.~(\ref{eq:phasevariablePCO}). A firing event occurs when the phase reaches its maximum value, after which the phase of the agent is reset and a pulse of strength $\epsilon$ is sent to the nearest neighbor oscillator, labeled as $nn$, that is:

\begin{equation}
	\phi_i(t^-)=1 \Rightarrow \left \{ \begin{array}{l} \phi_i(t^+)=0  \\
		\phi_{nn}(t^+)=\min \{ (1+\epsilon)\phi_{nn}(t^-),1\}
	\end{array} \right.
\end{equation}

In addition, upon the firing event, the heading of the nearest neighbor agent, $\theta_{nn}(t)$, is also randomly updated according to the random walker model~(\ref{eq:randomwalkers}).

\begin{figure}
	\centering
	\includegraphics[scale=0.3]{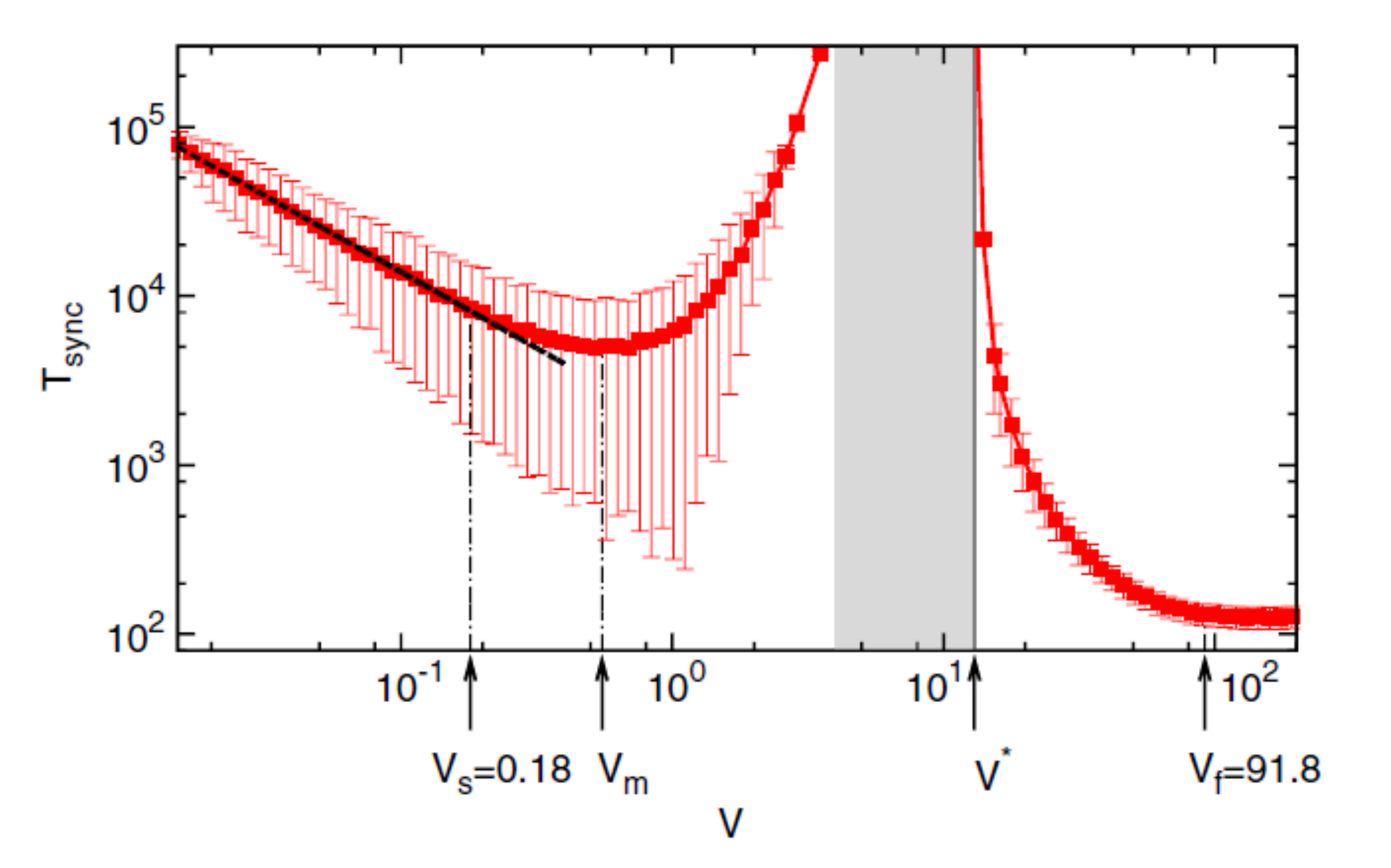}
	\caption{\label{fig:M4prig} Synchronization time $T_{sync}$ vs. agent speed $v$ for a system of mobile pulse-coupled integrate-and-fire oscillators with nearest neighbor connectivity. A number of $N=20$ agents moving in a square plane of size $L=400$, with coupling strength $\epsilon=0.1$ and firing period $T=1$ is considered. Results are averaged over 2000 realizations.
	\\ {\it Source}: Reprinted figure with permission from Ref.~\cite{prignano2013tuning} \textcopyright~2013 by the American Physical Society.}
\end{figure}

As in this system synchronization occurs through a series of firing events at discrete times, the Authors of Ref.~\cite{prignano2013tuning} define the synchronization time $T_{sync}$ as the number of cycles that a reference oscillator requires to enter the synchronized state, a definition in practice analogous to the parameter $n_T$ for continuously-coupled phase oscillators (Sec.~\ref{sec:secDiazGuilera}). By monitoring this parameter
as a function of the agent speed $v$, three regions are found (see Fig.~\ref{fig:M4prig} for an example): two regions of synchronization separated by an interval of values of $v$ for which synchronization is impossible. The behavior of $T_{sync}$ with $v$ is, therefore, highly non-monotonic with a region where increasing the agent speed, rather than favoring synchronization, on the contrary hinders it.

In the regions of synchronization, however, two different mechanisms are at work. In the one on the left, for $v<v_s$, oscillators move slowly and, hence, tend to have the same neighbors for a long time. Under these conditions local synchronization is favored, and global synchronization requires a large number of firings and changes of topology. In the region on the right, for $v>v_f$, the fast switching approximation holds, and $T_{sync}$ becomes independent of $v$. Here, the oscillators fire at
quickly-changing neighbors in such a way that they promote synchronization simultaneously with all the units of the system. Close to the boundaries with the other region, these mechanisms become less effective, until they become totally ineffective and synchronization is prohibited.

The presence of a region in which the rate of change of the topology prevents synchronization is also found when a more general interaction scheme is adopted, as in Ref.~\cite{perez2017control}. Similarly to Ref.~\cite{prignano2013tuning}, the model investigated in Ref.~\cite{perez2017control} also considers mobile pulse-coupled integrate-and-fire oscillators. However, at variance of Ref.~\cite{prignano2013tuning}, interactions are not limited to a single neighbor, but account for a number $K$ of nearest neighbors. In addition, also cone-vision connectivity is studied. In this case, agent $j$ is considered to be in contact with agent $i$ if $\| \mathbf{y}_i -\mathbf{y}_j\| \leq R $ and $\frac{\mathbf{y}_i -\mathbf{y}_j}{\| \mathbf{y}_i -\mathbf{y}_j\|}\frac{\mathbf{v}_j}{\| \mathbf{v}_j\|} \leq \cos \left(\frac{\Theta}{2}\right)$, where $\Theta$ is the cone vision angle. Finally, in view of a practical implementation of the system, the model does not assume cyclic boundary conditions, but a bounded environment, where each unit moves in a straight line until it reaches the arena boundary where it changes its heading to a new random value.

With this general setup, several important features of moving pulse-coupled oscillators are revealed. First, the type of phase-response curve used for coupling has an important effect on the behavior of the synchronization time as a function of agent speed. While a non-monotonic behavior appears for a multiplicative phase-response curve, for other types of coupling functions the relations seems to be monotonic.

\begin{figure}
	\centering
	\includegraphics[scale=0.3]{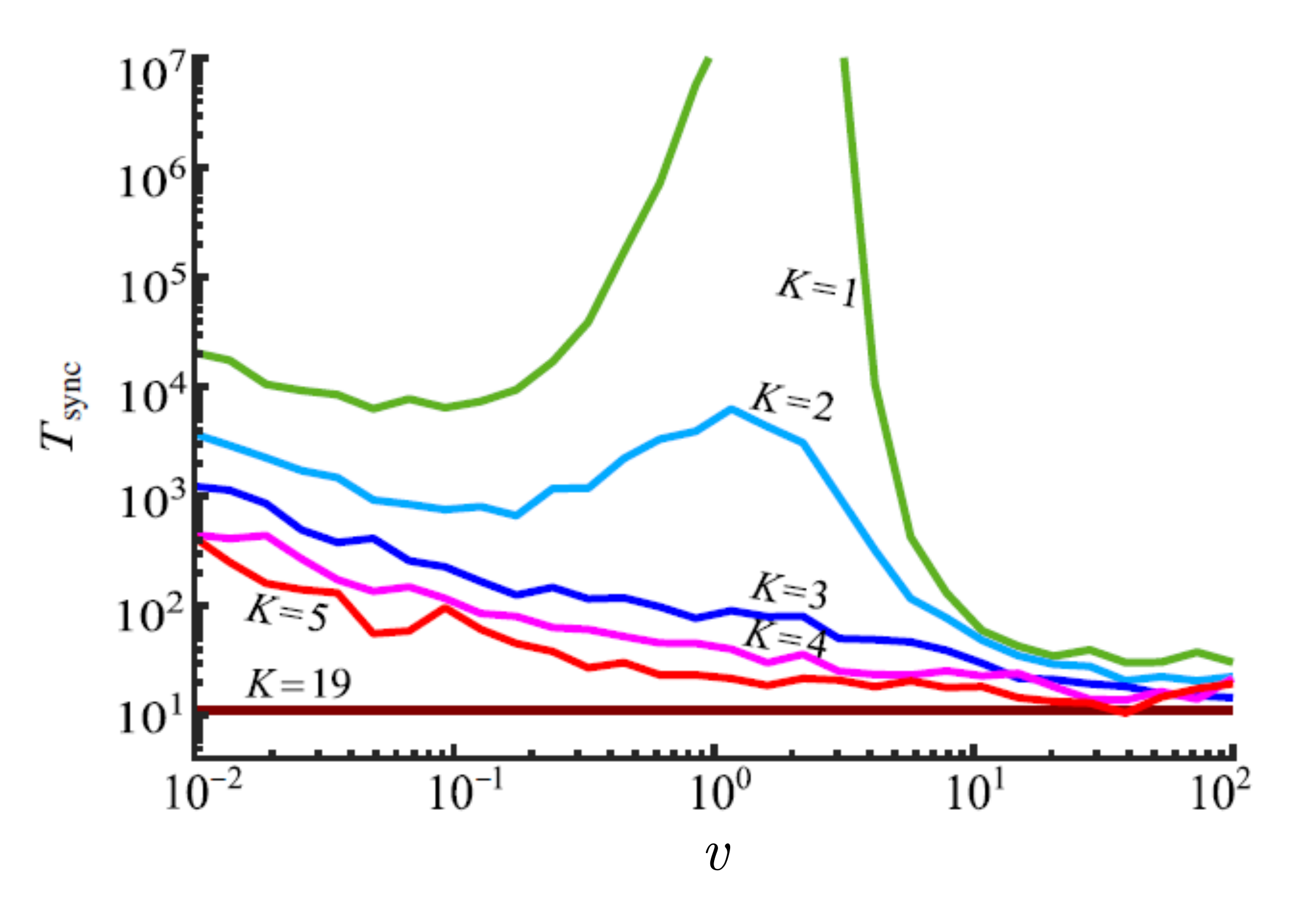}
	\caption{\label{fig:M5perez} Synchronization time $T_{sync}$ vs. agent speed $v$ for a system of mobile pulse-coupled integrate-and-fire oscillators with K-nearest neighbor connectivity. The system is formed by $N=20$ agents moving in a square plane of size $L=100$.
	\\ {\it Source}: Reprinted figure with permission from Ref.~\cite{perez2017control}.}
\end{figure}

The type of the neighborhood model, as well as the exact number of neighbors considered, i.e., $K$ in the K-nearest neighborhood interaction scheme, are also able to modulate the dependence of $T_{sync}$ on $v$. An example is shown in Fig.~\ref{fig:M5perez} which shows the synchronization time as a function of the agent speed for a system of $N=20$ agents at different values of $K$. Notice that the region where synchronization is totally inhibited by the mobility, found for $K=1$, disappears for the other values of $K$. For $K=2$ the behavior of $T_{sync}$ with $v$ is still non-monotonic, but for larger values it becomes monotonic. Similarly, for the cone-vision interaction scheme one finds either regions with a monotonic or a non-monotonic behavior, the latter typically arising for small sizes of the cone of vision, with a non-trivial interplay between the radius and the angle of the cone.

An explanation of the non-monotonic behavior in terms of the typical timescales of the system is given in Ref.~\cite{beardo2017influence}. The system is, in fact, characterized by two time scales, that of the motion of the units and that of the synchronization of local clusters. When the pattern of interactions is sparse and nonreciprocal and the coupling is highly nonlinear, these two processes can interfere in a destructive manner. In fact, when the typical time between two consecutive changes in the connectivity becomes comparable with the time scale of local synchronization, then, on the one hand, groups of neighboring units are broken before they may synchronize, and, on the other hand, the interactions are not rewired fast enough to promote the mechanism needed for global synchronization.

The models discussed in this section have been also validated experimentally. In particular, a practical implementation of the cone-vision interaction scheme based on mobile robots, communicating each other by emitting light pulses, has been proposed in Ref.~\cite{perez2017control}. The Authors have found experimental evidence of the non-monotonic behavior of the synchronization time, demonstrating that by controlling the agent speed the system may be switched from a synchronous state to a desynchronized state and viceversa. This may be of practical interest in multi-robot teams that are required to switch between different tasks, one associated to synchronization in which the activities of teams need to be performed at the same time, and one associated to desynchronization, in which the individual activities need to be distributed over time.

\subsection{Limit cycle and chaotic oscillators}
\label{sec:chaoticoscimovingagents}

\subsubsection{Limit cycle and chaotic oscillators in temporal proximity graphs}
\label{sec:basicmodel}

We now move to consider the case of multi-agent systems of limit cycle and chaotic oscillators that can be dealt with the same formalism. We will mostly refer to chaotic systems, as they represent the more general case, but keeping in mind that the results apply for both limit cycle and chaotic oscillators.

The model we start with (see Ref.~\cite{frasca2008synchronization}) consists of $N$ agents that are free to move in a two-dimensional space and interact with each other on the basis of a proximity graph, that is, two agents interact only if their reciprocal distance is smaller than the interaction radius $R$. Each agent $i=1,\ldots,N$ is characterized by a dynamical state $\mathbf{x}_i(t)\in \mathbb{R}^n$, which represents the state vector of the oscillator associated to the agent. This model may find application in the study of clock synchronization in mobile robots~\cite{buscarino2006dynamical} and in sensor networks with limited communication~\cite{sundararaman2005clock}, task coordination in swarming animals, or synchronized bulk oscillations in populations of yeast cells~\cite{dano1999sustained,dano2002synchronization,dano2007quantitative}.

The agent dynamics is described by
\begin{equation}
	\label{eq:eqTVadj}
	\dot{\mathbf{x}}_i=\mathbf{f}(\mathbf{x}_i)+\epsilon\sum_{j=1}^N\mathcal{A}_{ij}(t) \mathrm{B}(\mathbf{x}_j-\mathbf{x}_i)
\end{equation}

\noindent for $i=1,\ldots,N$. Here, $\mathbf{f}$ is the uncoupled dynamics, $\mathrm{B} \in \mathbb{R}^{n \times n}$ the inner coupling matrix, and $\epsilon$ the coupling strength. In Eq.~(\ref{eq:eqTVadj}), the time-varying pattern of interactions among agents is encoded in the coefficients $\mathcal{A}_{ij}(t)$ of the adjacency matrix $\mathcal{A}(t)$ that is function of time $t$. We suppose that the interactions are mutual, so that at each time instant the adjacency matrix is symmetric. The coefficients are, therefore, defined as $\mathcal{A}_{ij}(t)=\mathcal{A}_{ji}(t)=$ if $i$ and $j$ are connected by a link at time $t$, while $\mathcal{A}_{ij}(t)=\mathcal{A}_{ji}(t)=0$ otherwise, and $A_{ii}(t)=0$.

Equivalently, model~(\ref{eq:eqTVadj}) may be reformulated by using the Laplacian $\mathcal{L}(t)$ as follows:
\begin{equation}
	\label{eq:eqTVLapl}
	\dot{\mathbf{x}}_i=\mathbf{f}(\mathbf{x}_i)-\epsilon\sum_{j=1}^N\mathcal{L}_{ij}(t) \mathrm{B}\mathbf{x}_j.
\end{equation}

The time evolution of $\mathcal{A}(t)$ or $\mathcal{L}(t)$ is determined by the motion of the agents. They are considered to lie in a planar space of size $L$ and periodic boundary conditions. Each agent moves with velocity $\mathbf{v}_i(t)$ and direction of motion $\theta_i(t)$, i.e., $\mathbf{v}_i(t)=ve^{\iota \theta_i(t)}$. We assume that the modulus of the agent velocity $v$ is constant in time and equal for all the agents.

Let us now discuss the model for agent motion. Following the notation of Sec.~\ref{SecIISec:temporalproxgraph} we indicate with $\mathbf{y}_i(t)$ the position of the $i$-th agent in the plane at time $t$ and assume that agents perform a random walk on the plane, as in Eqs.~(\ref{eq:randomwalkers}) and incorporate in the model the possibility of long-distance jumps by including the parameter $p_{j}$, representing the probability for an individual to jump into a completely random new position. Thus, the position of each agent is updated according integrating Eqs.~(\ref{eq:randomwalkers}) with probability $1-p_{j}$, or to random coordinates in the plane with probability $p_{j}$. We will show below that the parameter $p_{j}$ plays a fundamental role for the system behavior.

From the positions of the agents at each time instant the coefficients of the time-varying adjacency matrix of the proximity graph are calculated. In particular, we have that $\mathcal{A}_{ij}(t)=1$ if $\| \mathbf{y}_i(t)-\mathbf{y}_i(t)\| \leq R$, and $\mathcal{A}_{ij}(t)=0$, otherwise.

It is here useful to begin the analysis of the model by considering the assumption that the switching among the possible topologies occurs at a fast time scale. Under the hypothesis of fast switching, the stability of the synchronization manifold of the time-varying network can be studied by applying Lemma 1 of Sec.~\ref{sec:mathtools} and, hence, by inspecting the properties of its time average.

According to this result (Ref.~\cite{stilwell2006sufficient}), when the switching is fast, the synchronous manifold in model~(\ref{eq:eqTVLapl}) is stable if the  system of coupled oscillators described by
\begin{equation}
	\label{eq:average}
	\dot{\mathbf{x}}_i=\mathbf{f}(\mathbf{x}_i)-\epsilon\sum_{j=1}^N
	\bar{\mathcal{L}}_{ij} \mathrm{B}\mathbf{x}_j
\end{equation}

\noindent supports a stable synchronization manifold and there exists a constant $\bar{T}$ such that $\frac{1}{\bar{T}}\int_t^{t+\bar{T}}\mathcal{L}(\tau)d\tau=\bar{\mathcal{L}}$.

The network in Eqs.~(\ref{eq:average}) is static and, therefore, its synchronization properties can be studied with the classical MSF approach. Thus, to perform the study of the stability of synchronization of the time-varying network inherited by mobile agents, it suffices to calculate $\bar{\mathcal{L}}$.

This calculation (detailed in Ref.~\cite{frasca2008synchronization}) gives $\bar{\mathcal{L}}=\frac{\pi r^2 \rho}{N}\mathcal{L}_{\mathcal{K}}$, where $\rho=\frac{N}{L^2}$ is the agent density and $\mathcal{L}_{\mathcal{K}}$ the Laplacian matrix of a complete graph $\mathcal{K}$. The term $\frac{\pi r^2 \rho}{N}$ represents the probability, in the limit of fast switching, that two agents are neighbors. In fact, in this limit the agents at each time instant occupy random positions in the space and the probability that a link exists is simply the fraction of the total area that is overlapped by the sensing region, that is, $\frac{\pi r^2}{L^2}=\frac{\pi r^2 \rho}{N}$.

Suppose now to consider a system having type III MSF (see Sec.~\ref{sec:mathtools} and Ref.~\cite{boccaletti2006complex}) with thresholds $\alpha_1$ and $\alpha_2$. Since the eigenvalues of the Laplacian matrix of the complete graph are $\gamma_i=N$ for $i=2,\ldots,N$, one can derive that
\begin{equation}
	\alpha_1 < \sigma \pi r^2 \rho < \alpha_2
\end{equation}

\noindent and thus
\begin{equation}
	\label{eq:msfpredictionMovingChaoticOsc}
	\frac{\alpha_1}{\pi r^2 \sigma} <  \rho < \frac{\alpha_2}{\pi r^2 \sigma}
\end{equation}

\noindent which expresses the fact that for type III MSF systems synchronization is attained if the density is in the interval $[\frac{\alpha_1}{\pi r^2\sigma},\frac{\alpha_2}{\pi r^2\sigma}]$.

With similar arguments, the condition for synchronization in type II systems can be obtained. In this case, one has that $\rho > \frac{\alpha_1}{\pi r^2 \sigma}$. This prediction by the model is particularly interesting as it is in agreement with what observed in some real systems, for instance in yeast cell populations where sustained oscillations emerge provided that the cell density is sufficiently high~\cite{dano1999sustained,dano2002synchronization}.

\begin{figure}
	\centering
	\includegraphics[scale=0.5]{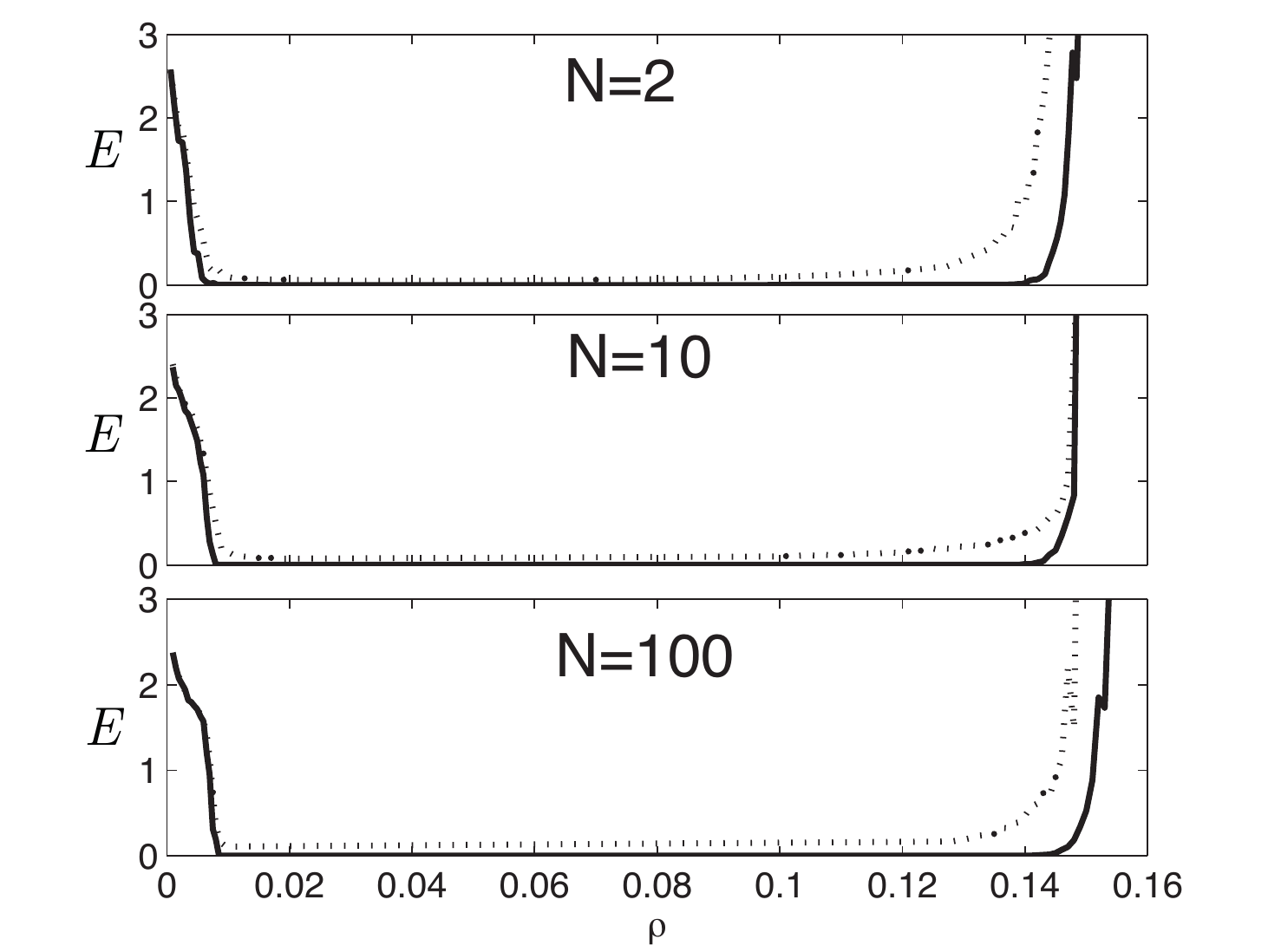}
	\caption{\label{fig:PRL08fig2}
		Synchronization error $E=\langle \sum_{i=2}^N{\frac{(|x^i_1-x^1_1|+|x^i_2-x^1_2|+|x^i_2-x^1_3|)}{3(N-1)}} \rangle_T$ vs. density $\rho$ for identical (continuous line) and non-identical (dotted line) R\"ossler oscillators with $N = 2$, $N = 10$, and $N = 100$. The coupling is fixed to $\epsilon=10$, and the other parameters to $\tau_M=10^{-3}$, $v = 1$, $R = 1$. Results are averaged over 50 realizations.
		\\ {\it Source}: Reprinted figure with permission from Ref.~\cite{frasca2008synchronization} \textcopyright~2008 by the American Physical Society.}
\end{figure}

Figure~\ref{fig:PRL08fig2} illustrates an example of synchronization in a system of moving R\"ossler type III oscillators in the regime of fast switching, showing that, as predicted by Eq.~(\ref{eq:msfpredictionMovingChaoticOsc}), the thresholds for synchronization do not depend on $N$.  Figure~\ref{fig:PRL08fig2} also shows that, while the theoretical predictions rigorously apply only to the case of identical oscillators, they still capture the behavior for non-identical agents.

According to the previous analysis, the density is a fundamental parameter to determine the conditions for synchronization. However, the other parameters are also important as they determine how fast is the switching among the possible topologies of interactions. Both the agent velocity $\nu$ and the jumping probability $p_{j}$, for instance, increase the likelihood of the fast switching assumption, as they both increase the average distance covered at each motion step, resulting in a faster switching among the possible interaction configurations.

The model discussed in this section has been extended in a three-dimensional space in Ref.~\cite{majhi2017synchronization}. Interestingly, under the hypothesis of fast switching, arguments similar to the two-dimensional case can be applied to derive the expression for the time-average Laplacian matrix $\bar{\mathcal{L}}$ that generalizes the one holding for motion in a plane. In fact, the Authors of Ref.~\cite{majhi2017synchronization} show that it is given by:

\begin{equation}
	\bar{\mathcal{L}}=\frac{V_R}{V_S}\mathcal{L}_\mathcal{K},
\end{equation}

\noindent where $V_R$ is the volume associated to the sensing region and $V_S$ the volume of the whole space. Once again, the ratio $\frac{V_R}{V_S}$ represents the probability, in the fast switching limit, that two agents are neighbors and, thus, interacting.

Another important remark is that the analysis based on the MSF, being based on linearization, provides a \emph{local} condition for stability. The study of the behavior for initial conditions far from the synchronization manifold requires the use of other techniques. A possibility, relying on numerical simulations, is the basin stability analysis~\cite{menck2013basin}, which is applied to the case of moving chaotic oscillators in Ref.~\cite{majhi2017synchronization}. The main idea is to simulate a large number of initial conditions, say $N_{sims}$, and count how many of them converge to the synchronous manifold, say $N_{sync}$. The basin stability is then calculated as $BS=N_{sync}/N_{sims}$. For large $N_{sims}$, the basin stability provides a measure of the volume of the basin of attraction of the synchronization manifold. For a time-varying network of moving R\"ossler oscillators with type II MSF, it is found, for instance, that $BS$ is typically low for coupling strength close to the transition to synchronization, and then approaches one for larger values.

When the unit dynamics are chaotic, but discrete-time~\cite{fujiwara2016synchronization}, the qualitative scenario observed does not significantly differ from continuous-time oscillators. In particular, the analysis based on the interplay between the diverse time scales in the system (see Ref.~\cite{fujiwara2011synchronization} and Sec.~\ref{sec:secDiazGuilera}) can still be applied, with the different synchronization mechanisms at work when the ratio between the time scales changes. In addition, in systems of moving chaotic maps (in particular, in Ref.~\cite{fujiwara2016synchronization} the case of the tent map is dealt with), the inclusion of noise induces a transition
between synchronization and desynchronized states for any motion rate. However, when the agent speed is low, close to the transition a switching between the quasi-synchronized and desynchronized states is found. This switching is due to the large fluctuations of the
transverse Lyapunov exponent arising because of the slow changes in the interaction topology. On the contrary, when the agent speed is large, these fluctuations are small and the Lyapunov exponent converges to that predicted by the fast switching approximation.

\subsubsection{Spatial pinning control of synchronization in moving oscillators}

As mentioned in Sec.~\ref{sec:KuramotoTN} a key result in network control is the idea of \emph{pinning control}~\cite{li2004pinning}, a technique that makes possible steering the dynamics of the entire network towards a target behavior without requiring the application of the control action to each single node of the structure. For networks with time-varying links, such as those deriving from the interaction of mobile agents, implementing a selective pinning would require, on the one hand, a non-trivial ranking on the agent topological features, and, on the other hand, controllers moving with the pinned agents.

These considerations motivated the introduction of the concept of \emph{spatial pinning control of multi-agent systems}~\cite{frasca2012spatial}. While in static networks feedback control is only applied to a fraction of nodes, in networks of multi-agent systems the control is only applied to a limited fraction of the available space.

In more detail, let us consider again the multi-agent system~(\ref{eq:eqTVLapl}) and define the control region as the square $\Gamma_c=\{(y_1,y_2):0 \leq y_1 \leq L_c, 0 \leq y_2 \leq L_c \}$ with $L_c \leq L$. We assume that control acts only on the agents that at time $t$ enter the control region, such that the dynamics of the controlled system is given by:
\begin{equation}
	\label{eq:eqTVLaplPlusControl}
	\dot{\mathbf{x}}_i=\mathbf{f}(\mathbf{x}_i)-\epsilon\sum_{j=1}^N\mathcal{L}_{ij}(t) \mathrm{B}\mathbf{x}_j -\epsilon \xi_i(t)\kappa \mathrm{B}(\mathbf{x}_i-\mathbf{s}),
\end{equation}

\noindent where $\kappa$ is the strength of the control action (measured in units of the coupling coefficient), and $\mathbf{s}(t)$ is the reference trajectory, which obeys to $\dot{\mathbf{s}}=\mathbf{f(\mathbf{s})}$. The binary variable $\xi_i(t)$ encodes the information on which agents are pinned at time $t$, that is, $\xi_i(t)=1$ if $\mathbf{y}_{i}(t)\in \Gamma_c$, and $\xi_i(t)=0$ otherwise. The control goal is to steer the agent trajectories towards the reference one, that is, the manifold defined by $\mathbf{x}_1(t)=\ldots=\mathbf{x}_N(t)=\mathbf{s}(t)$ has to be exponentially stable. To study this problem, we can define an augmented network with $N+1$ units, where agent $N+1$ is defined as $\mathbf{x}_{N+1}(t)=\mathbf{s}(t)$, see Ref.~\cite{sorrentino2007controllability}. This extra, virtual agent is added to the original system to represent the dynamics of the pinned controller. The equations for the augmented network read:

\begin{equation}
		\dot{\mathbf{x}}_i=\mathbf{f}(\mathbf{x}_i)- \epsilon \sum_{j=1}^{N+1}
		m_{ij}(t) \mathrm{B} \mathbf{x}^j,
	\label{eq:extended_system}
\end{equation}

\noindent where $i=1,\ldots,N+1$ and $M(t)=\{m_{ij}(t)\} \in \mathbb{R}^{(N+1) \times (N+1)}$ is defined as:

\begin{equation}
	M(t)=\left[
	\begin{array}{c c c c c}
		\mathcal{L}_{11} (t) + \xi_1(t) \kappa & \mathcal{L}_{12} (t) & \cdots & \mathcal{L}_{1N}(t) & -\xi_1(t) \kappa \\
		\mathcal{L}_{21} (t) & \mathcal{L}_{22} (t) + \xi_2(t) \kappa  & \cdots & \mathcal{L}_{2N}(t) & -\xi_2(t) \kappa \\
		\vdots & & \ddots & & \vdots \\
		\mathcal{L}_{N1} (t) & \mathcal{L}_{N2} (t) & \cdots & \mathcal{L}_{NN} (t) + \xi_N(t) \kappa  &  -\xi_N(t) \kappa \\
		0 & 0 & \cdots & 0 & 0
	\end{array}
	\right].
\end{equation}

Solving the control problem in the original multi-agent system is equivalent to find a stable synchronous manifold in the augmented network, such that the method illustrated in Sec.~\ref{sec:basicmodel} can be used. Here, we skip the calculations (which are detailed in Ref.~\cite{frasca2012spatial}) and illustrate the condition for pinning control for a system having type III MSF. In this case the reference evolution $\mathbf{s}(t)$ will be stable, that is, the control goal will be reached, if

\begin{equation}
	\begin{array}{l}
		\epsilon \kappa L^2_c/L^2 > \alpha_1, \\
		\epsilon (\pi R^2 \rho + \kappa L^2_c/L^2) < \alpha_2.
	\end{array}
	\label{eq:pinning_conditions}
\end{equation}

\begin{figure}
	\centering
	\includegraphics[scale=0.5]{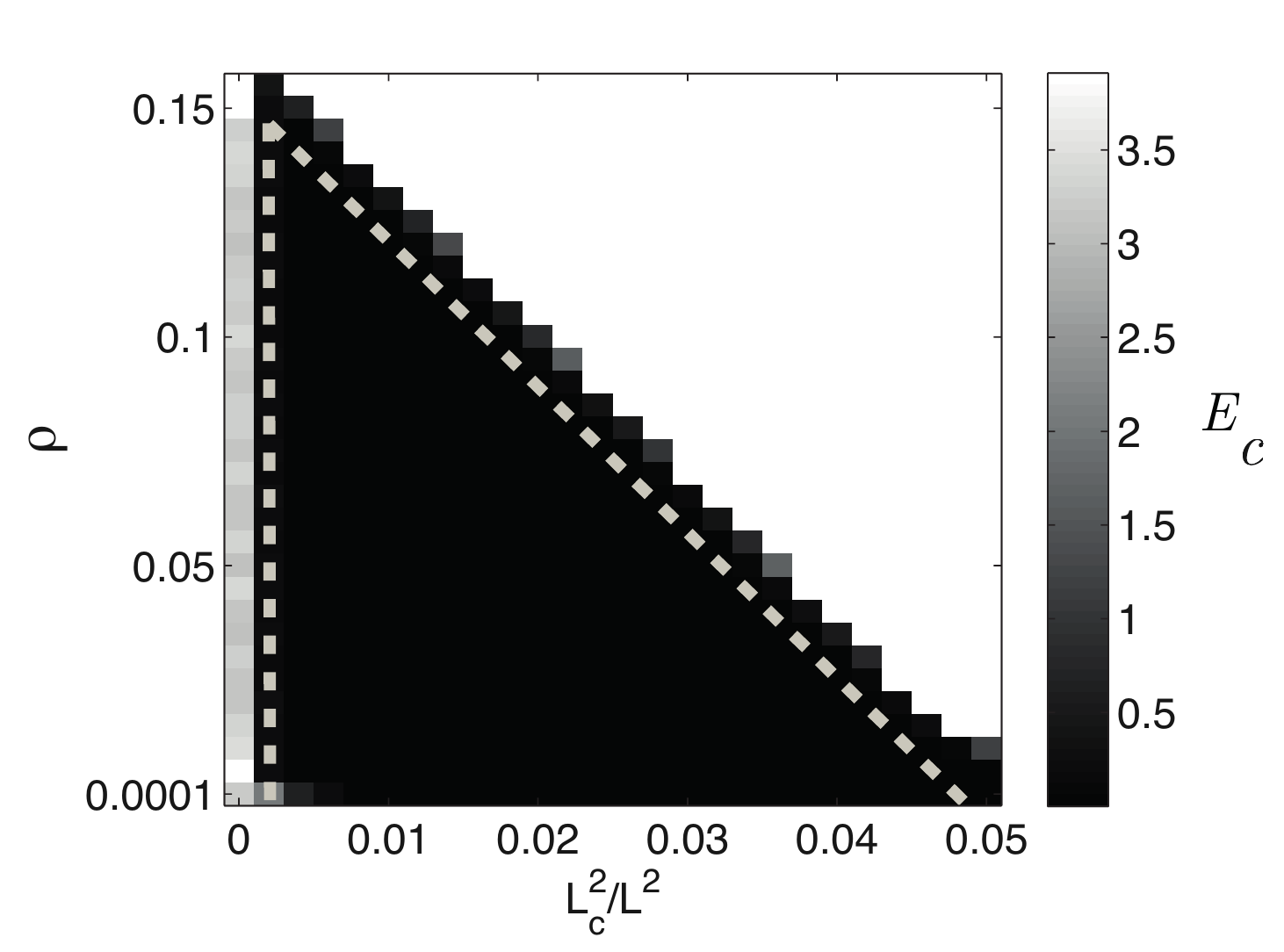}
	\caption{\label{fig:PRL12fig2} Control error $E_c=\langle\Sigma _{i=1} ^{N} \frac{(|x_{1}^{i}-\bar{s}_{1}|+|x_{2}^{i}-\bar{s}_{2}|+|x_{3}^{i}-\bar{s}_{3}|)}{3N}\rangle_T$ (in gray scale) in the plane $(L^2_c/L^2, \rho)$ for a system of moving R\"ossler oscillators under pinning control. The dashed lines indicate the theoretical prediction by Eq.~(\ref{eq:pinning_conditions}) for the stability of the target equilibrium point. The other simulation parameters have been set as $R=1$, $\epsilon=10$, $\kappa=10$, $T=500$, $\tau_M= 0.001$, $v=20$.
	\\ {\it Source}: Reprinted figure with permission from Ref.~\cite{frasca2012spatial} \textcopyright~2012 by the American Physical Society.}
\end{figure}

Suppose that the interaction radius $R$ is constant, as it is limited by the communication system equipped in the agents, then one can act on the two parameters $L^2_c/L^2$ and $\rho$ to achieve pinning control. Eq.~(\ref{eq:pinning_conditions}) shows that this objective can be reached if the control region is large enough and the agent density not too high. To illustrate this result, let us consider again a system of moving R\"ossler oscillators and assume that the goal of the control is to steer the system dynamics towards the equilibrium point $\bar{s}= \left [ \begin{array}{ccc} 0.0057 & -0.0286 & 0.0286 \end{array} \right]^T$. As shown in Fig.~\ref{fig:PRL12fig2}, there exist suitable values of the parameters (size of the control region, density, and implicitly agent speed) such that the network can be fully controlled towards a desired trajectory (in this case an equilibrium point). Furthermore, pinning control is achieved in a region of the parameter space which well corresponds to the theoretical predictions by Eq.~(\ref{eq:pinning_conditions}). This region shrinks for increasing values of the density. Finally, notice that, due to the fact that the system has a type III MSF, there are portions of the diagram where increasing the area of the control region is detrimental for stability of the target equilibrium point.

\subsubsection{Connection adaptive control}

In the previous subsection, we have illustrated a technique to steer a system of moving chaotic oscillators towards the trajectory generated by a reference extra node. In particular, when this extra node is in chaotic regime, then synchronization on a reference chaotic motion is achieved. The same control goal can be reached by using another strategy~\cite{zhou2017connection} that originates from a different assumption on the multi-agent system. In fact, in the previous section the communication network among agents is given, whereas, as we will see below, the communication network is here supposed to be adaptive, and, therefore, tuned by the control strategy. The strategy belongs to the more general class of adaptive control methods that have been successfully applied to the control of both synchronization in networks with time-independent links~\cite{de2008adaptive,de2008synchronization} and consensus in systems of moving oscillators~\cite{su2011adaptive,wang2016fully}.

The main idea of the connection adaptive control is to reconfigure at each time step the interaction matrix with the following algorithm:

\begin{itemize}
	\item Select an arbitrary, fixed in time, position for the extra node/agent.
	
	\item For each agent $i$, at each time $t$, fix a radius $R$ and calculate the agents that are at a distance smaller than $R$.
	
	\item Select, among the agents found at the previous step, the nearest one (labeled as $j$) to the extra node/agent and set $\mathcal{L}_{ij}(t)=-1$, only if the distance between agent $j$ and the extra node is smaller than that between agent $i$ and the extra node.
	
	\item If some of the previous steps cannot be accomplished, repeat them by selecting a larger radius.
	
\end{itemize}

This algorithm produces a network, which is time-varying as $v\neq 0$ and has a structure characterized by a set of directed trees. In addition, the network Laplacian matrix $\mathcal{L}(t) \in \mathbb{R}^{(N+1)\times (N+1)}$ has some special features. In fact, it can be demonstrated  that at any time its spectrum is given by: $\lambda_1=0$ and $\lambda_i=1$ for $i=2,\ldots,N+1$, see Ref.~\cite{zhou2017connection}. This property is particularly important to determine the condition for obtaining a stable synchronization manifold. In fact, the Authors  demonstrate that the condition
of stable synchronization for this time-varying structure is similar to that for a static network with the same spectrum. More specifically, for a type III system it is given by:

\begin{equation}
	\label{eq:conditionconnectionadaptivecontrol}
	\alpha_1 < \epsilon < \alpha_2
\end{equation}

\begin{figure}
	\centering
	\includegraphics[scale=0.4]{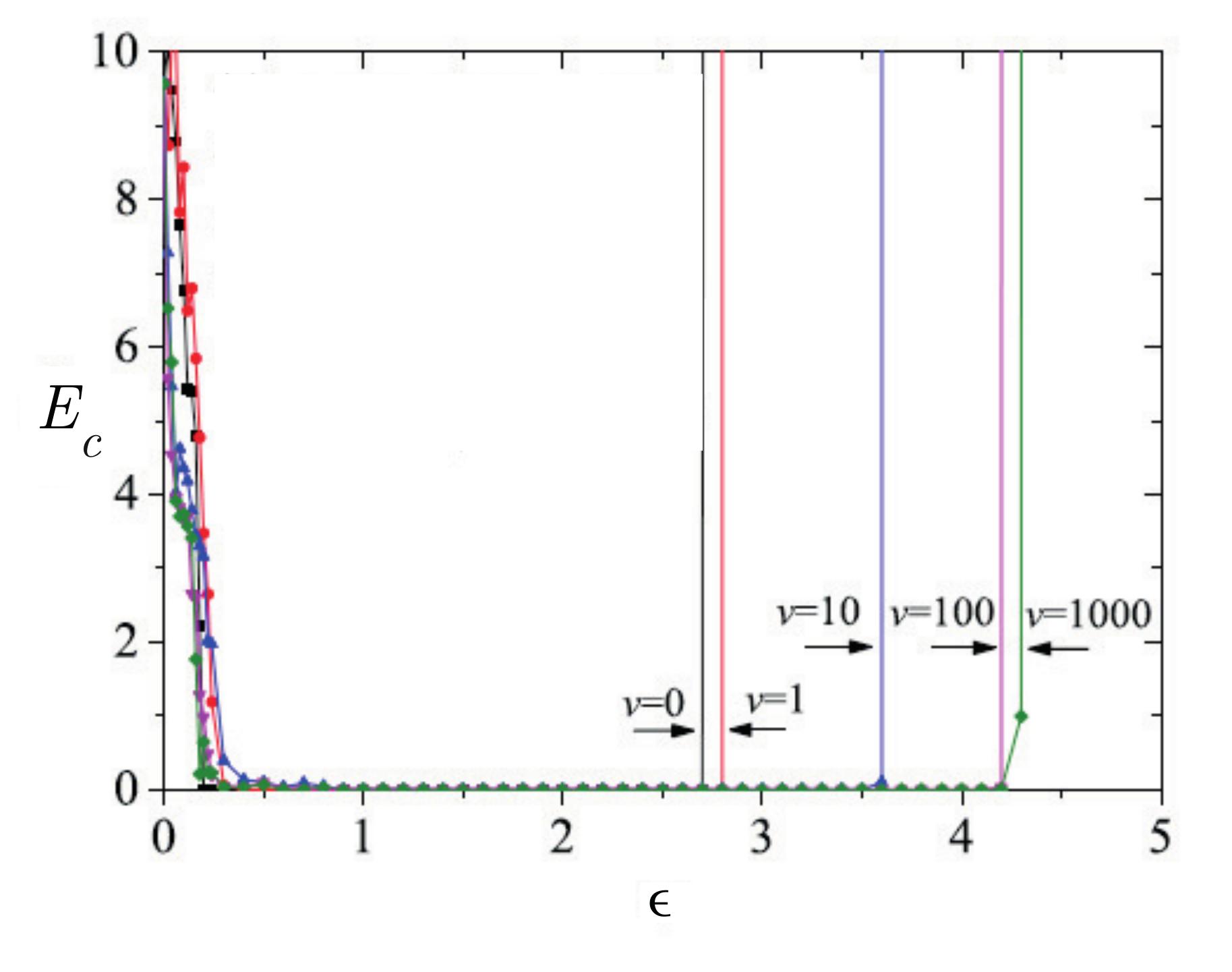}
	\caption{\label{fig:M7Zhou}
		Control error $E_c=\langle\Sigma _{i=1} ^{N} \frac{(|x_{1}^{i}-\bar{x}_1^{N+1}|+|x_{2}^{i}-{x}_{2}^{N+1}|+|x_{3}^{i}-x_{3}^{N+1}|)}{3N}\rangle_T$ vs. coupling strength for a system of moving R\"ossler chaotic oscillators with different values of agent speed $v$, after application of connection adaptive control. The extra agent is indexed as node $N+1$.
		\\ {\it Source}: Reprinted with permission from Ref.~\cite{zhou2017connection}.}
\end{figure}

From these considerations, it follows that there exists a range of the coupling strength where the multi-agent system can be controlled for any agent speed. This interval of values is clearly visible in Fig.~\ref{fig:M7Zhou} where the control error for a type III system is illustrated for different values of the agent speed. It is here interesting to note the effect of the velocity $v$. If this parameter is increased from zero, then, the range of the coupling strength yielding synchronization widens, up to the point where the prediction by Eq.~(\ref{eq:conditionconnectionadaptivecontrol}) becomes exact and a further increase of $v$ does not change anymore the thresholds for synchronization.

We can conclude that increasing the agent speed is, in general, beneficial for control. As the analysis carried out in Ref.~\cite{zhou2017connection} clarifies, this is due to the fact that the chains appearing in the time-varying interaction network are continuously broken and reformed by agent motion. Since shorter chains are easier to control and increasing the velocity is equivalent to reduce their maximum length, then agent mobility generally favors synchronization control.

Notice that in the connection adaptive control strategy, under particular circumstances (specifically, low density) it may occur that within the sensing radius there are no agents meeting the required criterion to be linked with. The solution discussed above requires to increase the radius, which, depending on the context, may not be always practical or convenient. An alternative solution is investigated in Ref.~\cite{zhou2020control}, where the Authors propose to tackle this case by allowing agent $i$ to perform a jump into a randomly chosen position where it can start again the search for a node to connect with.

Another way in which the basic control strategy may be enhanced is to partition the space where the agents move and instantiate an extra agent, i.e., a controller, in each of the cells of the partition \cite{zhu2019enhanced}. In this way, the length of the chains appearing in the interaction matrix will be reduced, favoring synchronization as lengthy chains are more difficult to control.

\subsubsection{Restricted interactions}

To move forward the idea of agents moving in a space that is not homogeneous, but, on the contrary, is characterized by regions where different dynamics take place (as it occurs, for instance, in the case of spatial pinning control), we now examine two interesting models proposed in Refs.~\cite{kim2013emergence} and~\cite{chowdhury2019synchronization}.

We begin with the model in Ref.~\cite{kim2013emergence}, where the agents are allowed to interact with each other only in a set of fixed zones in the space. Outside these zones, their dynamics is isolated. In other words, indicating with $\Gamma_h$ with $h=1,\ldots,m_z$ the $m_z$ zones of interactions, then $\mathcal{L}_{ij}(t)=-1$ whenever agents $i$ and $j$ lie in some of these zones (which do not necessarily need to be the same), i.e., whenever there exist $h_1$ and $h_2$ such that $\mathbf{y}_i(t) \in \Gamma_{h_1}$ and $\mathbf{y}_j(t) \in \Gamma_{h_2}$, where eventually (but not necessarily) $h_1$ may be equal to $h_2$. This model mimics a scenario where interactions among agents are restricted, as, for instance, it may occur in a wireless communication system where there are 'blind' areas that communication signals cannot reach~\cite{kim2013emergence}.

Remarkably,  the approach delineated in Sec.~\ref{sec:basicmodel} can still be used to investigate the question whether synchronization can be attained even under the limitations imposed by these restrictive interactions. Here, the key parameter is the \emph{probability of interaction} $p_I$ which is defined considering the ratio between the total area of the interaction zones, indicated as $A_z$, and the area of the space where the agents move, given by $L^2$, i.e., $p_I=\frac{A_z}{L^2}$. Let us, then, consider that switching among the possible interaction topologies occurs at high rate, then it can be demonstrated (see Ref.~\cite{kim2013emergence}) that, for this model, the time-average Laplacian matrix $\bar{\mathcal{L}}$ is given by:

\begin{equation}
	\label{eq:movingchaoticwithrestrictions}
	\bar{\mathcal{L}}=p_I^2 \mathcal{L}_\mathcal{K}.
\end{equation}

The expression holds for any space of dimension $d$, after replacing in the definition of $p_i$ the areas with the volumes in dimension $d$. From Eq.~(\ref{eq:movingchaoticwithrestrictions}), following the same arguments of Sec.~\ref{sec:basicmodel} one derives that for a type III system synchronization requires that

\begin{equation}
	\label{eq:pImovingchaoticwithrestrictions}
	p_I^l \leq p_I \leq p_I^u,
\end{equation}

\noindent where $p_I^l=\sqrt{\frac{\alpha_1}{\epsilon N}}$ and $p_I^u=\sqrt{\frac{\alpha_2}{\epsilon N}}$. It is noteworthy that, while in the model of Sec.~\ref{sec:basicmodel} the density
interval where synchronization can be reached does not depend on the system size $N$, in contrast, in this model the ability to achieve synchronization strongly depends on the system size as $p_I^l$ and $p_I^u$ scales with $N^{-1/2}$. Synchronization is, hence, more difficult for larger size systems. Equation~(\ref{eq:pImovingchaoticwithrestrictions}) points out another interesting feature of this model: increasing the agent capability to interact with the other units is necessarily beneficial for synchronization. Instead, the synchronization region is independent on the number of interacting regions, provided that the total volume is kept the same and the agent speed is high enough to guarantee the validity of the fast switching assumption.

At variance with the models discussed so far, in Refs.~\cite{chowdhury2019synchronizationth} and~\cite{chowdhury2019synchronization} the scenario studied requires that to interact the agents need to be in the same zone. In more detail, the model discussed in Ref.~\cite{chowdhury2019synchronizationth} encompasses a single interaction region, whereas in Ref.~\cite{chowdhury2019synchronization} a more general scenario, with $m_z$ zones, is dealt with. In this case, thus, $\mathcal{L}_{ij}(t)=-1$ whenever there exists an integer $\bar{h}$ such that $\mathbf{y}_i(t), \mathbf{y}_j(t) \in \Gamma_{\bar{h}}$.
Once again, under the hypothesis of fast switching, the method of Sec.~\ref{sec:basicmodel} may be applied to find the expression for the time-average Laplacian matrix $\bar{\mathcal{L}}$, that in this case is given by:

\begin{equation}
	\label{eq:movingchaoticwithrestrictions2}
	\bar{\mathcal{L}}=m_z \left ( \frac{A_I}{L^2}\right)^2 \mathcal{L}_\mathcal{K},
\end{equation}

\noindent where $A_I$ is the area of the interacting zone.

A very rich repertoire of dynamical features arise in this model when one further differentiates among zones with attractive interactions, i.e., with positive $\epsilon$, indicated as $\epsilon_a$, and repulsive ones, i.e., with negative $\epsilon$, indicated as $\epsilon_r$. A minimal scenario, represented by the presence of a single attractive and a single repulsive region, is investigated in Ref.~\cite{chowdhury2019synchronization} for a network of moving Lorenz chaotic systems.

\begin{figure}
	\centering
	\includegraphics[scale=0.4]{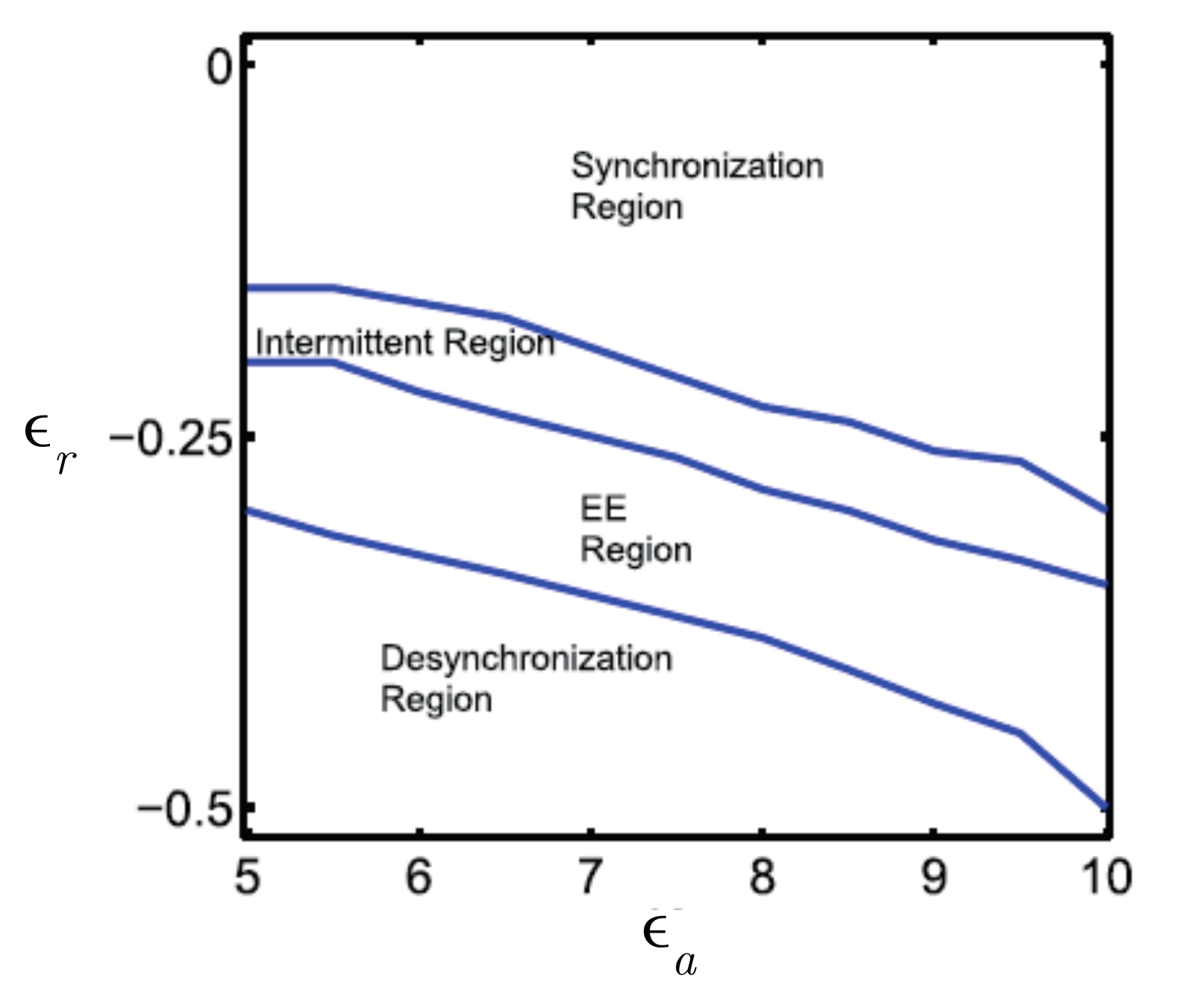}
	\caption{\label{fig:M6Perc} Phase diagram in the plane $(\epsilon_a,\epsilon_r)$ for a network of $N=100$ Lorenz chaotic oscillators moving in an area with both attractive and repulsive interaction regions. EE region indicates the region where extreme events occur.
	\\ {\it Source}: Reprinted with permission from Ref.~\cite{chowdhury2019synchronization}.}
\end{figure}

For this system four distinct dynamical regimes are observed when the two coupling parameters are varied (Fig.~\ref{fig:M6Perc}). The region of synchronization appears for small $\epsilon_r$ and large $\epsilon_a$, whereas for very negative $\epsilon_r$ the system is desynchronized. Two other, very interesting, regions are found in between the synchronization and the desynchronization region. In more details, starting from an high enough coupling strength $\epsilon_a$ and a value of $\epsilon_r$ close to zero, one may observe that decreasing $\epsilon_r$ synchronization becomes intermittent due to the effect of the repulsion zone. If $\epsilon_r$ is further decreased, then the dynamics may generate \emph{extreme events}~\cite{chowdhury2021extreme}. These are short-lasting events that appear at unpredictable times and are characterized by a very large synchronization error. Their appearance occurs at very low probability, but higher than that it would be observed if the event distribution were Gaussian.

These extreme events are associated with the sudden changes of states in the underlying complex systems, and the occurrence of extreme events often results in large social impact. They crucially appear in a variety of diverse fields like share market crashes~\cite{feigenbaum2001statistical}, electric power transmission system~\cite{dobson2007complex}, earthquakes~\cite{sornette2012dragon}, and even epileptic seizures in the human brain~\cite{albeverio2006extreme}. The model discussed in this section demonstrates that they can also occur in a system of oscillators dynamically coupled via a temporal network induced by the agent motion.

\subsubsection{Effects of motion on synchronization of moving chaotic oscillators}

So far, synchronization of chaotic oscillators in temporal networks induced by agents moving as random walkers has been dealt with. An interesting aspect to study is the effect of the type of motion on synchronization~\cite{buscarino2016interaction}. This issue can be conveniently investigated considering the Vicsek's model~\cite{vicsek1995novel}, which as a function of a single parameter can tune the agent motion from disordered (e.g., random walkers) to ordered (e.g., platoons). In the Vicsek's model the positions of the agents are updated as follows:

\begin{equation}
	\label{eq:vicsekmodel}
	\left. \begin{array}{l} \mathbf{y}_i(t_k+\tau_M)=\mathbf{y}_i(t_k)+\tau_M\mathbf{v}_i(t_k)
		\\
		\theta_i(t_k+\tau_M)=\langle \theta_i(t_k) \rangle_R + \Delta \theta_i \end{array}
	\right.
\end{equation}

\noindent where, as usual, $\mathbf{v}_i=e^{\iota \theta_i}$ indicates the agent velocity. The equation for the update of the agent heading includes the noise term $\Delta \theta_i$ , which is generated at each time step with uniform probability in the interval $[-\frac{\eta}{2},\frac{\eta}{2}]$ (being $\eta$ the noise level). The term $\langle \theta_i(t_k) \rangle_R$ represents the average direction of the units in the disc of radius $R$ of agent $i$ and is calculated as follows:

\begin{equation}
	\label{eq:mediaangolo}
	\langle \theta_i(t_k) \rangle_R = \arctan \left ( \frac{\langle \sin (\theta_i(t_k)) \rangle_R}{\langle \cos (\theta_i(t_k)) \rangle_R} \right )
\end{equation}

In this model the noise level $\eta$ modulates the type of motion, which is ordered for low values of $\eta$, and disordered for high values. To elucidate the effect of motion on synchronization of moving chaotic agents, Eqs.~(\ref{eq:eqTVLapl}) can be used with agent motion ruled by the Vicsek's model~(\ref{eq:vicsekmodel}). In fact, the motion model determines how the Laplacian $\mathcal{L}(t)$ appearing in Eqs.~(\ref{eq:eqTVLapl}) evolves in time, and so whether and how synchronization is attained.

Once again, setting an high speed $v$ entitles the application of the MSF formalism under the fast switching scenario, yielding the following condition for synchronization in the more general case of type III systems:
\begin{equation}
	\label{eq:syncconditionVicsekmodelChaotic}
	\frac{\alpha_1}{\langle k \rangle} \approx \epsilon_{c1} \leq \epsilon \leq \epsilon_{c2} \approx \frac{\alpha_2}{\langle k \rangle}
\end{equation}

\noindent where $\langle k \rangle$ denotes the average degree in the networks representing interactions in the Vicsek’s model. Note that this parameter is a function of the noise level, agent density and speed, such that the two synchronization thresholds are influenced by the motion characteristics via this parameter.

\begin{figure}
	\centering
	\includegraphics[scale=0.4]{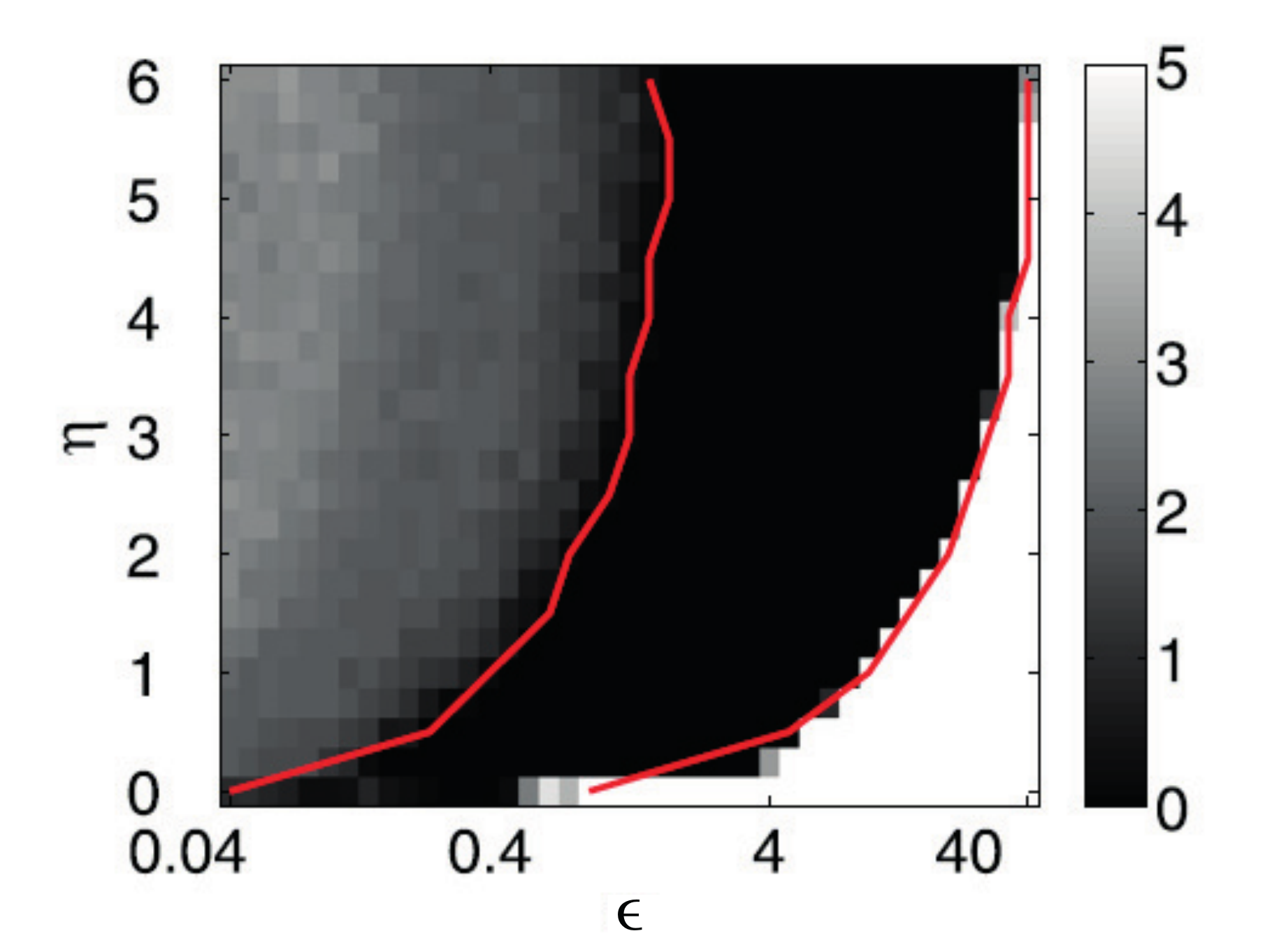}
	\caption{\label{fig:fris}
		Synchronization error $E=\langle \sum_{i=2}^N{\frac{(|x^i_1-x^1_1|+|x^i_2-x^1_2|+|x^i_2-x^1_3|)}{3(N-1)}} \rangle_T$
		as a function of the coupling strength $\epsilon$ and noise level $\eta$ in a system of R\"ossler oscillators moving according to the Vicsek's model~(\ref{eq:vicsekmodel}). The system is formed by $N=10$ agents. Other parameters are: $\rho=0.04$, $R=1$, and $v=1$. The predictions of the synchronization thresholds, represented with a red continuous line, are derived from Eq.~(\ref{eq:syncconditionVicsekmodelChaotic}), under the hypothesis of fast switching.
		\\ {\it Source}: Reprinted figure from Ref.~\cite{buscarino2016interaction}, with the permission of AIP Publishing.}
\end{figure}

The analysis of this model reveals interesting and unexpected features of the interplay between motion and synchronization dynamics. In fact, depending on the system parameters, synchronization of all chaotic oscillators may be induced either by a coordinated motion or, at the opposite, by a disordered motion. An example is shown in Fig.~\ref{fig:fris} where three regions are visible: one, occurring for small coupling strength, $\epsilon < 0.4$, where noise deteriorates synchronization, which is therefore lost in the transition from ordered to disordered motion; a second region for $0.4 < \epsilon < 40$ where noise favors synchronization, which is then associated to a disordered motion; and, finally, a third region, for $\epsilon>40$ where synchronization is not possible for any value of $\eta$.

\subsubsection{Other relevant settings}

For brevity we have not discussed the case of chaotic oscillators moving as random walkers in a finite lattice dealt with in Ref.~\cite{porfiri2006random}, but we mention here that this important study was one of the first papers posing the synchronization problem in a stochastic dynamic framework.

Several works have investigated the effect of the simultaneous presence of competitive interactions in complex networks with links fixed in time, revealing diverse stationary regimes, chimera states, and solitary waves~\cite{maistrenko2014solitary,sathiyadevi2018distinct,sathiyadevi2018stable,majhi2019solitary}. In the context of temporal networks induced by agent mobility, this issue has been investigated in Ref.~\cite{chowdhury2020distance}. In more details, the Authors of Ref.~\cite{chowdhury2020distance} have introduced a model where attractive interactions (i.e., associated with a positive coupling strength) are set for agents at a reciprocal distance greater than a threshold $d_R$, whereas repulsive interactions (i.e., associated with a negative coupling strength) for agents at a reciprocal distance smaller than the threshold $d_R$. In particular, they have studied a system of Stuart-Landau oscillators and have found that the coexistence of competing interactions can prevent the synchronous behavior that is observed when only attractive coupling is used. Instead, as a function of its parameters, the system typically displays inhomogeneous small oscillations, intermittent dynamics, and even states that can be classified as extreme events.

Another relevant configuration of moving oscillators considers the presence of interdependencies among different networked structures~\cite{majhi2019emergence}. The model consists of two layers of oscillators, where in each layer agents perform a random walk on a two-dimensional lattice, and interact with their neighbors and their replicas in the other layer. For a small intra-layer coupling strength, inter-layer synchronization without intra-layer synchronization may be observed, such that the replicas are synchronized even if each individual layer is not synchronous. Instead, for larger intra-layer coupling strength, complete synchronization is found. Quite interestingly, the phenomenon of inter-layer synchronization is found to be robust with respect to the presence of static nodes, i.e., nodes for which the mobility is inhibited in both layers.

\section{Open problems and perspectives}
\label{sec:secVII}
After this long journey inside synchronization of temporal networks, it is clear to us that this area of investigation will become crucial and attract more and more attention in nonlinear science within the years to come.
We therefore end our report with a brief discussion on what are, in our opinion, the challenges that still need to be addressed and the noteworthy routes for future research.

First of all, despite the many advances summarized in our Section \ref{sec:monolayer}, it is clear that the necessary and sufficient conditions (in terms of the critical coupling strength) for the stability of synchronization (even the least complicated complete synchronization states) in temporal networks have remained elusive so far in the general case in which one considers an arbitrary switching frequency. This, together with the more complicated case of a multilayer network with several tiers of connections where intra- and inter-layer synchronization states may occur, needs therefore an extra effort from the side of the community of network scientists.
	
In the case of mobile oscillators, we have seen in Section~\ref{sec:secV} that tuning the agents' mobility is effective in controlling synchronization (and desynchronization) in the network. This property is particularly relevant, and its potentialities in practical applications have not yet been fully exploited. For instance, one may think of making explicit use of such a feature in man made systems, as a driving paradigm at the initial moment of engineering the system itself.

As for applications, biological and technological networks seem to be the realms where the predictions of the theory may be implemented, especially in view of the fact that adaptivity seems to be the key mechanism through which biological networks operate. Yet, many questions remain open about the role and the specific features of the different adaptive mechanisms in the onset of mesoscopic and macroscopic functional behavior.
At the same time, it is certainly to be pointed out the current lack of experimental realizations of the different synchronization patterns emerging in time-varying networks. This constitutes undoubtedly a big limitation, and we are convinced that future efforts should be spent to realize controlled laboratory setups (either with electronic circuits, or with moving robots carrying electronically implemented dynamical systems) by means of which a confirmation of the distinct collective and emerging dynamical organizations predicted by the theory can be obtained, and their robustness conveniently tested.
 
Finally, another very hot topic in network science is to consider the effects of higher-order interactions, i.e., to discuss the dynamics of hypergraphs and/or simplicial complexes. This is because, from biology to social science, the functioning of a wide range of systems is the result of interactions which involve more than two constituents.
While a lot of studies have recently focused on processes and dynamics of static networks with higher-order interplay, such group interactions have not yet been considered in the context of  temporal networks. We are convinced that this will be one of the major challenges that nonlinear scientists have to deal with within the years to come, starting from giving answers to fundamental questions such as: how higher-order interactions manifest themselves in time-varying structures, how they change the functioning of such systems, and how can they be of use for engineering and/or controlling the behavior of real world networks.


\section{Acknowledgments}

The Authors would like to gratefully acknowledge R. E. Amritkar, B. Barzel, I. Belykh, V. Belykh, E. M. Bollt, J. Bragard, J. M. Buld\'{u}, J. Burguete, A. Buscarino, A. Cardillo, T. Carroll, M. Clerc, R. Criado, S. K. Dana, P. De Lellis, M. Di Bernardo, A. Diaz Guilera, R. D'Souza, E. Estrada, S. Focardi, U. Feudel, L. Fortuna, N. Frolov, L.V. Gambuzza, J. Garcia-Ojalvo, A. Garcimartin, G. Giacomelli, J. G\'{o}mez-Garde\~{n}es, S. Havlin, J. Hizanidis, P. H\"{o}vel, A. E. Hramov, S. Jafari, S. Jalan, M. Jusup, T. Kapitaniak, K. Kovalenko, S. Kurkin, J. Kurths, M. Lakshmanan, S. Lepri, V. Latora, C. Masoller, H. Mancini, D. Maza, R. Meucci, L. Minati, G. Mindlin, Y. Moreno, D. Musatov, A. Raigorovskii, S. Olmi, I. Omelchenko, O. Omel'chenko, G. V. Osipov, P. Parmananda, L. M. Pecora, M. Perc, N. Perra, A. N. Pisarchik, A. Politi, M. Porfiri, A. Prasad, R. Ramaswamy, P.L. Ramazza, R.G. Rojas, M. Romance, E. Sch\"{o}ll, M. D. Shrimali, S. Sinha, F. Sorrentino, B. Tadic, A. Torcini, J. Tredicce, S. Yanchuk, A. Zakharova, I. Zuriguel, for the many  discussions on the various subjects covered in this review article, which greatly inspired our writing.

\vskip 0.4 truecm

Work partly supported by the Department of Science and Technology, Government of India (Project no. INT/RUS/RFBR/360), the Italian Ministry of Education, Universities and Research (Project PRIN 2017 “Advanced Network Control of Future Smart Grids”, project VECTORS), the Italian Ministry of Foreign Affairs and International Cooperation  (bilateral project "Mac2Mic" between Italy and Israel), the Compagnia di San Paolo, Torino, Italy, (Starting Grant).


\bibliographystyle{unsrtK}
\bibliography{manuscript}

\end{document}